\documentclass{ingosclass}
\bibliography{bib}

\usepackage{cleveref}
\crefformat{equation}{(#2#1#3)}
\crefmultiformat{equation}
  {(#2#1#3)}
  { and (#2#1#3)}
  {, (#2#1#3)}
  {, and (#2#1#3)}
\crefrangeformat{equation}{(#3#1#4)--(#5#2#6)}

\crefformat{figure}{Fig.~#2#1#3}

\graphicspath{{images}}
\usepackage{pgfplots}
\DeclareUnicodeCharacter{2212}{−}
\usepgfplotslibrary{groupplots,dateplot}
\usetikzlibrary{patterns,shapes.arrows,positioning,arrows.meta,decorations.text}
\pgfplotsset{compat=newest}
\usepackage{transparent}
\newcommand{\mathdefault}[1]{\displaystyle #1}
\pgfplotsset{
  every axis/.style={
    label style={font=\footnotesize},
    tick label style={font=\tiny},
    title style={font=\small\bfseries},
    legend style={font=\footnotesize},
  }
}

\newenvironment{sigstatement}
{\par\medskip\textbf{Significance:}}
{\par\medskip}

\newcommand{\wrt}{w.\,r.\,t.}
\newcommand{\eg}{e.\,g.}

\newcommand{\ie}{i.\,e.}

\newcommand{\resp}{resp.}

\newcommand{\formComma}{\,\text{,}}
\newcommand{\formPeriod}{\,\text{.}}

\newcommand{\R}{\mathbb{R}}

\newcommand{\surf}{\mathcal{S}}

\newcommand{\Tcal}{\mathcal{T}} 
\newcommand{\Acal}{\mathcal{A}} 
\newcommand{\Lcal}{\mathcal{L}} 

\newcommand{\tangent}[2][]{\tensor{\operatorname{T}\!}{#1}#2}
\newcommand{\tangentS}[1][]{\tangent[#1]{\surf}}
\newcommand{\tangentR}[1][]{\tangent[#1]{\R^3\vert_{\surf}}}
\newcommand{\tangentQS}{\tensor{\operatorname{\mathcal{Q}}\!}{^2}\surf}
\newcommand{\tangentQR}{\tensor{\operatorname{\mathcal{Q}}\!}{^2}\R^3\vert_{\surf}}

\newcommand{\tangentScal}{\tangentS[^0]}

\newcommand{\dbdot}{\operatorname{:}}

\newcommand{\Dt}[1][]{\operatorname{D}^{#1}_t \!}

\newcommand{\Dlow}{\Dt[\flat]}
\newcommand{\Djau}{\Dt[\jau]}

\newcommand{\Comp}{\mathsf{C}}
\newcommand{\nablaC}{\nabla_{\!\Comp}}

\newcommand{\Tr}{\operatorname{Tr}}
\renewcommand{\div}{\operatorname{div}}
\newcommand{\Div}{\operatorname{Div}}

\newcommand{\DivC}{\Div_{\!\Comp}}

\newcommand{\Grad}{\operatorname{Grad}}
\newcommand{\GradC}{\Grad_{\Comp}}

\newcommand{\hil}{\operatorname{L}^{\!2}}

\newcommand{\inner}[3][0]{\left\langle #3 \right\rangle_{\hspace{-#1pt}#2}}
\newcommand{\normsq}[3][0]{\left\| #3 \right\|_{\hspace{-#1pt}#2}^2}

\newcommand{\paraC}{X}
\newcommand{\para}{\boldsymbol{\paraC}}
\newcommand{\normalC}{\nu}
\newcommand{\normal}{\boldsymbol{\normalC}}
\newcommand{\shopC}{I\!I}
\newcommand{\shop}{\boldsymbol{\shopC}}
\newcommand{\meanc}{\mathcal{H}}
\newcommand{\gaussc}{\mathcal{K}}

\newcommand{\Id}{\boldsymbol{Id}}
\newcommand{\IdS}{\Id_{\surf}}

\newcommand{\ofrak}{\mathfrak{o}}
\newcommand{\jau}{\mathcal{J}}

\newcommand{\eb}{\boldsymbol{e}}
\newcommand{\fb}{\boldsymbol{f}}

\newcommand{\qb}{\boldsymbol{q}}

\newcommand{\vb}{\boldsymbol{v}}
\newcommand{\wb}{\boldsymbol{w}}
\newcommand{\xb}{\boldsymbol{x}}

\newcommand{\Qb}{\boldsymbol{Q}}
\newcommand{\Rb}{\boldsymbol{R}}

\newcommand{\Vb}{\boldsymbol{V}}
\newcommand{\Wb}{\boldsymbol{W}}

\newcommand{\etab}{\boldsymbol{\eta}}

\newcommand{\sigmab}{\boldsymbol{\sigma}}      

\newcommand{\tp}{\tilde{p}}
\newcommand{\tsigmab}{\tilde{\sigmab}}

\newcommand{\vnor}{v_{\bot}}
\newcommand{\wnor}{w_{\bot}}

\newcommand{\nullb}{\boldsymbol{0}}

\newcommand{\potenergy}{\mathfrak{U}}

\newcommand{\fluxpotential}{\mathfrak{R}}

\newcommand{\energyH}{\potenergy_{\textup{H}}}

\newcommand{\EL}{\textup{EL}}
\newcommand{\energyEL}{\potenergy_{\EL}}

\newcommand{\omegaEL}{\omega_{\EL}}

\newcommand{\fbEL}{\fb_{\EL}}

\newcommand{\THT}{\textup{TH}} 
\newcommand{\energyTH}{\potenergy_{\THT}}

\newcommand{\omegaTH}{\omega_{\THT}}

\newcommand{\fbTH}{\fb_{\THT}}

\newcommand{\coeffIF}{\mu}

\newcommand{\FD}{\textup{FD}}
\newcommand{\energyFD}{\fluxpotential_{\FD}}
\newcommand{\omegaFD}{\omega_{\FD}}
\newcommand{\fbFD}{\fb_{\FD}}

\newcommand{\IM}{\textup{IM}}
\newcommand{\energyIM}{\fluxpotential_{\IM}}
\newcommand{\omegaIM}{\omega_{\IM}}
\newcommand{\fbIM}{\fb_{\IM}}

\newcommand{\NV}{\textup{NV}}
\newcommand{\energyNV}{\fluxpotential_{\NV}}

\newcommand{\omegaNV}{\omega_{\NV}}

\newcommand{\fbNV}{\fb_{\NV}}

\newcommand{\IE}{\textup{IE}}
\newcommand{\energyIE}{\fluxpotential_{\IE}}
\newcommand{\fbIE}{\fb_{\IE}}
\newcommand{\omegaIE}{\omega_{\IE}}

\newcommand{\VP}{\textup{VP}}
\newcommand{\energyVP}{\fluxpotential_{\VP}}
\newcommand{\fbVP}{\fb_{\VP}}
\newcommand{\omegaVP}{\omega_{\VP}}

\newcommand{\deltafrac}[2]{\frac{\delta #1}{\delta #2}}
\newcommand{\ddfrac}[2][]{\frac{\textup{d} #1}{\textup{d} #2}}
\newcommand{\ddt}[1][]{\ddfrac[#1]{t}}

\newcommand{\dS}{\textup{d}\surf}
\newcommand{\dt}{\textup{d}t}

\newcommand{\meanelastic}{\kappa}
\newcommand{\gausselastic}{\kappa_{\textup{G}}}

\newcommand{\dwpotential}{u_{\THT}}

\newcommand{\DiscSpace}[1]{V_{\!#1}}
\newcommand{\DiscSpaceVector}[1]{\boldsymbol{V}_{\!#1}}
\newcommand{\Cell}{T}

\newcommand{\meshparam}{h}
\newcommand{\norm}[1]{\lVert #1 \rVert }

\newcommand{\Inner}[3][]{\left({#2}\,,\,{#3}\right)\ifthenelse{\equal{#1}{}}{}{_{#1}}}
\newcommand{\InnerApprox}[2]{\Inner[\meshparam]{#1}{#2}}

\newcommand{\paramUpdate}{\mathbf{Y}}
\newcommand{\paramUpdateTest}{\mathbf{Z}}
\newcommand{\amdis}{\texttt{AMDiS}}
\newcommand{\dune}{\textsc{Dune}}
\newcommand{\alugrid}{\textsc{Dune-alugrid}}
\newcommand{\curvedgrid}{\textsc{Dune-curvedgrid}}
\newcommand{\functions}{\textsc{Dune-functions}}
\newcommand{\eigen}{Eigen}

\title{Liquid crystalline order and its impact on shape evolution of fluid lipid membranes}

\author[1]{Ingo Nitschke}
\author[1]{Maik Porrmann}
\author[1,2,3,*]{Axel Voigt}

\affil[1]{Institut f{\"u}r Wissenschaftliches Rechnen, Technische Universit{\"a}t Dresden, 01062~Dresden, Germany}
\affil[2]{Center for Systems Biology Dresden (CSBD), Pfotenhauerstr.~108, 01307~Dresden, Germany}
\affil[3]{Cluster of Excellence Physics of Life (PoL), Technische Universität Dresden, 01062~Dresden, Germany}

\affil[*]{Correspondence: axel.voigt[at]tu-dresden.de}

\ihead{Ingo Nitschke, Maik Porrmann, and Axel Voigt}

\begin{document}

\maketitle


\begin{abstract}
We consider extensions of the classical curvature-elasticity free energy for a fluid lipid membrane, explicitly treating the liquid crystal structure of the membrane, which results from the lipid molecules that are, on average, oriented normal to the surface. Locally varying liquid crystal order leads to local variations in bending rigidities and surface viscosity and thus influences the shape evolution. This provides a multiscale coupling between the local membrane mechanics on the level of the lipids and the mesoscopic length and time scales of biological functions. Numerical simulations in the one-constant approximation demonstrate the impact of local membrane mechanics on the emerging equilibrium shapes. These shapes not only depend on the specified conserved surface area and enclosed volume, as for the classical curvature-elasticity free energy, but also the elastic parameter and are characterized by a reduced liquid crystal order at highly curved regions. The modified shapes and their varying liquid crystal order, as function of curvature, provide experimentally testable properties. Exploring the dynamic evolution demonstrate a tight coupling of tangential flow, shape evolution and liquid crystalline order. We further point to model extensions towards lipid phase separation and asymmetric lipid bilayers and biological implications.
\end{abstract}

\begin{sigstatement}
Multi-scale approaches for fluid lipid membranes couple liquid crystalline order and shape evolution and link the microscopic details of lipid order to the mesoscopic length and time scales of biological functions. The coupling of scales is realized by a mechanical feedback mechanism, in which geometric properties of the membrane have an impact on local liquid crystalline order and local liquid crystalline order can lead to local variations in bending rigidities and surface viscosity, and thus influence the shape evolution. Extrapolating the results for symmetric fluid lipid membranes to multi-component membranes, asymmetric lipid bilayers, and lipid-protein interactions will be key to understanding the role of local membrane mechanics in vital biological functions.
\end{sigstatement}

\section*{Introduction}

In \cite{Helfrich_ZfNC_1973} Helfrich proposed a curvature-elasticity free energy for a fluid lipid membrane which has become the standard continuum theory for describing such membrane shapes. Helfrich treated the membrane as a two-dimensional fluid with liquid crystalline order. This allowed him to apply the principles of liquid crystal elasticity to describe the membrane’s resistance to bending. The resulting free energy reads
\begin{equation} \label{eq:helfrichenergy}
\energyH = \int_\surf \frac{\meanelastic}{2} (\meanc - \meanc_0)^2 + \gausselastic \gaussc \,\dS
\end{equation}
with $\surf$ a closed surface representing the membrane, $\meanc$ its mean curvature, $\gaussc$ its Gaussian curvature, and $\meanc_0$ the spontaneous curvature, $\meanelastic$ the bending rigidity and $\gausselastic$ the Gaussian bending rigidity, which Helfrich refers to as the curvature-elastic moduli.
The $\frac{\meanelastic}{2}\meanc^2$ and $\gausselastic \gaussc$ energy contributions are analogous to the Frank--Oseen energy of the underlying liquid-crystal theory and represent the splay and saddle-splay terms, respectively. The spontaneous curvature $\meanc_0$ results from a possible asymmetry between inner and outer leaflet of the lipid bilayer.
Additional complexity arises from the inextensibility of the membrane and the osmotic pressure balance between the cell and its surroundings that are mathematically modeled by enforcing local inextensibility of $\surf$ directly or conservation of global area $A = \int_\surf \dS$, and conservation of enclosed volume $V = \frac{1}{3} \int_\surf \mathbf{x} \cdot \normal \dS$, where $\normal$ denotes the normal to the surface $\surf$. If $\gausselastic$ is constant, the Gauss--Bonnet theorem implies that the integral of $\gausselastic \gaussc$ is a topological constant. For symmetric lipid bilayers, \ie, in the absence of any local distinction between the two sides of the surface, it also holds that $\meanc_0 = 0$. In the resulting setting, in which also $\meanelastic$ is assumed to be constant, minimizing the energy under the constraints provides a shape equation, which has been extensively studied in the axisymmetric setting, see e.g.,  \cite{PhysRevE.47.461,PhysRevE.49.4728,seifertShapeTransformationsVesicles1991}. Numerical solutions lead to the classical Seifert shapes, which only depend on the reduced volume $V_r = {6 \sqrt{\pi} V(t)} / {A(t)^{3/2}}$. The existence of such minimizers can also be rigorously proven \cite{mondino2020existence,kusner2023canham}. The physical background, the theoretical foundations and the computational results contributed to the success of the continuum theory of \cite{Helfrich_ZfNC_1973} and its wide applicability towards describing biological functions on mesoscopic length and time scales, see \cite{deserno2015fluid} for a review.

\begin{figure}[tbp]
	\centering
	\newlength{\spu}
	\setlength{\spu}{0.142857\textwidth} 
	\setlength{\spu}{0.95\spu} 
	\begin{tikzpicture}[x=\spu,y=\spu,node distance=0,font=\small] 
		\coordinate (TK1) at (1,0);
		\coordinate (TK2) at (2,0);
		\coordinate (TK3) at (3,0);
		\coordinate (TK4) at (4,0);
		\coordinate (TK5) at (5,0);
		\coordinate (TK6) at (6,0);

		\draw (TK1) -- (TK6);
		\draw[dotted] (0,0) -- (TK1);
		\draw[dotted] (7,0) -- (TK6);
		\node[yshift=-0.15\spu,xshift=0.5\spu] at (TK3) {space [$\mu$m]};
		\foreach \tkpos/\tklabel in {TK1/10,TK2/1,TK3/0.1,TK4/0.01,TK5/0.001,TK6/0.0001}
		{
			\draw (\tkpos) -- ++(0,0.1);
			\node[yshift=0.2\spu] at (\tkpos) {$\tklabel$};
		}

    \node[yshift=0.38\spu] at (TK6) {\emph{covalent bonds}};
    \node[yshift=0.38\spu] at (TK5) {\emph{lipid molecules}};
    \node[yshift=0.38\spu,xshift=0.5\spu] at (TK3) {\emph{membranes}};
    \node[yshift=0.38\spu,xshift=0.5\spu] at (TK1) {\emph{cells/vesicles}};

    \node(PV) [above=0.6\spu of TK1,xshift=0.5\spu] {\includegraphics[width=2\spu]{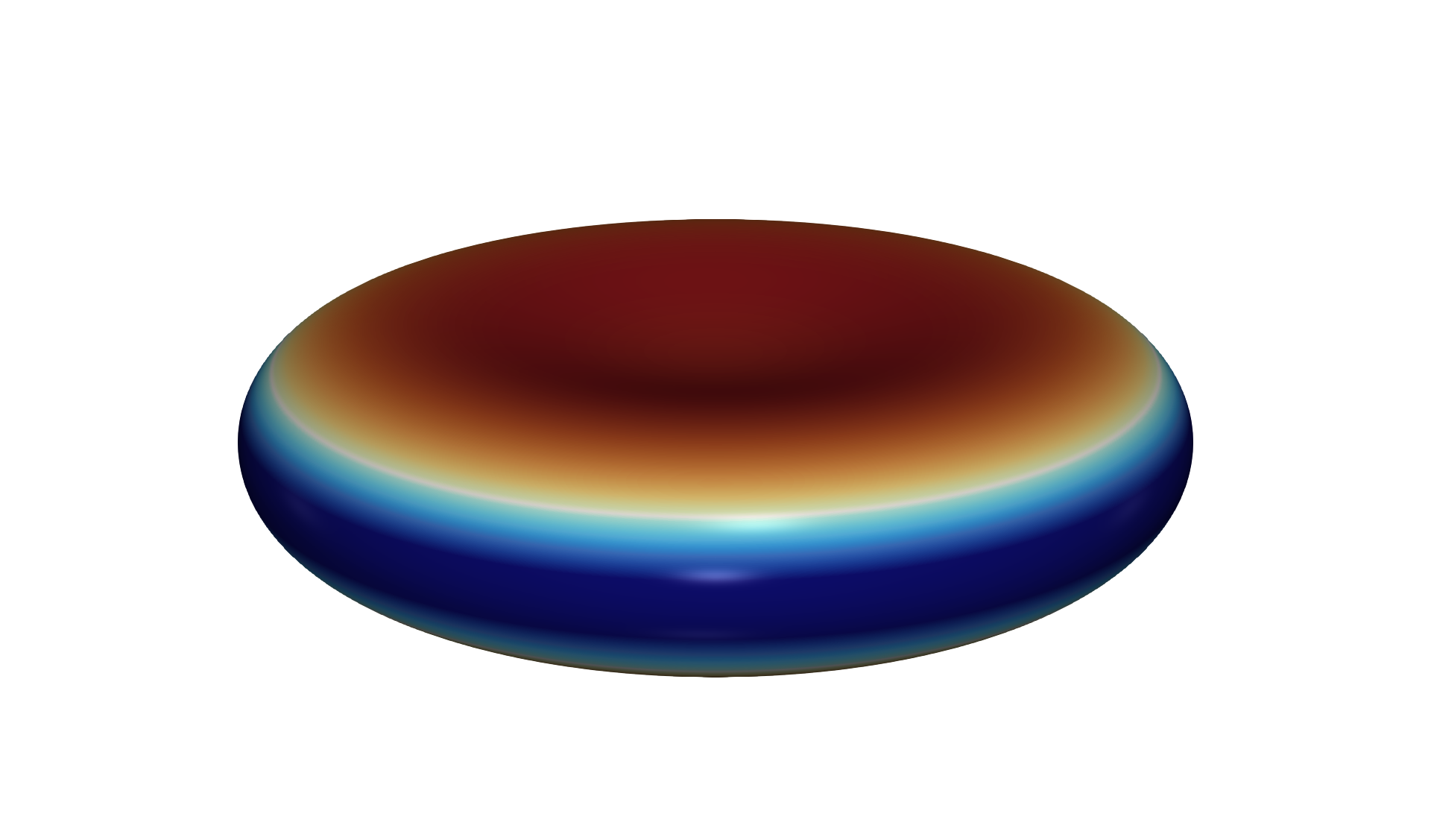}};
    \node(B) [above=0.1\spu of PV] {\includegraphics[width=1.4\spu]{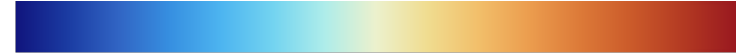}};
        \node[yshift=-0.03\spu] at (B.south) {$\beta$};
        \node[yshift=-0.03\spu,anchor=west] at (B.south west) {$0.5$};
        \node[yshift=-0.03\spu,anchor=east] at (B.south east) {$0.667$};
		\node[yshift=+0.03\spu, font=\footnotesize] at (B.north) {less ordered $\longleftrightarrow$ fully ordered};
        \draw ([xshift=-0.0\spu,yshift=0.12\spu]B.north west) rectangle ([xshift=0.0\spu,yshift=-0.12\spu]B.south east);

    \node(PCC) [above=0.6\spu of TK6,xshift=-0.05\spu] {\includegraphics[width=0.9\spu]{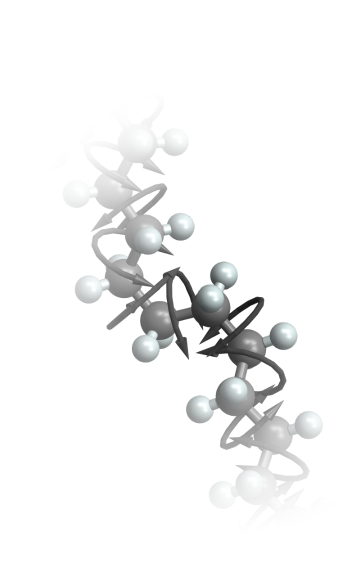}};

	\node(PMOL) [above=0.4\spu of TK5] {\includegraphics[width=1.05\spu]{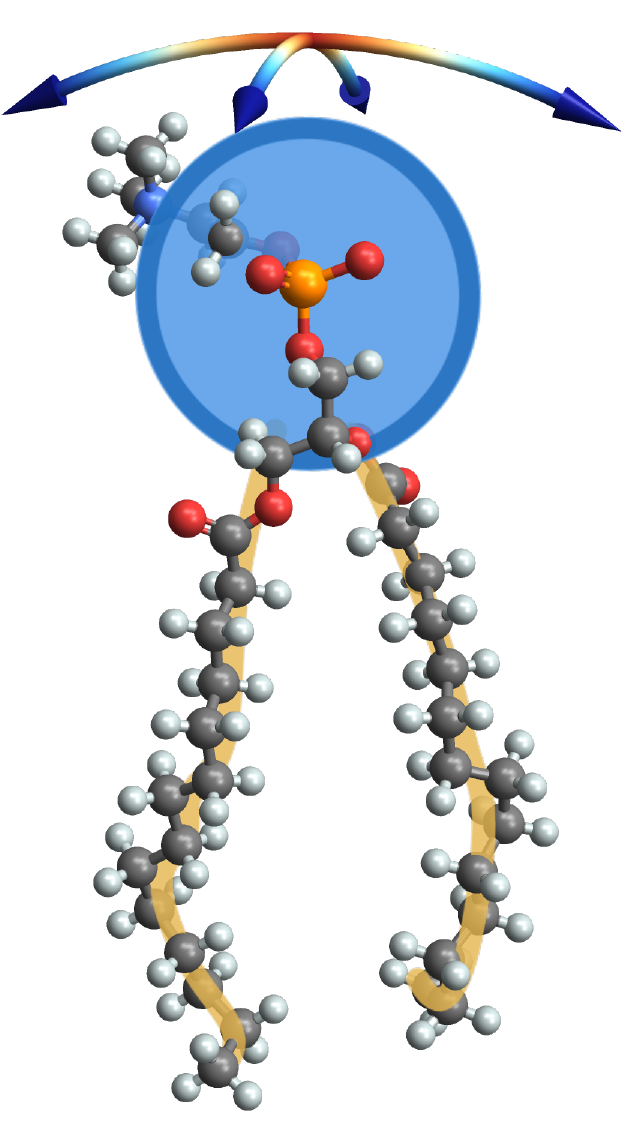}};
		\node at ([yshift=-0.05\spu]PMOL.north) {$\theta$};

    \node(PL) [above=0.33\spu of TK3,xshift=0.5\spu] {\includegraphics[width=2\spu]{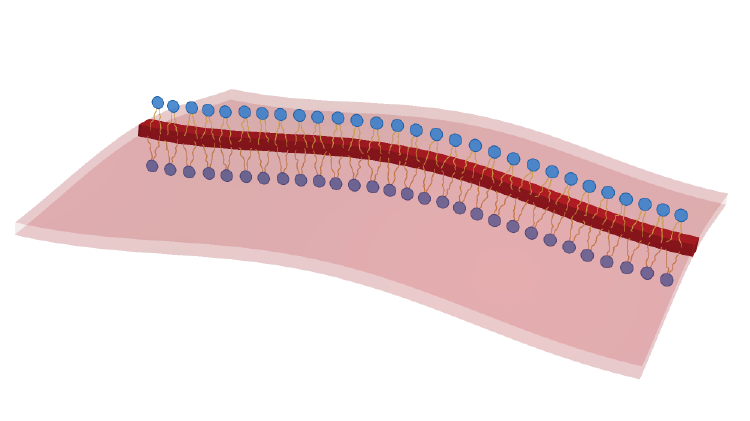}};
      \node[anchor=west,yshift=-0.25\spu,xshift=-0.1\spu] at (PL.west) {Helfrich};

    \node(PLB) [above=-0.19\spu of PL] {\includegraphics[width=2\spu]{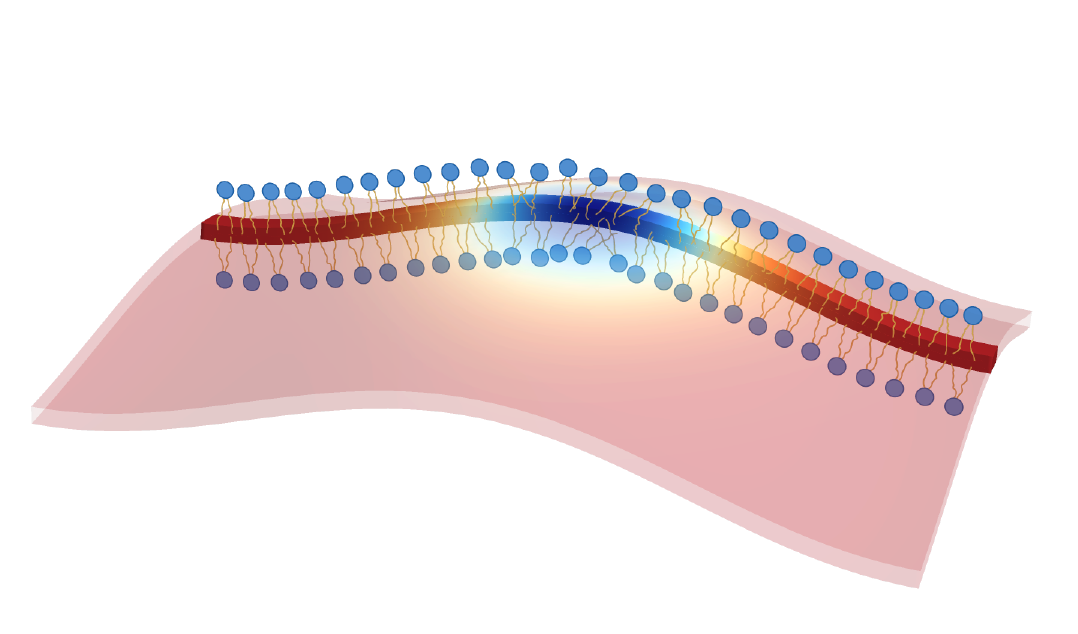}};
      \node[anchor=west,yshift=-0.35\spu,xshift=-0.1\spu] at (PLB.west) {Beris--Edwards--Helfrich};

    \draw[-{Stealth},thick]
      ([xshift=0.3\spu]PMOL.west) to[out=177, in=-90]  ([yshift=0.02\spu,xshift=0.2\spu]PLB.center);
    \node[anchor=south east,rotate=-11,font=\footnotesize] at  ([xshift=0.3\spu]PMOL.west) {coarse-graining};

    \draw[-{Stealth},thick]
    	([xshift=0.3\spu]PMOL.west) to[out=180, in=90]  ([yshift=0.3\spu,xshift=0.1\spu]PL.center);
    \node[anchor=north east,rotate=-7,font=\footnotesize] at  ([xshift=0.3\spu]PMOL.west) {(fully ordered)};

    \draw[dashdotted] ([xshift=0.05\spu,yshift=-0.1\spu]PCC) ellipse ({0.4\spu} and {0.8\spu});
    \draw[dashdotted] ([xshift=0.27\spu,yshift=-0.35\spu]PMOL) ellipse ({0.15\spu} and {0.3\spu});
    \draw[] ([xshift=-0.35\spu,yshift=-0.1\spu]PCC.center) -- ([xshift=0.42\spu,yshift=-0.35\spu]PMOL.center);
	\end{tikzpicture}
	\caption{Schematic description of spatial scales linking lipid orientation to membrane shape evolution.
			From right to left, the schematic spans the scales of covalent bonds, individual lipid molecules, fluid lipid membranes, and entire cells or vesicles.
      A DMPC molecule is used for visualization purposes only. The multiscale approach neglects the effects of atomic bonds, but account for the molecular orientation of individual lipids.
      Assuming that the molecular orientation is, on a local spatial average, perpendicular to the surface representing the lipid membrane,
      the angle $\theta$ describes the deviation from this average. This determines a local order parameter $\beta = \langle \cos^2 \theta - \frac{1}{3}\rangle$, where $\langle\cdot\rangle$ denotes the average \wrt\ a statistical distribution of $\theta$.
      Spatial coarse-graining yields the scalar order field $\beta$. We consider the same color bar for the local order parameter on the scale of the lipid molecules and the coarse-grained scalar order field on the scale of the membranes and cells or vesicles. The impact of the scalar order field $\beta$ on the bending rigidity and surface viscosity of the fluid lipid membrane, is shown by larger deformations for the Surface Beris-Edwards-Helfrich model, considered in this work, compared with the special case of a fully ordered state, $\beta\equiv\frac{2}{3}$, the ansatz adopted by Helfrich in \cite{Helfrich_ZfNC_1973} for a symmetric bilayer. These dependencies influence the global geometry and dynamics of cells or vesicles, here indicated only by an equilibrium shape with varying scalar order field $\beta$ obtained by solving the Surface Beris--Edwards--Helfrich (SBEH) model.
			}
	\label{fig:scales}
\end{figure}

However, the relation to biological functions on mesoscopic length and time scales remains qualitative. Various experimental techniques have been used to examine the dependence between structure and mechanics. While several studies confirm the dependence of the bending rigidity as a function of local membrane properties \cite{BASSEREAU201447,dimova2014recent}, others show striking deviations from the proposed behavior \cite{PhysRevE.80.021931,gracia2010effect,doktorova2017determination,chakraborty2020cholesterol}. In order to reveal this discrepancy, let us critically review the derivation considered in \cite{Helfrich_ZfNC_1973}. Helfrich has been very explicit on the assumptions he considered to derive the curvature-elasticity free energy \cref{eq:helfrichenergy}. For example, he assumes the liquid crystal to be in the fully ordered state with the director pointing in the direction of the surface normal. In the meantime various experimental studies have shown that lipids do not need to be fully ordered \cite{doktorova2017determination,chakraborty2020cholesterol} and that liquid crystalline order has an impact on the bending properties \cite{chakraborty2020cholesterol}. These effects are not considered in \cref{eq:helfrichenergy} and it remains open, whether and to which extent this additional degree of freedom influences the equilibrium shapes. Another strong assumption in \cite{Helfrich_ZfNC_1973} is the treatment of the fluidity of the lipid membrane. Helfrich assumes an unrestricted internal fluidity and simply considers a vanishing shear modulus. In the meantime not only bending properties have been measured but also surface viscosity in fluid lipid membranes \cite{honerkamp2013membrane,faizi2022vesicle}. While not important, if only the equilibrium shape is considered, experimental and computational studies \cite{faizi2022vesicle,arroyo2009relaxation,torres2019modelling,reuther2020numerical,olshanskii2023equilibrium,PBV25} have confirmed that surface viscosity has an impact on the shape evolution. However, these continuous models for the dynamic evolution of fluid lipid membranes are far from being predictive, which is also partially due to limited experimental data on the dynamics. A model which accounts for liquid crystalline order and surface hydrodynamics still remains an open problem. Given the significance of the bending rigidity and surface viscosity on mesoscopic length and time scales, a deeper understanding of the role of local membrane mechanics will be key to further exploring biological functions, that occur on mesoscopic scales. This paper addresses this by suggesting a multiscale approach, which accounts for the properties of the lipids on the shape evolution of fluid lipid membranes. The different scales and their coupling are illustrated in \cref{fig:scales}.

We will first review dynamic approaches based on the curvature-elasticity free energy \cref{eq:helfrichenergy}, including constrained gradient flows and models which explicitly account for the solid-fluid duality of fluid lipid membranes. While the last consider surface viscosity, these models still treat the fluid lipid membrane as an effectively isotropic thin material. In order to also consider varying liquid crystalline order, we here aim to combine these models with an explicit treatment of the liquid crystal structure of the fluid lipid membrane. We consider a Surface Beris-Edwards--Helfrich (SBEH) model, which accounts for the local membrane mechanics of the surface liquid crystal and the surface viscosity. The model results as a special case of a general Surface Beris--Edwards--Helfrich (SBEH) model \cite{nitschke2025beris,NV24}, where the unknowns are a surface Q-tensor $\Qb(\mathbf{x},t)$, the surface velocity $\Vb(\mathbf{x},t)$, the surface pressure $p(\mathbf{x},t)$ and the parametrization of the surface $\para(t)$. Considering $\Qb = \beta (\normal\otimes\normal - \frac{1}{2}\IdS)$ with $\beta$ the nematic order parameter for orientation in normal direction, allows to derive equations for $\beta(\mathbf{x},t)$, $\Vb(\mathbf{x},t)$, $p(\mathbf{x},t)$ and $\para(t)$. The resulting model contains a bending force similar to the Helfrich models, but its parameters are not merely determined by the elastic constants but scaled by $\beta$, and a nematic viscosity, again depending on $\beta$. The model thereby provides a multi-scale coupling, directly linking the microscopic details of lipid order to the mesoscopic length and time scales of biological functions. We relate the approach to the classical curvature-elasticity free energy for fluid lipid membranes \cite{Helfrich_ZfNC_1973} and explore the impact of the additional degrees of freedom on the dynamic evolution and the equilibrium shape, providing experimentally testable predictions.

\section*{Methods} \label{sec:methods}

\subsection*{Dynamic approaches based the curvature-elasticity free energy}

Besides solving the Euler--Lagrange equations to obtain the Seifert shapes \cite{PhysRevE.47.461,PhysRevE.49.4728,seifertShapeTransformationsVesicles1991}, dynamic approaches have been also considered. They address the $L^2$-gradient flow of the energy with the constraints incorporated by Lagrange multipliers. Under the above assumptions, and in case of local inextensibility and conservation of enclosed volume, the problem reads: Given an initial (closed, simply connected) surface $\surf(0)$ via a parametrization $\para(0)$, determine the surface velocity $\Vb=\Vb(\xb,t)$, the Lagrange function $p=p(\xb,t)$, the Lagrange parameters $\lambda=\lambda(t)$, and the parametrization $\para=\para(t)$ by solving
\begin{subequations}\label{eq:L2Helfrich}
\begin{align}
     \nabla p + p \meanc \normal + \lambda \normal + \Vb &= f_B \normal \label{eq:L2Helfrich_1}\\
     \DivC \Vb &= 0 \\
     \int_\surf \Vb \cdot \normal \, \dS &= 0
\end{align}
\end{subequations}
on $\surf=\surf(t)$ given by $\para(t)$ and $\partial_t\para = (\Vb \cdot \normal) \normal$, given an initial condition for $ \Vb $.
Above we have introduced the bending force $f_B$, obtained from variation of \cref{eq:helfrichenergy}, and explicitly given by
\begin{align} \label{eq:variationaderivative}
    f_B &= -\frac{\delta \energyH}{\delta \para} \cdot \normal = -\meanelastic (\Delta\meanc + \frac{1}{2}\meanc^3 - 2 \meanc \gaussc) \;.
\end{align}
Above, $\Delta = \div \nabla$ denotes the Laplace--Beltrami operator on $\surf$ and $\div$ and $\nabla$ the covariant derivatives, and $\DivC := \Tr\circ\nablaC$ is the componentwise trace divergence with $ \nablaC = (\IdS\nabla_{\R^3})\vert_{\surf} $, where $\IdS$ is the surface identity tensor field. For further details on the notation, we refer to the corresponding section of the Appendix, while for a detailed discussion on the variational derivative $\frac{\delta \bullet}{\delta \para}$, we refer to \cite{NITSCHKE2022104428,nitschke2023tensorial,nitschke2023tangential}. The volume constraint is realized through the Lagrange multiplier $\lambda$, and for local inextensibility, an equation for the Lagrange function $p$ needs to be solved. Approximating local inextensibility by global area conservation, the problem reduces to: Given an initial (closed, simply connected) surface $\surf(0)$ via a parametrization $\para(0)$, determine the surface velocity $\Vb=\Vb(\xb, t)$, the Lagrange parameters $p=p(t)$ and $\lambda=\lambda(t)$, and the parametrization $\para=\para(t)$ by solving
\begin{subequations}\label{eq:area}
\begin{align}
     p \meanc + \lambda  + \Vb \cdot \normal &= f_B  \label{eq:area+volume_1} \\
     \int_\surf \Vb \cdot \normal \, \meanc \, \dS &= 0 \label{eq:area+volume_2}\\
     \int_\surf \Vb \cdot \normal \, \dS &= 0 \label{eq:area+volume_3}
\end{align}
\end{subequations}
on $\surf=\surf(t)$ given by $\para(t)$ and $\partial_t \para = (\Vb \cdot \normal) \normal$.
Now both constraints are realized by Lagrange multipliers, $p$ for the area constraint and $\lambda$ for the constraint on the enclosed volume. Both formulations significantly differ, as the latter only considers the geometric evolution of the surface, neglecting tangential motions of the material. While the resulting equilibrium shapes are the same, the evolutions differ. For the last problem, \cref{eq:area,eq:variationaderivative}, which is a classical geometric evolution equation with non-local components, short and long time existence of solutions and convergence towards the equilibrium shapes can be mathematically proven \cite{rupp2026gradient}. Similar results for the first problem \cref{eq:L2Helfrich,eq:variationaderivative}, which is a coupled geometric evolution and surface partial differential equation problem, are not known. However, a huge literature exists on numerical schemes solving both problems, see e.g., \cite{du2004phase,barrett2008parametric,elliott2010modeling,aland2014diffuse}. 

A model, which accounts for the solid-fluid duality of fluid lipid membranes more explicitly, is the Surface Stokes--Helfrich (SSH) model \cite{arroyo2009relaxation,torres2019modelling}: Given an initial (closed, simply connected) surface $\surf(0)$ via a parametrization $\para(0)$, the unknowns are the surface velocity $\Vb(\xb, t)$, the surface pressure $p(\xb, t)$, the Lagrange multiplier $\lambda(t)$, and the parametrization $\para(t)$, which are obtained by solving
\begin{subequations}\label{eq:u-p}
\begin{align}
    \nabla p + p \meanc \normal -  \mu\DivC \sigmab + \lambda \normal + \gamma \Vb&= f_B \normal  \label{eq:u-p-mb} \\
    \DivC \Vb &= 0 \label{eq:u-p-inext}\\
    \int_\surf \Vb \cdot \normal \, \dS &= 0 \label{eq:volume_conservation}
\end{align}
\end{subequations}
on $\surf = \surf(t)$ given by $\para(t)$ and $\partial_t \para = (\Vb \cdot \normal) \normal$, given an initial condition for $ \Vb $. Here, $\sigmab := \IdS\nablaC\Vb + (\IdS\nablaC\Vb)^T = \nabla\vb + \nabla^T\vb - 2\vnor\shop$, with shape operator $ \shop = -\nablaC\normal $, is the surface rate of deformation tensor, where $\Vb  = \vb +  \vnor \normal$ has been decomposed into its tangential $\vb = \IdS \Vb$ and normal $\vnor = \Vb \cdot \normal$ part. Furthermore $\mu > 0$ denotes the surface viscosity, and $\gamma > 0$ is the friction coefficient. The volume constraint is realized through the Lagrange multiplier $\lambda$. The area constraint is established by the inextensibility condition \cref{eq:u-p-inext} and the pressure $p$, serving as the Lagrange function. The bending force $f_B \normal$ still results from \cref{eq:variationaderivative}. These equations provide a more realistic approach for the dynamic evolution of fluid lipid membranes. However, the equilibrium shape is still determined by minimizers of \cref{eq:helfrichenergy} under the discussed constraints. In the friction-dominated limit, the previously considered $L^2$-gradient flow \cref{eq:L2Helfrich,eq:variationaderivative} is obtained after appropriate rescaling, see \cite{bachini2023derivation} for a detailed analysis. Theoretical results for the Surface Stokes--Helfrich (SSH) model are not yet known, and also numerical schemes to solve the equations are more recent \cite{reuther2020numerical,krause2023numerical,olshanskii2023equilibrium,sauer2025curvilinear,sahu2025arbitrary,PBV25,garcke2025parametric,igel2026streamfunctionformulationsurfacestokeshelfrich}.

All models discussed so far are based on the curvature-elasticity free energy \cref{eq:helfrichenergy} and therefore consider bending properties which result from the assumption of an underlying fully ordered liquid crystal.

\subsection*{Surface Beris--Edwards--Helfrich (SBEH) models}

In order to weaken this assumption and allow for a dynamically varying liquid crystalline structure, we derive a more general model, which contains \cref{eq:helfrichenergy} and the above dynamic models as special cases. Instead of the surface Frank--Oseen energy considered in \cite{Helfrich_ZfNC_1973}, we build our derivation on a surface Landau-de Gennes energy and account for an additional dissipation mechanism. Instead of the full Q-tensor, we only consider $\Qb = \beta (\normal\otimes\normal - \frac{1}{2}\IdS)$ with $\beta$ the nematic order parameter for orientation in normal direction. The free energy $\potenergy
= \energyEL + \energyTH$ consists of an elastic and a thermotropic energy, which read
\begin{subequations}\label{eq:energy}
  \begin{align}
\energyEL
		&= \int_{\surf} \frac{l_1}{2} \normsq{}{\nabla\beta} + \frac{\meanelastic(\beta)}{2}\meanc^2 + \gausselastic(\beta)\gaussc \dS \label{eq:energya}\\
  \energyTH
        	&= \int_{\surf} \frac{\beta^2}{8}\left( 12a + 4b\beta + 9c\beta^2 \right) \dS \formComma \label{eq:energyb}
  \end{align}
\end{subequations}
where
\begin{align*}
	l_1
		&= \frac{1}{4}\left(6 L_1 + L_2 + L_3  \right) \formComma
	&\meanelastic(\beta)
		&= \frac{9}{4}\beta^2 (2 L_1 + L_2 + L_3) \formComma
	&\gausselastic(\beta)
		&= - \frac{9}{4}\beta^2 (2 L_1  + L_3)
\end{align*}
with elastic parameters $L_1$, $L_2$, and $L_3$, thermotropic parameters $a$, $b$, and $c$, and liquid crystalline order dependent bending and Gaussian bending rigidities $\meanelastic(\beta)$ and $\gausselastic(\beta)$, respectively. The free energy $\potenergy$ thus reflects the experimentally proposed \cite{chakraborty2020cholesterol} liquid crystalline order dependent bending properties. The elastic splay and saddle-splay parameters $L_\bullet$ are in principle experimentally measurable or can be obtained by molecular dynamics simulations. The functional dependency in the thermotropic part should be considered as one example of a general local energy $\energyTH
= \int_{\surf} \dwpotential(\beta) \dS$. This provides a simple mechanism to extend the proposed model towards more tailored applications, like lyotropic liquid crystals. The full derivation of the model, based on the Lagrange-d'Alembert principle following \cite{bachini2023derivation,nitschke2025beris}, is provided in the corresponding sections of the Appendix. The resulting model reads: Given an initial (closed, simply connected) surface $\surf(0)$ via a parametrization $\para(0)$, the unknowns are the surface velocity field $ \Vb(\mathbf{x},t)$, the scalar field $ \beta(\mathbf{x},t)$,
the generalized pressure field $ \tp(\mathbf{x},t)$, the Lagrange multiplier $\lambda(t)$, and the parametrization $\para(t)$, which are obtained by solving
\begin{subequations}\label{eq:model}
  \begin{align}
    \nabla\tp + \tp\meanc\normal -\DivC\tsigmab + \lambda\normal + \gamma\Vb
       &=  \left(\frac{\meanelastic'(\beta)}{2}\meanc^2 + \gausselastic'(\beta)\gaussc \right)\nabla\beta + \tilde{f}_{B}\normal\formComma\\
    \frac{3}{2}\left( M + \frac{\coeffIF\xi^2}{2} \right)\dot{\beta}\label{eq:model_order}
      &= l_1\Delta\beta - \frac{\meanelastic'(\beta)}{2}\meanc^2 - \gausselastic'(\beta)\gaussc - \dwpotential'(\beta)
          \formComma\\
    \DivC\Vb &= 0\formComma\\
    \int_{\surf} \Vb\cdot\normal \dS &= 0\formComma
  \end{align}
\end{subequations}
on $\surf = \surf(t)$ given by $\para(t)$ and $\partial_t \para(t) = (\Vb \cdot \normal) \normal$, and given initial condition for $ \beta $.
The tangential stress tensor field $\tsigmab\in\tangentS[^2]$ and normal bending force field $\tilde{f}_B \in\tangentScal$ are given by
\begin{subequations}\label{eq:model2}
  \begin{align}
    \tsigmab
      &= \coeffIF\left( 1 + \frac{\xi}{2}\beta \right)^2 \left(\IdS\nablaC\Vb + (\IdS\nablaC\Vb)^T \right) - l_1\nabla\beta\otimes\nabla\beta = \coeffIF\left( 1 + \frac{\xi}{2}\beta \right)^2 \left( \nabla\vb + \nabla^T\vb - 2\vnor \shop \right) - l_1\nabla\beta\otimes\nabla\beta \formComma\\
    \tilde{f}_{B}
    	&=  - \Delta(\meanelastic(\beta)\meanc) - \frac{\meanelastic(\beta)}{2}\meanc\left( \meanc^2 - 4\gaussc \right)
    		-\div\left( \meanc\nabla\gausselastic(\beta) - \shop\nabla\gausselastic(\beta) \right) \formPeriod
  \end{align}
\end{subequations}
Instead of the Surface Beris--Edwards--Helfrich model considered in \cite{nitschke2025beris,NV24}, which involves an equation for the surface Q-tensor field $\Qb$, and thus requires solving a tensor-valued surface partial differential equation, \cref{eq:model,eq:model2} only require solving a scalar-valued surface partial differential equation for $\beta$, to account for the liquid crystalline order. We still call \cref{eq:model,eq:model2} Surface Beris--Edwards--Helfrich (SBEH) model, to highlight the hydrodynamic nature, the underlying Q-tensor approach and the resulting curvature-elasticity. The structure of the equations is similar to the Surface Stokes--Helfrich (SSH) model, but with an additional equation for $\beta$, and additional terms depending on $\nabla \beta$ in the momentum equation, the rate-of-deformation tensor $\tsigmab$ and the bending force $\tilde{f}_{B}$. Since only inextensible solutions are considered, the particular composition of the pressure $\tp$ is immaterial.
For completeness, however, we provide it here:
\begin{align*}
    \tp
      &= p
      	 -\frac{l_1}{2} \normsq{}{\nabla\beta}
         - \dwpotential(\beta)
         - \frac{\coeffIF\xi}{2} \left( \left( 1 + \frac{\xi}{2}\beta \right)\dot{\beta} \right)
         \formPeriod
\end{align*}

Setting $\xi=0$ and fixing $\beta$ constant, the order parameter equation \cref{eq:model_order} can be omitted by this constraint, and we recover the Surface Stokes--Helfrich model \cref{eq:u-p,eq:variationaderivative}.
Alternatively, if the elastic moduli do not depend on $\beta$, \ie, $\meanelastic(\beta)\equiv\meanelastic$ and $\gausselastic(\beta)\equiv\gausselastic$,
and we set  $\xi=l_1=0$, then the order parameter equation \cref{eq:model_order} decouples from the remaining equations, which also yield the Surface Stokes--Helfrich (SSH) model \cref{eq:u-p,eq:variationaderivative}. In the friction-dominated limit, e.g. formally setting in addition $\mu=0$ and $\gamma=1$, we obtain the $L^2$-gradient flow \cref{eq:L2Helfrich,eq:variationaderivative}.

\subsection*{Numerical approach to solve the dynamic equations}

\cref{eq:model,eq:model2} are discretized using an Arbitrary-Lagrangian-Eulerian (ALE) Surface Finite Element Method (SFEM) combining the approaches considered in \cite{bachini2023derivation,sischka2025two, NitschkeSischkaVoigt_JoFM_2026_Hlcmflb}. For the resulting special cases, e.g. \cref{eq:u-p,eq:variationaderivative} we consider the approaches in \cite{krause2023numerical, bachini2023derivation}. A detailed description is provided in the corresponding section of the Appendix. The implementation is realized within the finite element toolbox \amdis{} \cite{vey2007amdis,witkowski2015software}, available at \cite{AMDiS:2.10}, which is a high level module in the \dune{} \cite{Dune2.10} ecosystem. For an overview of \dune{} we refer to \cite{duneBook}. Moreover, we use the modules \alugrid{} \cite{dune-alugrid} to manage a grid with sphere topology, \curvedgrid{} \cite{dune-curvedGrid} to map this grid onto the surface by the parametrization $\para_\meshparam$. This parametrization, as well as all other discrete functions and their bases are represented by means of \functions{} \cite{dune-functions}. We use \eigen{}  \cite{eigenweb} as a linear algebra backend, and the direct solver PARDISO as well as the multithreading framework TBB from Intel's oneAPI \cite{intel2025oneapi}. Finally, all $3$D visualizations have been created with ParaView \cite{Paraview}.

\section*{Results}

\subsection*{Numerical experiments}
We present several numerical experiments, aiming to highlight the impact of the additional degrees of freedom in the proposed Surface Beris--Edwards--Helfrich model \cref{eq:model,eq:model2} on mesoscopic length and time scales. We will therefore compare the results with the Surface Stokes--Helfrich model \cref{eq:u-p,eq:variationaderivative}, which builds on \cref{eq:helfrichenergy}. Due to the high dimensionality of the parameter space, we restrict ourselves to the one-constant approximation $L_1 = L, L_2 = L_3 = 0$.
Moreover, we set $\xi = 0$.
For the simplicity of presentation, we keep $\coeffIF = 1, M = 1$, and $\gamma = 0.1$ fixed. The thermotropic parameters are taken as $a = 0, b = -27, c = 13.5$, which gives an energy density with minimum at $ \beta = \frac{2}{3}$ and a metastable critical point at $\beta = 0$. These parameters have already been considered in \cite{NitschkeSischkaVoigt_JoFM_2026_Hlcmflb}. All shapes have the area of a unit sphere. We focus on variations of the elastic constant $L$ and the enclosed volume $V$, or equivalently the reduced volume $V_{r} = \frac{3V}{4\pi}$, as in \cite{seifertShapeTransformationsVesicles1991}.

\subsection*{Stationary shapes}
We consider the relaxation from rotationally symmetric ellipsoid shapes with $\beta(\xb,0) = \frac{2}{3}$. The shapes are already close to the classical Seifert shapes. We consider oblate and prolate shapes and vary the reduced volume $V_r$ and the elastic parameter $L$. We solve the Surface Beris--Edwards--Helfrich (SBEH) model, \cref{eq:model,eq:model2}, and the Surface Stokes--Helfrich (SSH) model, \cref{eq:u-p,eq:variationaderivative}, towards convergence.
\begin{figure}[tbp]
    \def\svgwidth{\linewidth}
    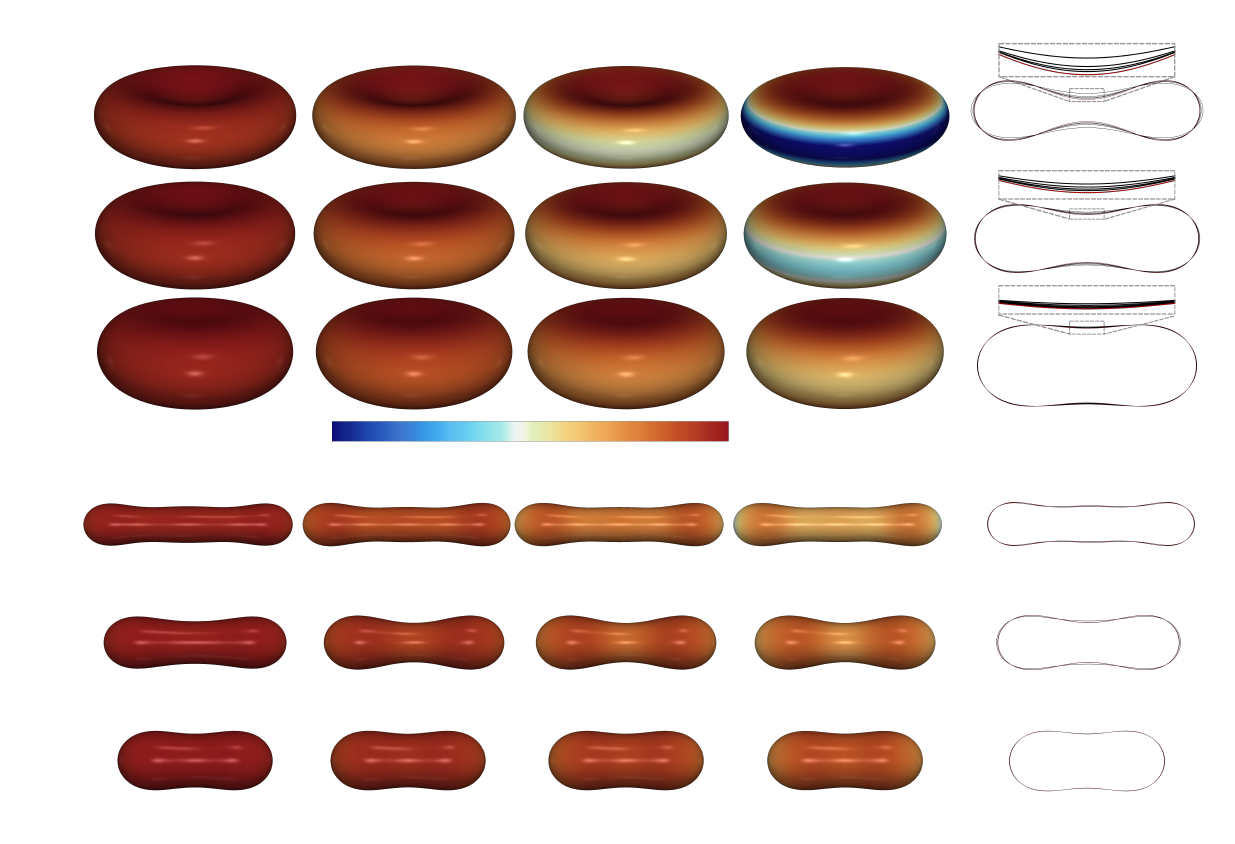
    \caption{Stationary shapes of the Surface Beris--Edwards--Helfrich (SBEH) model. (top) Oblate-like shapes. (bottom) Prolate-like shapes. We vary $L \in \{0.01, 0.03, 0.05, 0.07\}$ and $V_r \in \{0.6, 0.7, 0.8\}$. The right column shows cuts through each of the stationary shapes (black) in the corresponding row, together with a cut through the minimizer of the Helfrich energy of the same reduced volume $V_r$ (red), obtained by solving the Surface Stokes--Helfrich (SSH) model. For the oblate-like shapes the differences in shape enlarge by increasing $L$. The prolate-like shapes do not significantly depend on $L$ and are almost identical to the corresponding Seifert shapes obtained by solving the Surface Stokes--Helfrich (SSH) model.}
    \label{fig:Shapes}
\end{figure}

In Fig. \ref{fig:Shapes} the stationary shapes are shown together with $\beta$ and compared with respect to each other. The stationary shapes for the Surface Beris--Edwards--Helfrich (SBEH) model show a dependence on $L$. For the Surface Stokes--Helfrich (SSH) model the stationary shapes for different $L$ are identical and correspond to the classical Seifert shapes. As a consequence, the shapes differ between both models. We observe a clear trend, namely, stronger deviations between the Surface Beris-Edwards-Helfrich (SBEH) model and the Surface Stokes--Helfrich (SSH) model are observed for decreasing $V_r$ and increasing $L$. These differences in shape also correspond to differences in the order parameter $\beta$, for which we observe an increase in variance for decreasing $V_r$ and increasing $L$. Significantly lower values for $\beta$ can be observed at regions of larger curvature $\norm{\shop}^2$, essentially at the rim of the oblate-like shapes and the caps of the prolate-like shapes. The differences are more severe for the oblate-like shapes and less pronounced for the prolate-like shapes. We argue that this is a consequence of the different curvature distributions. In order to confirm this, we sample points at all equilibrium shapes, measure the curvature $\norm{\shop}^2$ at these points and plot the value of $\beta$ against $\norm{\shop}^2$. In \cref{fig:PhaseDiagram} these plots are shown for the oblate- and prolate-like shapes. They clearly show larger curvature values for the oblate-like shapes. The plots also indicate a linear dependency of $\beta$ on $\norm{\shop}^2$. The slope only depends on $L$, but not on the geometry, as can be seen in the insets, in which the extracted slope is shown. We conclude, that larger $\norm{\shop}^2$ in the geometry leads to lower $\beta$, which explains the different properties between the oblate-like and prolate-like shapes.
The different equilibrium shapes provide an experimentally testable effect of microscopic details of lipid order on mesoscopic shapes. Also the dependency of $\beta$ on $\norm{\shop}^2$ could be experimentally confirmed.

\begin{figure}[tbp]
  \centering
  \noindent
\begin{tikzpicture}
\def\canvasW{\linewidth}
\definecolor{crimson2143940}{RGB}{214,39,40}
\definecolor{darkgray176}{RGB}{176,176,176}
\definecolor{darkorange25512714}{RGB}{255,127,14}
\definecolor{forestgreen4416044}{RGB}{44,160,44}
\definecolor{steelblue31119180}{RGB}{31,119,180}
  \input{images/Oblate_scatter_B.tex}


\begin{axis}[
at={(oblate.south west)},
anchor=south west,
width={0.11*\canvasW},
xshift=0.065*\canvasW,
yshift=0.04*\canvasW,
xlabel shift=-5pt,
ylabel shift=-7pt,
label style={font=\tiny},
tick style={font=\tiny},
scale only axis,
scaled y ticks=false,
yticklabel style={
    /pgf/number format/fixed,
    /pgf/number format/fixed zerofill,
    /pgf/number format/precision=3
},
scaled x ticks=false,
xticklabel style={
  /pgf/number format/fixed,
  /pgf/number format/fixed zerofill,
  /pgf/number format/precision=2
},
legend cell align={left},
legend style={
  fill opacity=0.8,
  draw opacity=1,
  text opacity=1,
  at={(1.1,1.1)},
  anchor=south west,
  draw=none,
  font=\footnotesize
},
tick align=outside,
tick pos=left,
x grid style={darkgray176},
xlabel={\(\displaystyle L\)},
xmin=0.007, xmax=0.073,
xtick style={color=black},
y grid style={darkgray176},
ylabel={Slope},
ymin=-0.00909042737125516, ymax=-0.000740089072374173,
ytick style={color=black}
]
\addplot [
  mark=o,
  only marks,
  scatter,
  scatter/@post marker code/.code={%
  \endscope
},
  scatter/@pre marker code/.code={%
  \expanded{%
  \noexpand\definecolor{thispointdrawcolor}{RGB}{\drawcolor}%
  }%
  \scope[draw=thispointdrawcolor]%
},
  visualization depends on={value \thisrow{draw} \as \drawcolor}
]
table{%
x  y  draw
0.01 -0.00112776668804332 31,119,180
0.03 -0.00349816424701118 255,127,14
0.05 -0.00606188997746067 44,160,44
0.07 -0.00871086653948784 214,39,40
};
\addplot [
  mark=square,
  only marks,
  scatter,
  scatter/@post marker code/.code={%
  \endscope
},
  scatter/@pre marker code/.code={%
  \expanded{%
  \noexpand\definecolor{thispointdrawcolor}{RGB}{\drawcolor}%
  }%
  \scope[draw=thispointdrawcolor]%
},
  visualization depends on={value \thisrow{draw} \as \drawcolor}
]
table{%
x  y  draw
0.01 -0.00112348559340223 31,119,180
0.03 -0.00345265809887224 255,127,14
0.05 -0.00591042218622982 44,160,44
0.07 -0.00851498601606659 214,39,40
};
\addplot [
  mark=triangle,
  only marks,
  scatter,
  scatter/@post marker code/.code={%
  \endscope
},
  scatter/@pre marker code/.code={%
  \expanded{%
  \noexpand\definecolor{thispointdrawcolor}{RGB}{\drawcolor}%
  }%
  \scope[draw=thispointdrawcolor]%
},
  visualization depends on={value \thisrow{draw} \as \drawcolor}
  ]
table{%
x  y  draw
0.01 -0.00111964990414149 31,119,180
0.03 -0.00341369716758985 255,127,14
0.05 -0.00578889655011859 44,160,44
0.07 -0.00825395663507089 214,39,40
};
\end{axis}

  \input{images/Prolate_scatter_B.tex}


\begin{axis}[
at={(prolate.south west)},
anchor=south west,
width={0.11*\canvasW},
xshift=0.065*\canvasW,
yshift=0.04*\canvasW,
xlabel shift=-5pt,
ylabel shift=-7pt,
label style={font=\tiny},
tick style={font=\tiny},
scale only axis,
scaled y ticks=false,
yticklabel style={
    /pgf/number format/fixed,
    /pgf/number format/fixed zerofill,
    /pgf/number format/precision=3
},
scaled x ticks=false,
xticklabel style={
  /pgf/number format/fixed,
  /pgf/number format/fixed zerofill,
  /pgf/number format/precision=2
},
legend cell align={left},
legend style={
  fill opacity=0.8,
  draw opacity=1,
  text opacity=1,
  at={(1.1,1.1)},
  anchor=south west,
  draw=none,
  font=\footnotesize
},
tick align=outside,
tick pos=left,
x grid style={darkgray176},
xlabel={\(\displaystyle L\)},
xmin=0.007, xmax=0.073,
xtick style={color=black},
y grid style={darkgray176},
ylabel={Slope},
ymin=-0.00947223390758069, ymax=-0.000726677141376329,
ytick style={color=black}
]
\addplot [
  mark=o,
  only marks,
  scatter,
  scatter/@post marker code/.code={%
  \endscope
},
  scatter/@pre marker code/.code={%
  \expanded{%
  \noexpand\definecolor{thispointdrawcolor}{RGB}{\drawcolor}
  }%
  \scope[draw=thispointdrawcolor]%
},
  visualization depends on={value \thisrow{draw} \as \drawcolor}]
table{%
x  y  draw
0.01 -0.00113515547151503 31,119,180
0.03 -0.00353626446852518 255,127,14
0.05 -0.00616268748958186 44,160,44
0.07 -0.00907470860002595 214,39,40
};
\addplot [
  mark=square,
  only marks,
  scatter,
  scatter/@post marker code/.code={%
  \endscope
},
  scatter/@pre marker code/.code={%
  \expanded{%
  \noexpand\definecolor{thispointdrawcolor}{RGB}{\drawcolor}%
  }%
  \scope[draw=thispointdrawcolor]%
},
  visualization depends on={value \thisrow{draw} \as \drawcolor}
]
table{%
x  y  draw
0.01 -0.00113004959154777 31,119,180
0.03 -0.00347216668935892 255,127,14
0.05 -0.00595149784263572 44,160,44
0.07 -0.00859155082245416 214,39,40
};
\addplot [
  mark=triangle,
  only marks,
  scatter,
  scatter/@post marker code/.code={%
  \endscope
},
  scatter/@pre marker code/.code={%
  \expanded{%
  \noexpand\definecolor{thispointdrawcolor}{RGB}{\drawcolor}%
  }%
  \scope[draw=thispointdrawcolor]%
},
  visualization depends on={value \thisrow{draw} \as \drawcolor}
]
table{%
x  y  draw
0.01 -0.00112420244893107 31,119,180
0.03 -0.00345155377370551 255,127,14
0.05 -0.00589920731429908 44,160,44
0.07 -0.00848316949635081 214,39,40
};
\end{axis}

\end{tikzpicture}
  \caption{Order parameter $\beta$ as a function of the curvature measure $\norm{\shop}^2$. We consider the equilibrium shapes of Figure \ref{fig:Shapes} and sample $50$ points. (left) Oblate-like shapes. (right) Prolate-like shapes. (Inset) Slope of the fitted lines. Color-coding is according to the elastic parameter $L$, symbols vary with $V_r$. We observe a clear linear dependency of $\beta$ on $\norm{\shop}^2$, which does not depend on the reduced volume $V_r$, but only on the elastic parameter $L$.}
  \label{fig:PhaseDiagram}
\end{figure}

A closer inspection shows small deviations for $L = 0.07$ and the high curvature values of the oblate-like shape for $V_r = 0.6$, see the red circles in \cref{fig:PhaseDiagram}. Here, we already see the onset towards a qualitatively different behavior. For the oblate-like shapes, for even larger values of $L$ (not shown), $\beta$ is further reduced at the rim. As a consequence $\kappa(\beta)$ is locally reduced at this region, allowing for high curvature at reduced cost, which further decreases $\beta$. This feedback mechanism can lead to the formation of a cusp with $\beta \approx 0$ and $\kappa(\beta) \approx 0$. In this parameter regime the linear dependency of $\beta$ on $\norm{\shop}^2$ no longer holds. We also observe a similar phenomenon for prolate-like shapes, albeit for larger values of $L$, and concentrated around the maximum of curvature, which for prolate-like ellipsoids are the two tips. This general behavior, corresponding to a melting of the liquid crystal, is a consequence of the considered thermotropic energy and would warrant a modification of the considered state potential. We refrain from doing so here, as options are manifold, but point out that the general structure of the model allows such modifications in a simple manner.

\subsection*{Dynamic evolution}

\begin{figure}[tbp]
    \centering
\begin{tikzpicture}

\definecolor{darkgray176}{RGB}{176,176,176}
\definecolor{darkorange25512714}{RGB}{255,127,14}
\definecolor{lightgray204}{RGB}{204,204,204}
\definecolor{steelblue31119180}{RGB}{31,119,180}

\begin{groupplot}[group style={group size=3 by 1},
scale only axis,
width=0.2\textwidth,
height=0.2\textwidth,
xlabel={$t$},
xtick style={color=black},
xtick={0.001,0.01,0.1,1, 10},
xticklabels={
  $\mathdefault{10^{-3}}$,
  $\mathdefault{10^{-2}}$,
  $\mathdefault{10^{-1}}$,
  $\mathdefault{10^{0}}$,
  $\mathdefault{10^1}$
},
ylabel={$\potenergy$},
ylabel style={
    xshift=0pt, 
    yshift=-10pt    
  },
ytick style={color=black},
ytick={0.5,0.75,1,2,3,4,5},
yticklabels={
  $\mathdefault{0.5}$,
  $\mathdefault{0.75}$,
  $\mathdefault{1}$,
  $\mathdefault{2}$,
  $\mathdefault{3}$,
  $\mathdefault{4}$,
  $\mathdefault{5}$
},
unbounded coords=jump,
x grid style={darkgray176},
y grid style={darkgray176},
log basis x={10},
log basis y={10},
tick align=outside,
tick pos=left,
xmode=log,
ymode=log,
xmin=0.000670207339231113, xmax=10,
ymin=0.4, ymax=8
]
\nextgroupplot[
legend cell align={left},
legend style={
  font=\small,
  fill opacity=0.8,
  draw opacity=1,
  text opacity=1,
  at={(3.8,0)},
  anchor=south west,
  draw=lightgray204
},
title={$V_r = 0.6$}
]
\addplot [thick, darkorange25512714]
table {images/data/Oblate_F_tot-000.dat};
\addlegendentry{$L=0.01$-SSH}
\addplot [thick, steelblue31119180]
table {images/data/Oblate_F_tot-001.dat};
\addlegendentry{$L=0.01$-SBEH}
\addplot [thick, darkorange25512714, dash pattern=on 2pt off 3.3pt]
table {images/data/Oblate_F_tot-002.dat};
\addlegendentry{$L=0.05$-SSH}
\addplot [thick, steelblue31119180, dash pattern=on 2pt off 3.3pt]
table {images/data/Oblate_F_tot-003.dat};
\addlegendentry{$L=0.05$-SBEH}
\addplot [thick, darkorange25512714, dash pattern=on 7.4pt off 3.2pt]
table {images/data/Oblate_F_tot-004.dat};
\addlegendentry{$L=0.03$-SSH}
\addplot [thick, steelblue31119180, dash pattern=on 7.4pt off 3.2pt]
table {images/data/Oblate_F_tot-005.dat};
\addlegendentry{$L=0.03$-SBEH}
\addplot [thick, darkorange25512714, dash pattern=on 12.8pt off 3.2pt on 2pt off 3.2pt]
table {images/data/Oblate_F_tot-006.dat};
\addlegendentry{$L=0.07$-SSH}
\addplot [thick, steelblue31119180, dash pattern=on 12.8pt off 3.2pt on 2pt off 3.2pt]
table {images/data/Oblate_F_tot-007.dat};
\addlegendentry{$L=0.07$-SBEH}

\nextgroupplot[
title={$V_r = 0.7$}
]
\addplot [thick, darkorange25512714]
table {images/data/Oblate_F_tot-008.dat};
\addplot [thick, steelblue31119180]
table {images/data/Oblate_F_tot-009.dat};
\addplot [thick, darkorange25512714, dash pattern=on 2pt off 3.3pt]
table {images/data/Oblate_F_tot-010.dat};
\addplot [thick, steelblue31119180, dash pattern=on 2pt off 3.3pt]
table {images/data/Oblate_F_tot-011.dat};
\addplot [thick, darkorange25512714, dash pattern=on 7.4pt off 3.2pt]
table {images/data/Oblate_F_tot-012.dat};
\addplot [thick, steelblue31119180, dash pattern=on 7.4pt off 3.2pt]
table {images/data/Oblate_F_tot-013.dat};
\addplot [thick, darkorange25512714, dash pattern=on 12.8pt off 3.2pt on 2pt off 3.2pt]
table {images/data/Oblate_F_tot-014.dat};
\addplot [thick, steelblue31119180, dash pattern=on 12.8pt off 3.2pt on 2pt off 3.2pt]
table {images/data/Oblate_F_tot-015.dat};
\nextgroupplot[
title={$V_r = 0.8$}
]
\addplot [thick, darkorange25512714]
table {images/data/Oblate_F_tot-016.dat};
\addplot [thick, steelblue31119180]
table {images/data/Oblate_F_tot-017.dat};
\addplot [thick, darkorange25512714, dash pattern=on 2pt off 3.3pt]
table {images/data/Oblate_F_tot-018.dat};
\addplot [thick, steelblue31119180, dash pattern=on 2pt off 3.3pt]
table {images/data/Oblate_F_tot-019.dat};
\addplot [thick, darkorange25512714, dash pattern=on 7.4pt off 3.2pt]
table {images/data/Oblate_F_tot-020.dat};
\addplot [thick, steelblue31119180, dash pattern=on 7.4pt off 3.2pt]
table {images/data/Oblate_F_tot-021.dat};
\addplot [thick, darkorange25512714, dash pattern=on 12.8pt off 3.2pt on 2pt off 3.2pt]
table {images/data/Oblate_F_tot-022.dat};
\addplot [thick, steelblue31119180, dash pattern=on 12.8pt off 3.2pt on 2pt off 3.2pt]
table {images/data/Oblate_F_tot-023.dat};
\end{groupplot}

\end{tikzpicture}

\begin{tikzpicture}

\definecolor{darkgray176}{RGB}{176,176,176}
\definecolor{darkorange25512714}{RGB}{255,127,14}
\definecolor{lightgray204}{RGB}{204,204,204}
\definecolor{steelblue31119180}{RGB}{31,119,180}

\begin{groupplot}[group style={group size=3 by 1},
scale only axis,
width=0.2\textwidth,
height=0.2\textwidth,
xlabel={$t$},
xtick style={color=black},
xtick={0.001,0.01,0.1,1,10},
xticklabels={
  $\mathdefault{10^{-3}}$,
  $\mathdefault{10^{-2}}$,
  $\mathdefault{10^{-1}}$,
  $\mathdefault{10^{0}}$,
  $\mathdefault{10^1}$
},
ylabel={$\potenergy$},
ylabel style={
    xshift=0pt, 
    yshift=-10pt    
  },
ytick style={color=black},
ytick={0.5,0.75,1,2,3,4,5},
yticklabels={
  $\mathdefault{0.5}$,
  $\mathdefault{0.75}$,
  $\mathdefault{1}$,
  $\mathdefault{2}$,
  $\mathdefault{3}$,
  $\mathdefault{4}$,
  $\mathdefault{5}$
},
unbounded coords=jump,
x grid style={darkgray176},
y grid style={darkgray176},
log basis x={10},
log basis y={10},
tick align=outside,
tick pos=left,
xmode=log,
ymode=log,
xmin=0.000670207339231113, xmax=10,
ymin=0.4, ymax=8
]
\nextgroupplot[
legend cell align={left},
legend style={
  font=\small,
  fill opacity=0,
  draw opacity=0,
  text opacity=0,
  at={(3.8,0)},
  anchor=south west,
  draw=lightgray204
},
]
\addplot [thick, darkorange25512714]
table {images/data/Prolate_F_tot-000.dat};
\addlegendentry{$L=0.01$-SSH}
\addplot [thick, steelblue31119180]
table {images/data/Prolate_F_tot-001.dat};
\addlegendentry{$L=0.01$-SBEH}
\addplot [thick, darkorange25512714, dash pattern=on 2pt off 3.3pt]
table {images/data/Prolate_F_tot-002.dat};
\addlegendentry{$L=0.05$-SSH}
\addplot [thick, steelblue31119180, dash pattern=on 2pt off 3.3pt]
table {images/data/Prolate_F_tot-003.dat};
\addlegendentry{$L=0.05$-SBEH}
\addplot [thick, darkorange25512714, dash pattern=on 7.4pt off 3.2pt]
table {images/data/Prolate_F_tot-004.dat};
\addlegendentry{$L=0.03$-SSH}
\addplot [thick, steelblue31119180, dash pattern=on 7.4pt off 3.2pt]
table {images/data/Prolate_F_tot-005.dat};
\addlegendentry{$L=0.03$-SBEH}
\addplot [thick, darkorange25512714, dash pattern=on 12.8pt off 3.2pt on 2pt off 3.2pt]
table {images/data/Prolate_F_tot-006.dat};
\addlegendentry{$L=0.07$-SSH}
\addplot [thick, steelblue31119180, dash pattern=on 12.8pt off 3.2pt on 2pt off 3.2pt]
table {images/data/Prolate_F_tot-007.dat};
\addlegendentry{$L=0.07$-SBEH}
\nextgroupplot[
]
\addplot [thick, darkorange25512714]
table {images/data/Prolate_F_tot-008.dat};
\addplot [thick, steelblue31119180]
table {images/data/Prolate_F_tot-009.dat};
\addplot [thick, darkorange25512714, dash pattern=on 2pt off 3.3pt]
table {images/data/Prolate_F_tot-010.dat};
\addplot [thick, steelblue31119180, dash pattern=on 2pt off 3.3pt]
table {images/data/Prolate_F_tot-011.dat};
\addplot [thick, darkorange25512714, dash pattern=on 7.4pt off 3.2pt]
table {images/data/Prolate_F_tot-012.dat};
\addplot [thick, steelblue31119180, dash pattern=on 7.4pt off 3.2pt]
table {images/data/Prolate_F_tot-013.dat};
\addplot [thick, darkorange25512714, dash pattern=on 12.8pt off 3.2pt on 2pt off 3.2pt]
table {images/data/Prolate_F_tot-014.dat};
\addplot [thick, steelblue31119180, dash pattern=on 12.8pt off 3.2pt on 2pt off 3.2pt]
table {images/data/Prolate_F_tot-015.dat};

\nextgroupplot[
]
\addplot [thick, darkorange25512714]
table {images/data/Prolate_F_tot-016.dat};
\addplot [thick, steelblue31119180]
table {images/data/Prolate_F_tot-017.dat};
\addplot [thick, darkorange25512714, dash pattern=on 2pt off 3.3pt]
table {images/data/Prolate_F_tot-018.dat};
\addplot [thick, steelblue31119180, dash pattern=on 2pt off 3.3pt]
table {images/data/Prolate_F_tot-019.dat};
\addplot [thick, darkorange25512714, dash pattern=on 7.4pt off 3.2pt]
table {images/data/Prolate_F_tot-020.dat};
\addplot [thick, steelblue31119180, dash pattern=on 7.4pt off 3.2pt]
table {images/data/Prolate_F_tot-021.dat};
\addplot [thick, darkorange25512714, dash pattern=on 12.8pt off 3.2pt on 2pt off 3.2pt]
table {images/data/Prolate_F_tot-022.dat};
\addplot [thick, steelblue31119180, dash pattern=on 12.8pt off 3.2pt on 2pt off 3.2pt]
table {images/data/Prolate_F_tot-023.dat};
\end{groupplot}

\end{tikzpicture}
    \caption{Energy evolutions of the Surface--Beris--Edwards--Helfrich (SBEH) model compared to the Surface--Stokes--Helfrich (SSH) model for various setups in log-log scale. Different columns correspond to different values of the reduced volume $V_r$. In each plot we compare the evolution of $\potenergy$ for four different values of $L$. (top) Oblate-like shapes. (bottom) Prolate-like shapes. }
    \label{fig:EnergyEvolution}
\end{figure}

In \cref{fig:EnergyEvolution} the evolution of the free energy $\potenergy$ is shown for both models, the Surface Stokes--Helfrich (SSH) model and the Surface Beris--Edwards--Helfrich (SBEH) model. The evolution corresponds to the simulations considered in the previous section. For the Surface Stokes--Helfrich (SSH) model the free energy is the sum of $\energyH$, with $\kappa = \kappa(\frac{2}{3})$, $\kappa_G = \kappa_G(\frac{2}{3})$ and $\meanc_0 = 0$, and $\energyTH$ with $\beta = \frac{2}{3}$. All curves show a decreasing energy and convergence to the stationary shape. The free energy $\potenergy$ is always lower for the Surface Beris--Edwards--Helfrich (SBEH) model. The difference between the free energies for the Surface Beris--Edwards--Helfrich (SBEH) model and the Surface Stokes--Helfrich (SSH) model increases with decreasing $V_r$ and increasing $L$. As expected from \cref{fig:Shapes} the differences are larger for the oblate-like shapes, than for the prolate-like shapes. While the demonstrated decrease in energy provides a validation of the models and the considered numerical algorithms, the lower energy curves for the Surface Beris--Edwards--Helfrich (SBEH) model also demonstrate the significance of the additional degrees of freedom on the shape evolution.

\begin{figure}[tbp]
    \centering
    \includegraphics[width=0.18\linewidth]{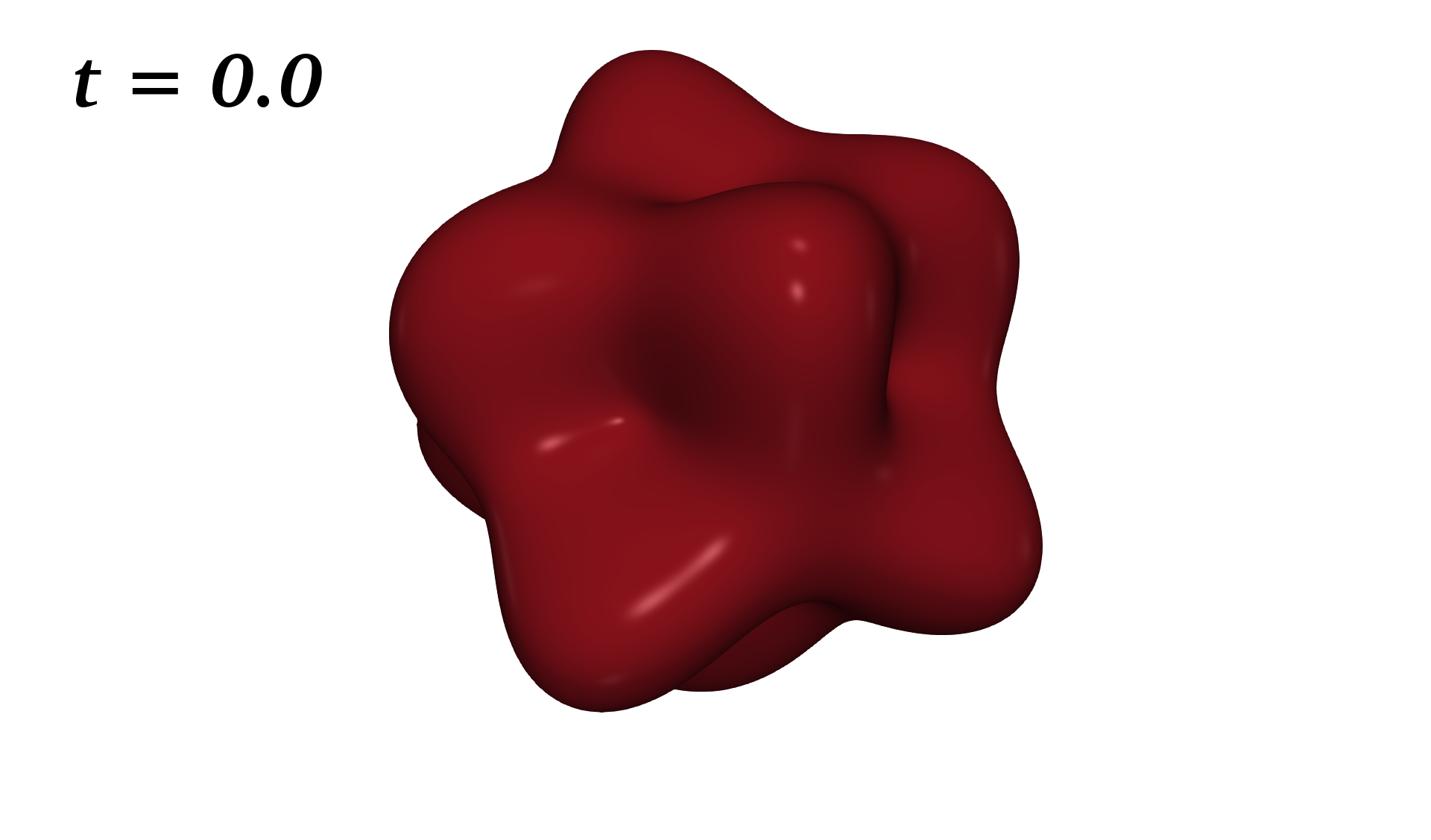}
    \includegraphics[width=0.18\linewidth]{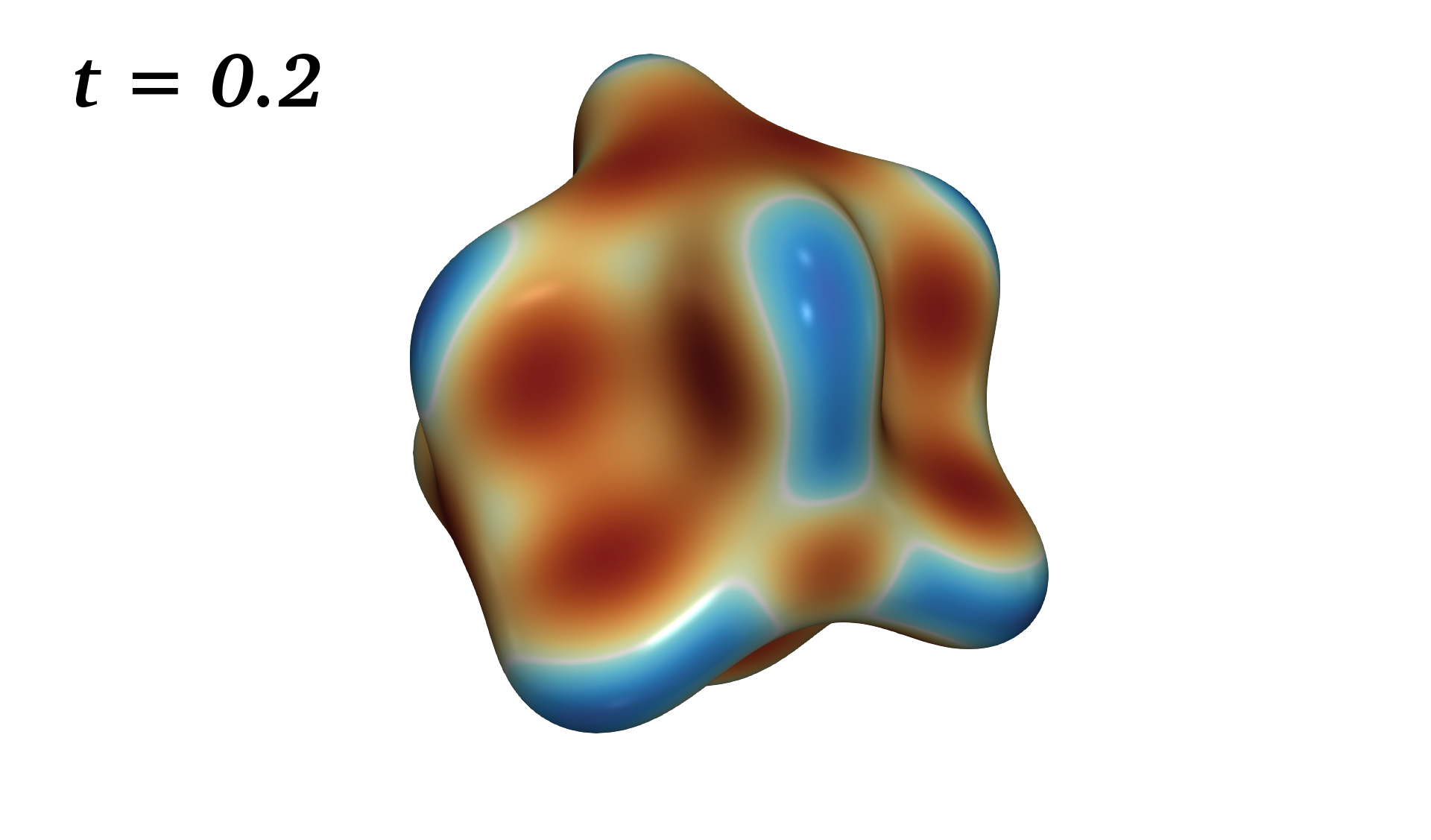}
    \includegraphics[width=0.18\linewidth]{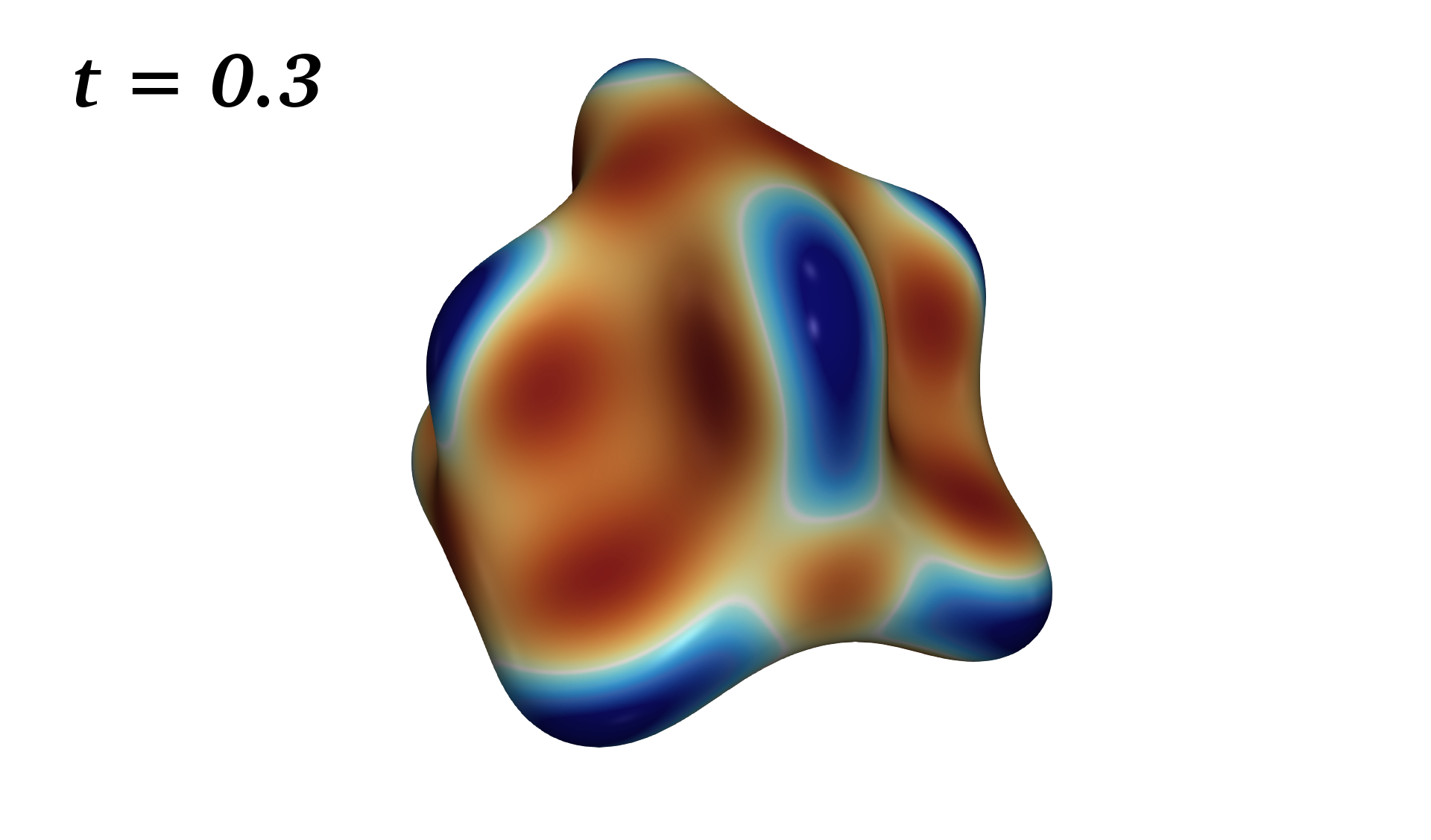}
    \includegraphics[width=0.18\linewidth]{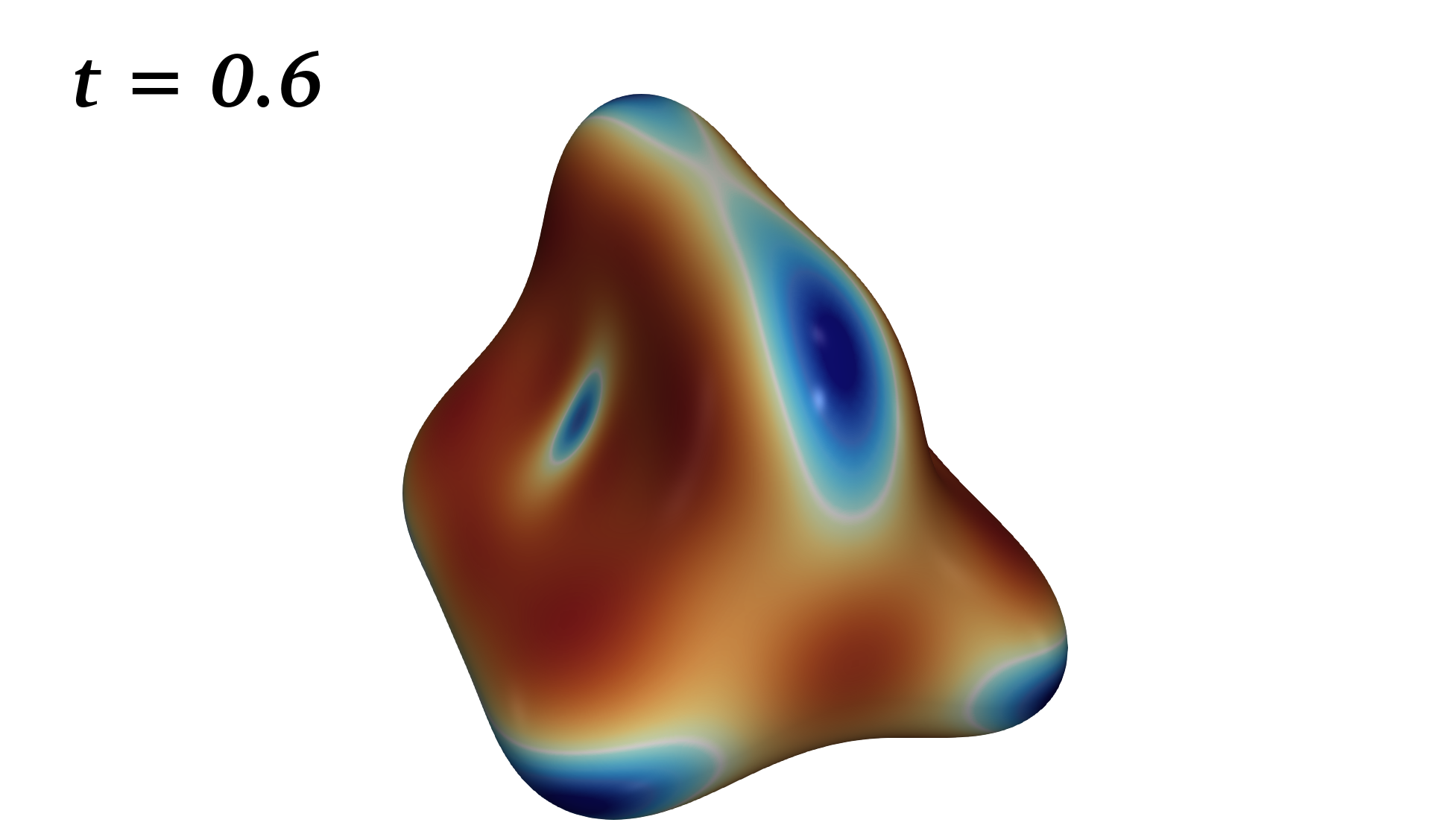}
    \includegraphics[width=0.18\linewidth]{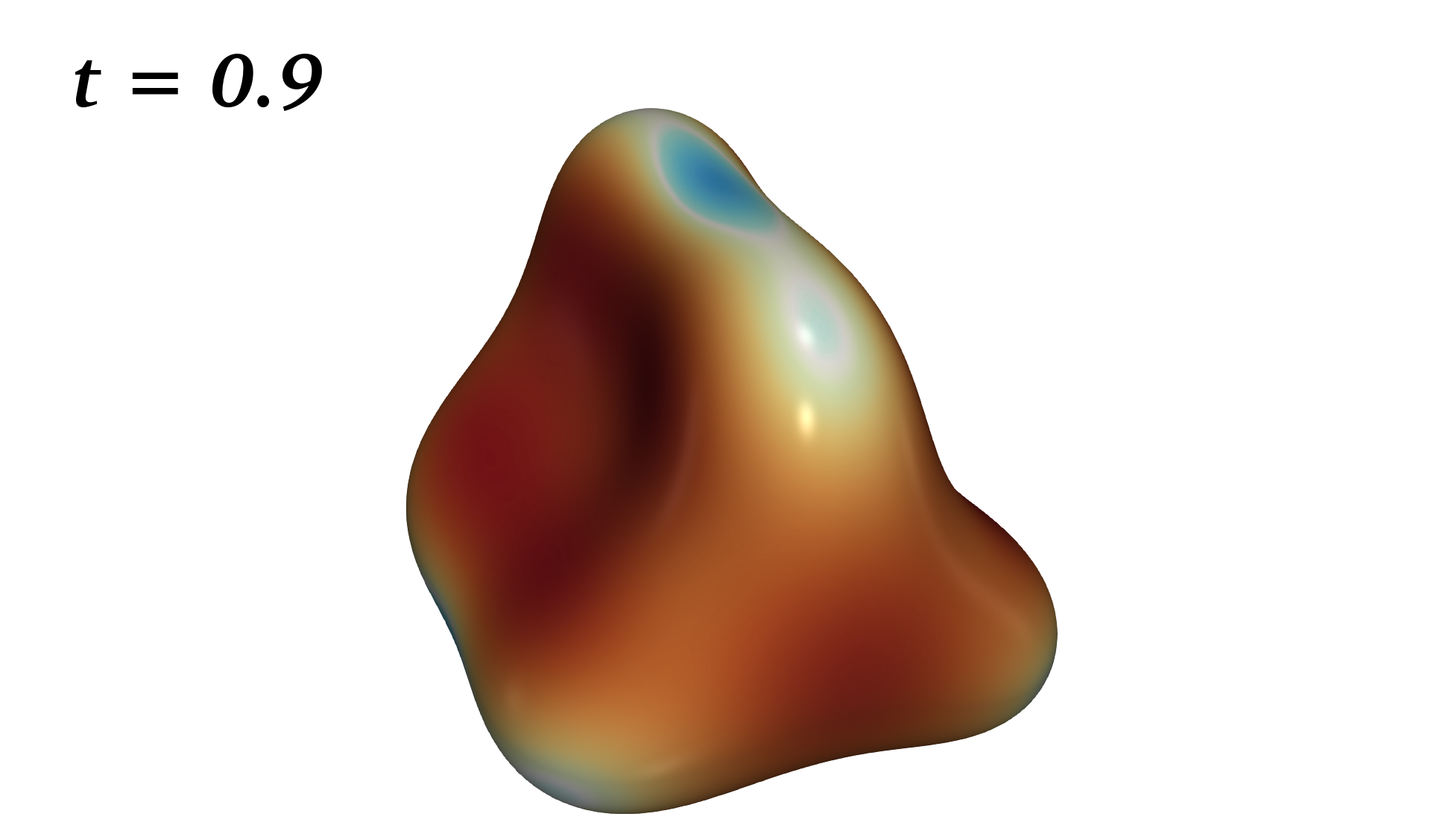}

    \includegraphics[width=0.18\linewidth]{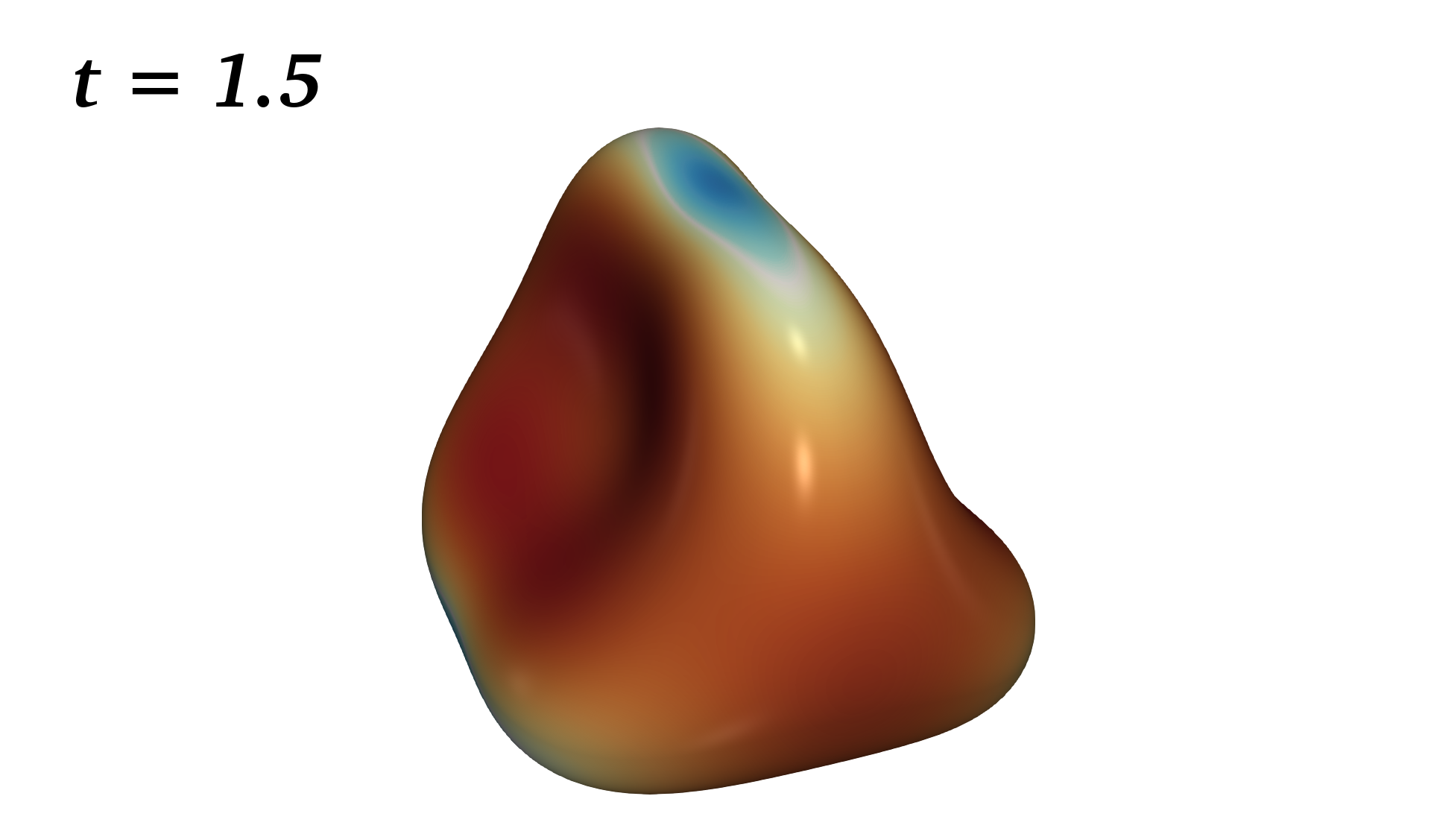}
    \includegraphics[width=0.18\linewidth]{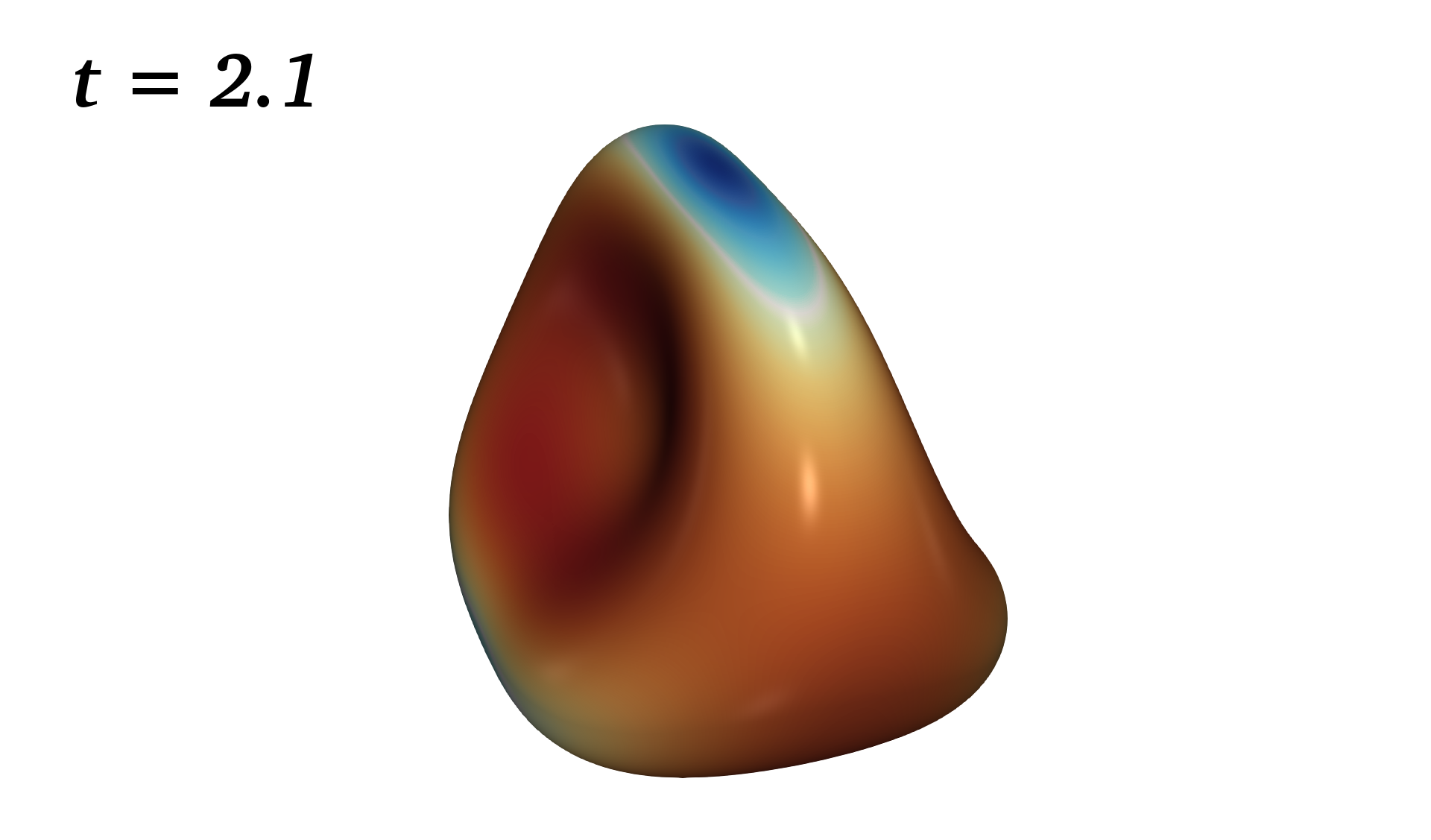}
    \includegraphics[width=0.18\linewidth]{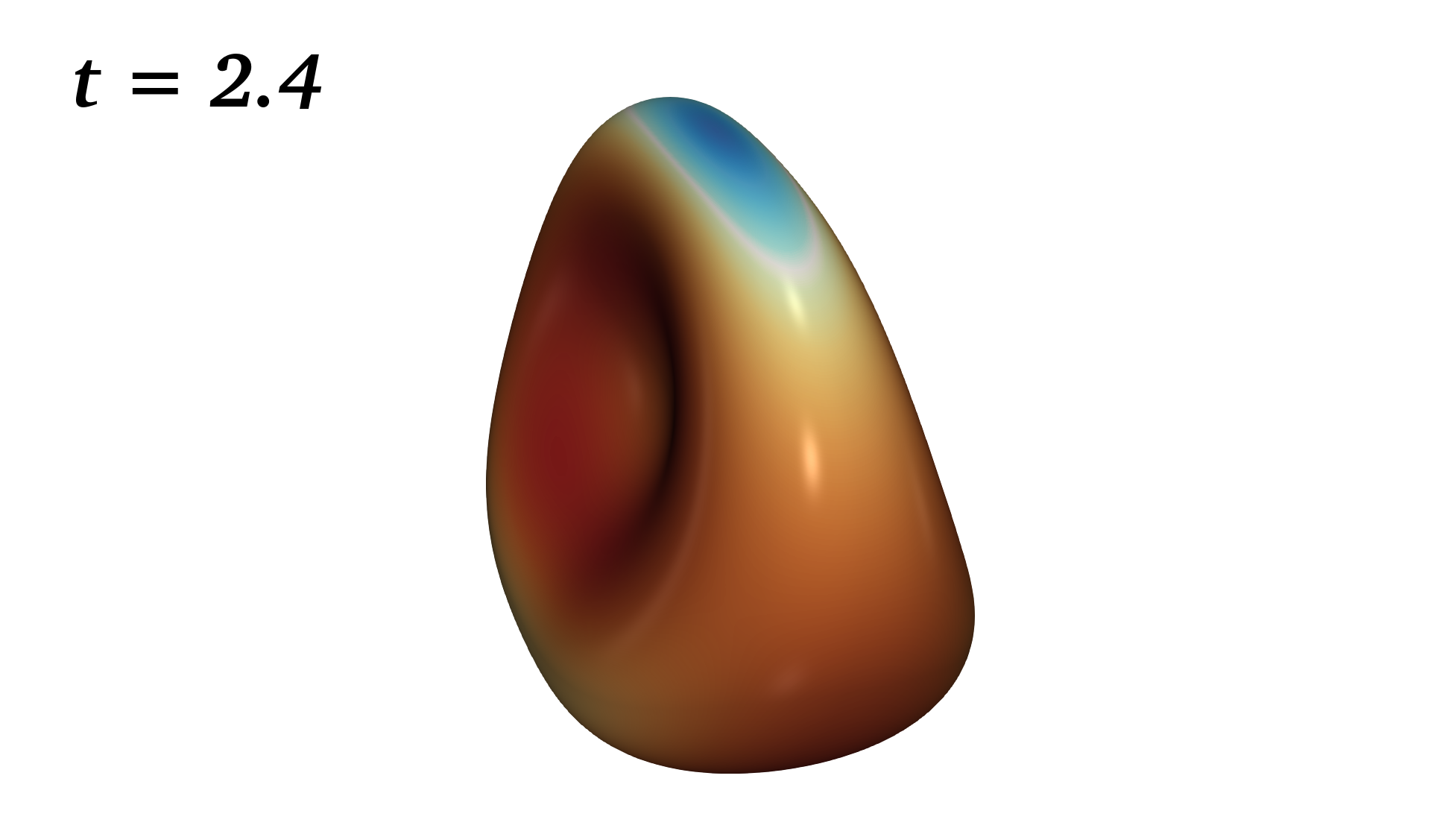}
    \includegraphics[width=0.18\linewidth]{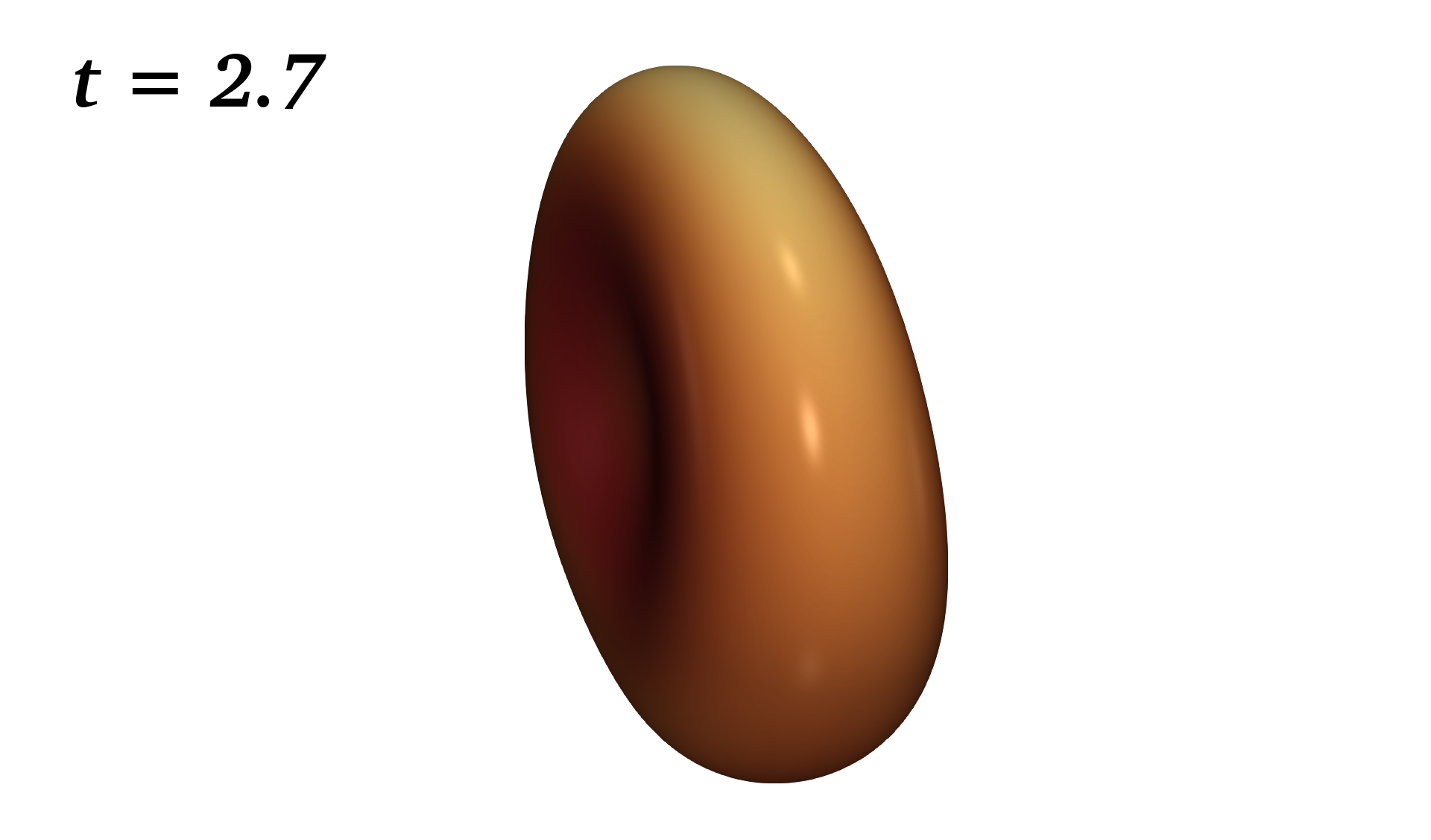}
    \includegraphics[width=0.18\linewidth]{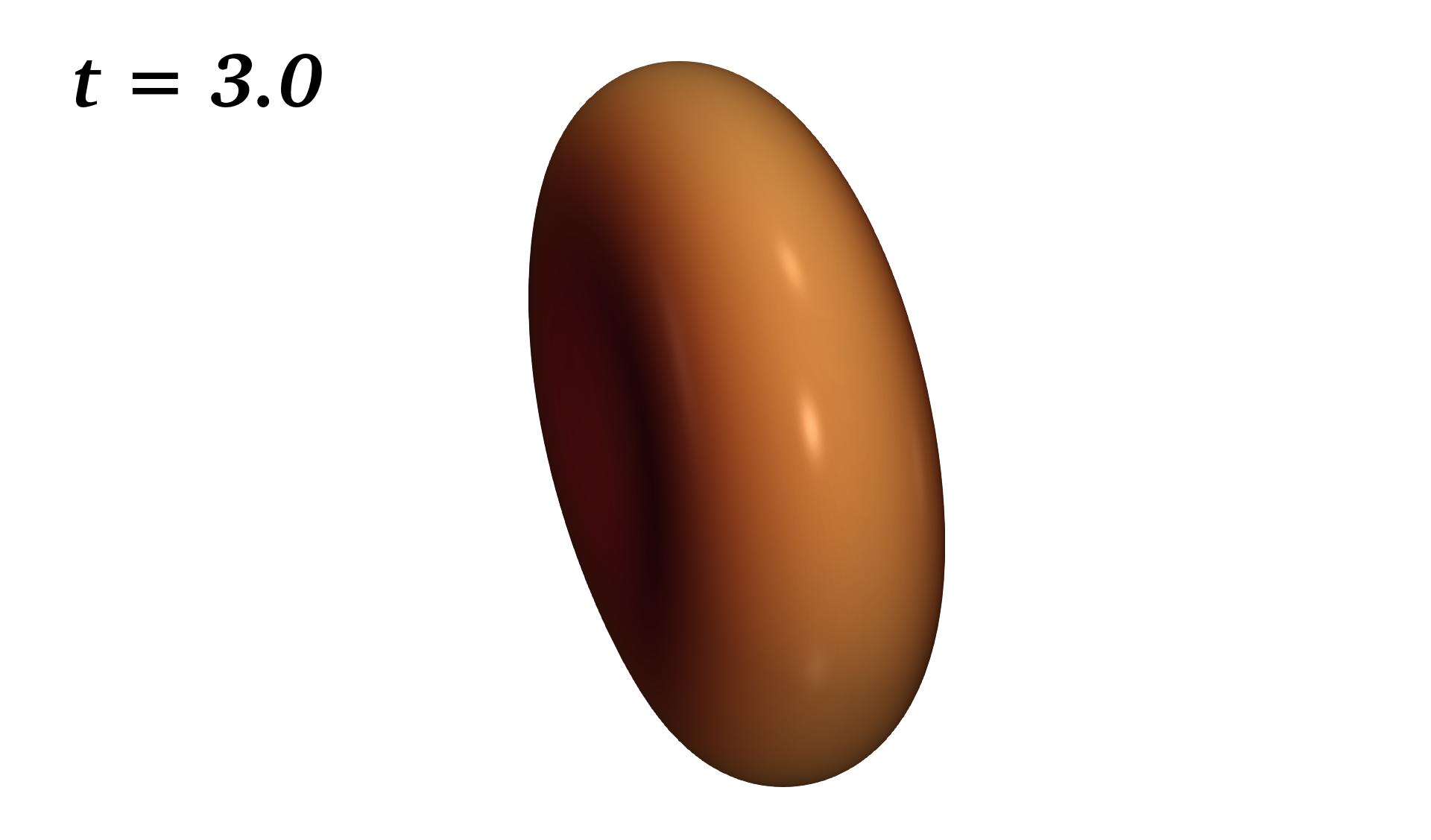}

    \includegraphics[width=0.18\linewidth]{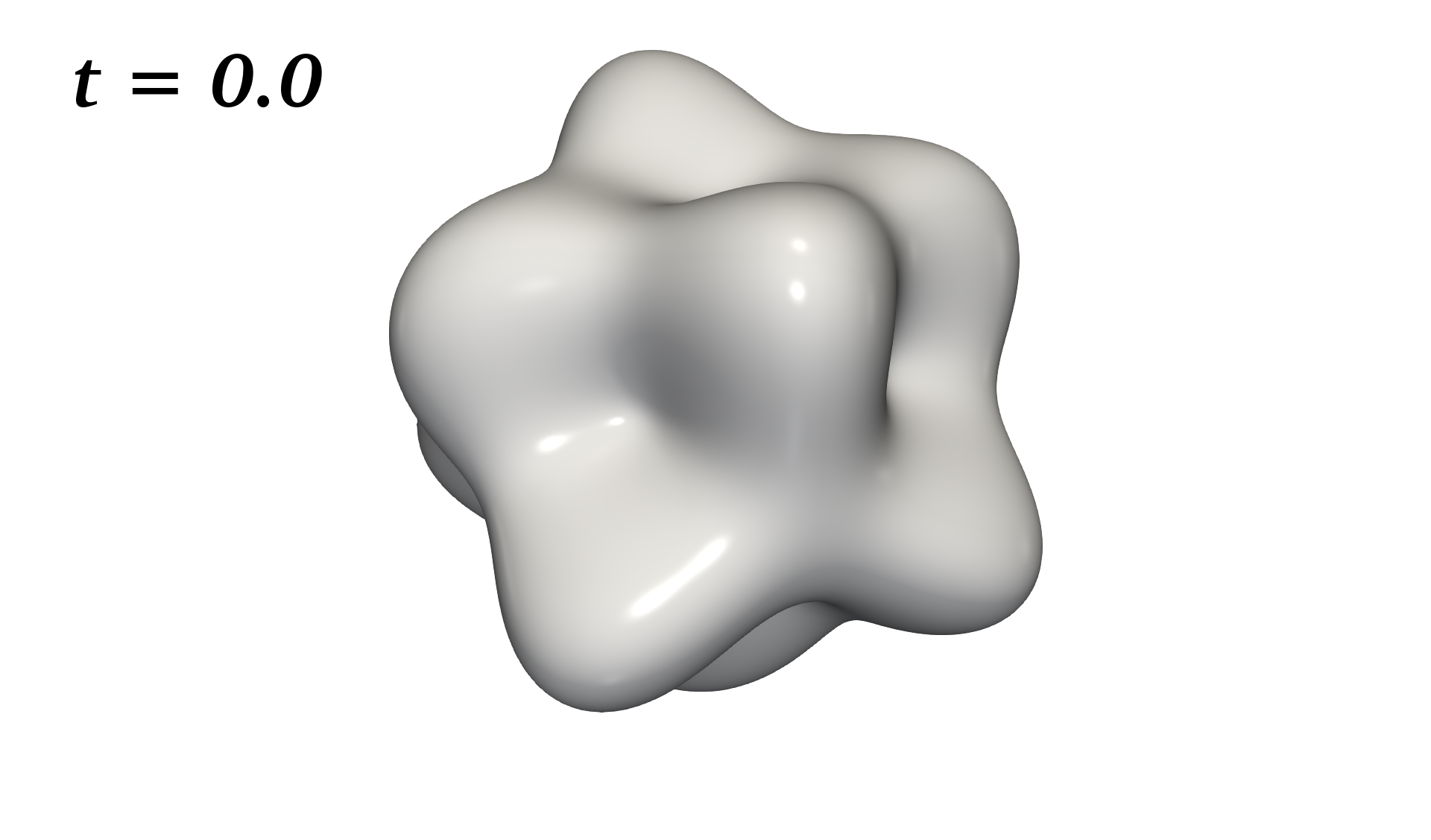}
    \includegraphics[width=0.18\linewidth]{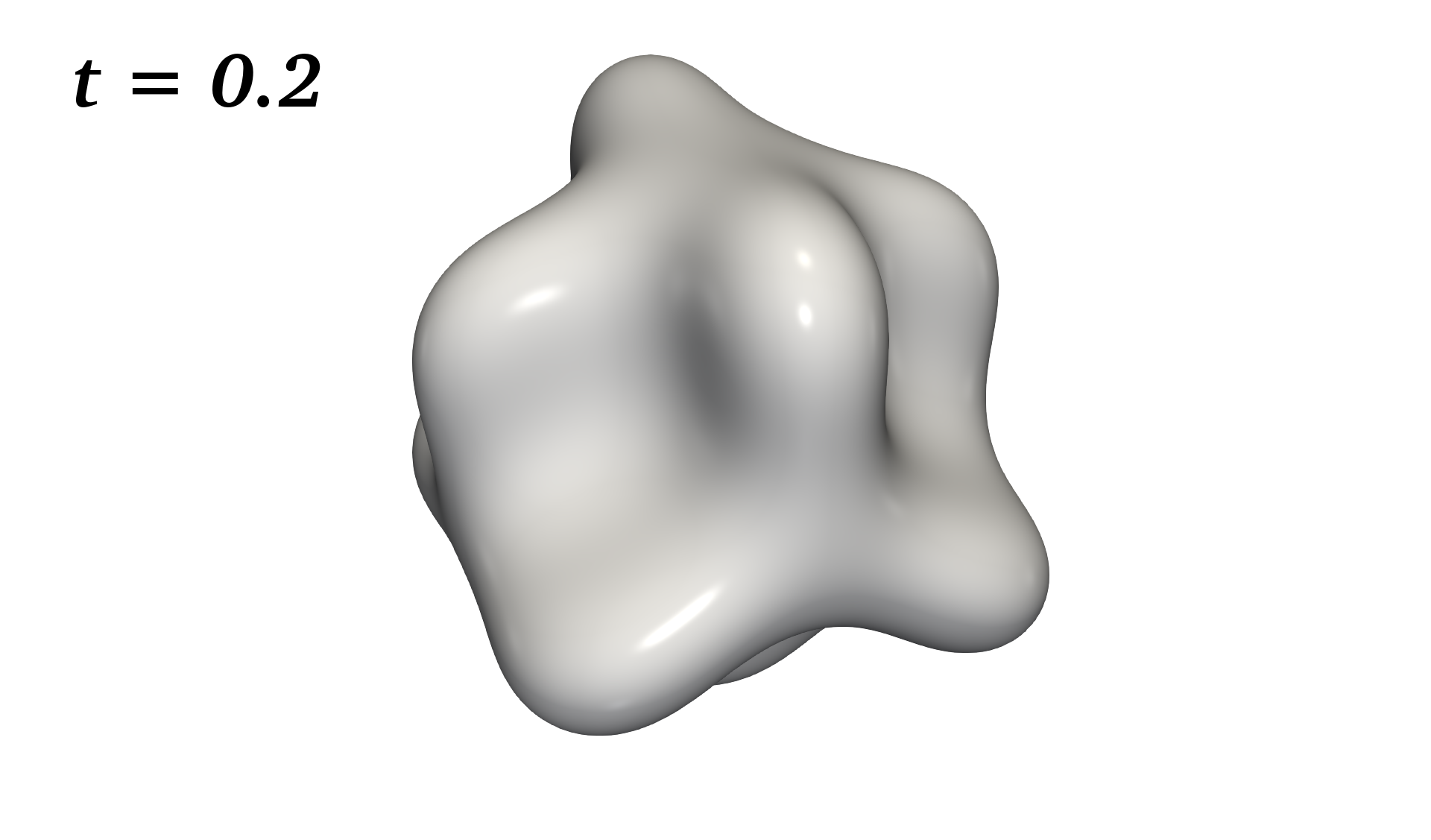}
    \includegraphics[width=0.18\linewidth]{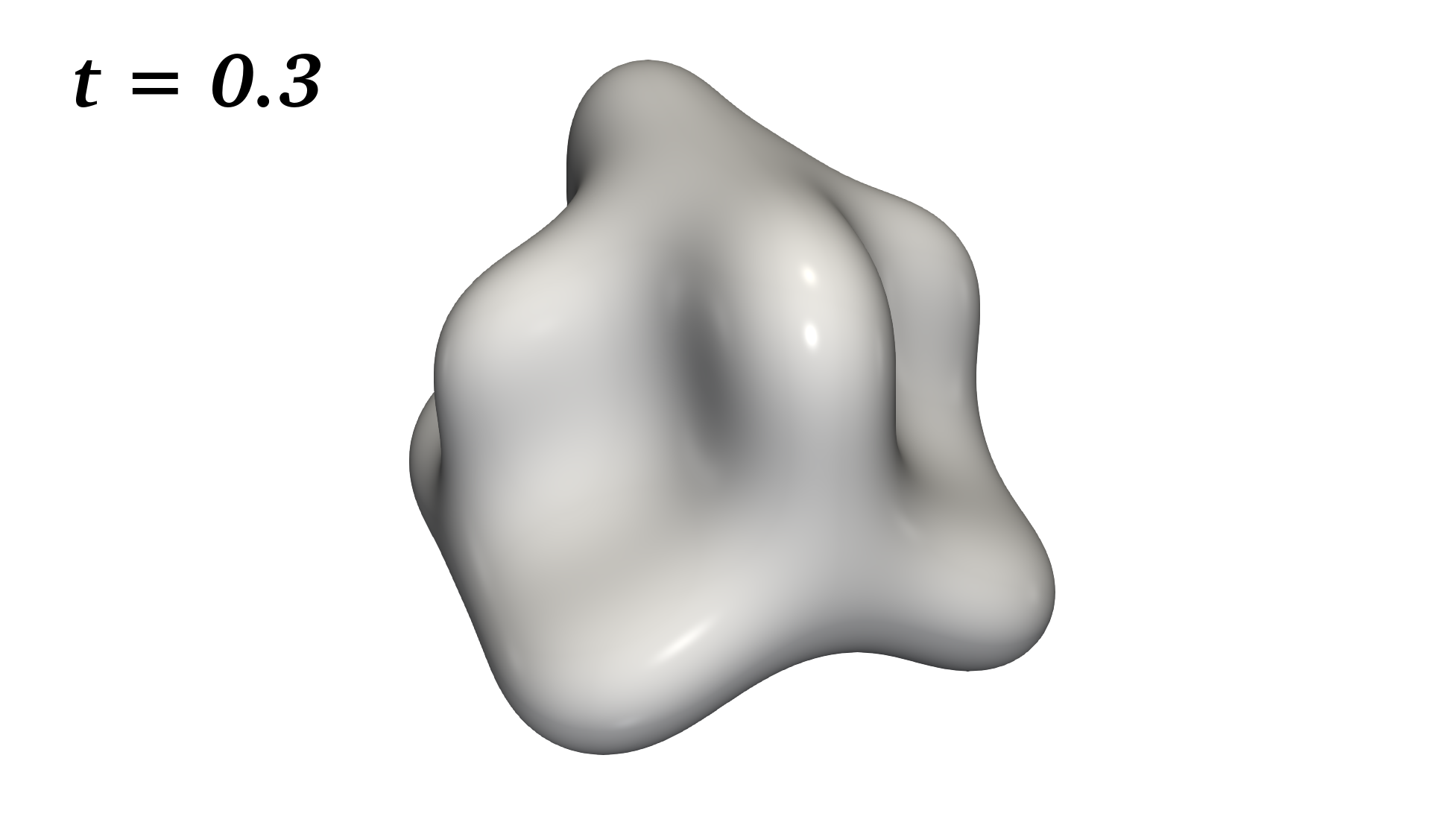}
    \includegraphics[width=0.18\linewidth]{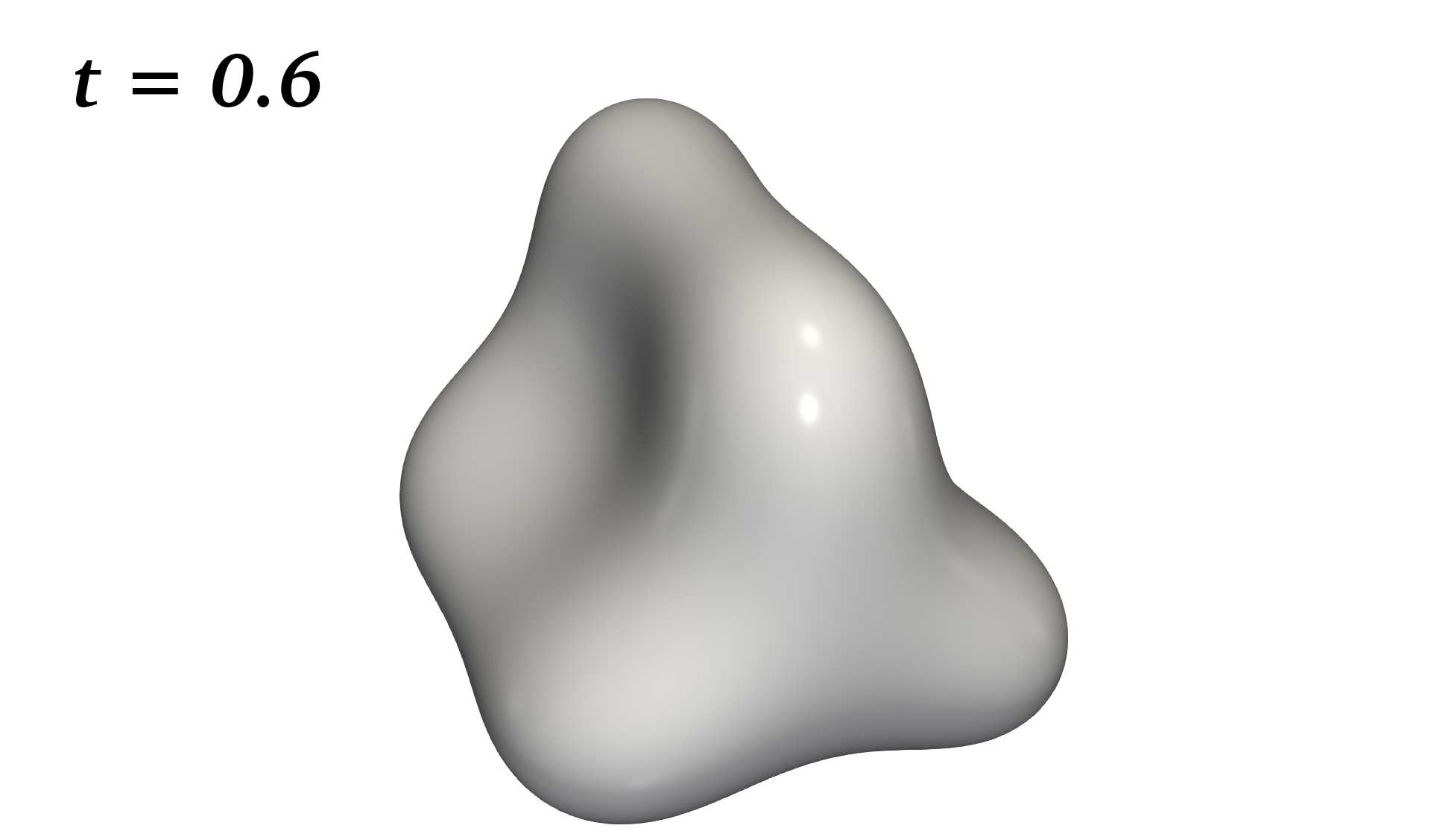}
    \includegraphics[width=0.18\linewidth]{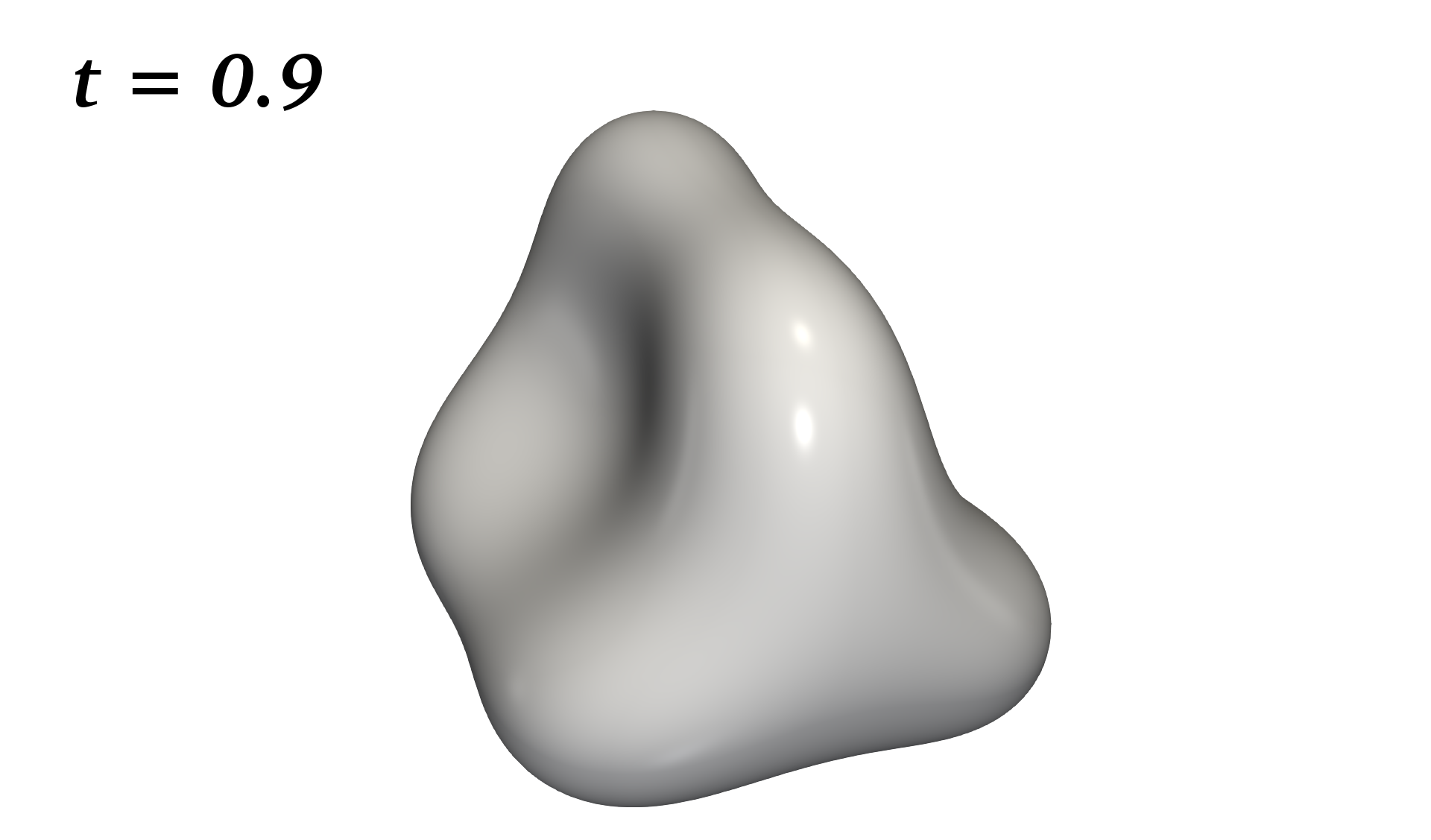}

    \includegraphics[width=0.18\linewidth]{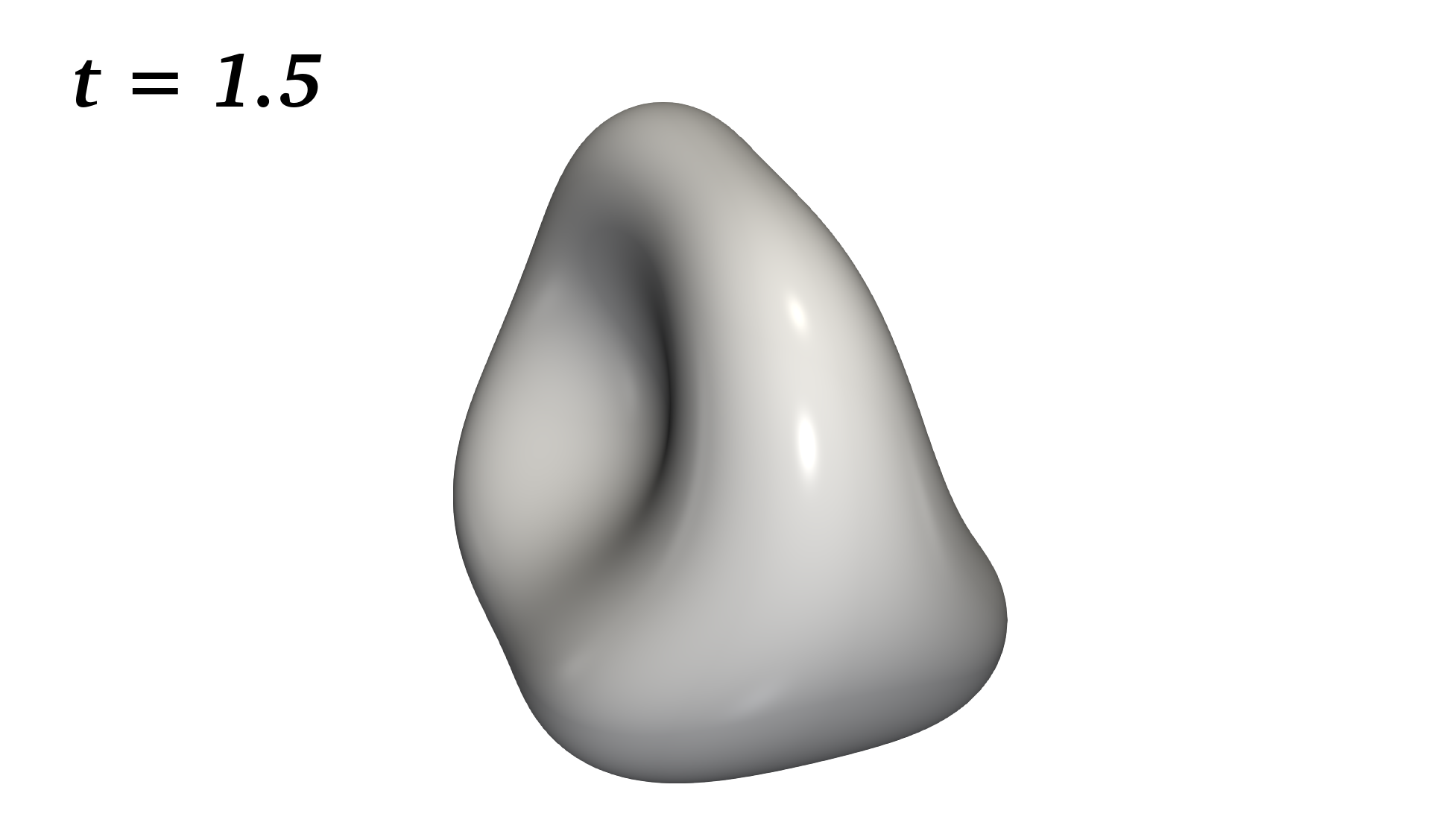}
    \includegraphics[width=0.18\linewidth]{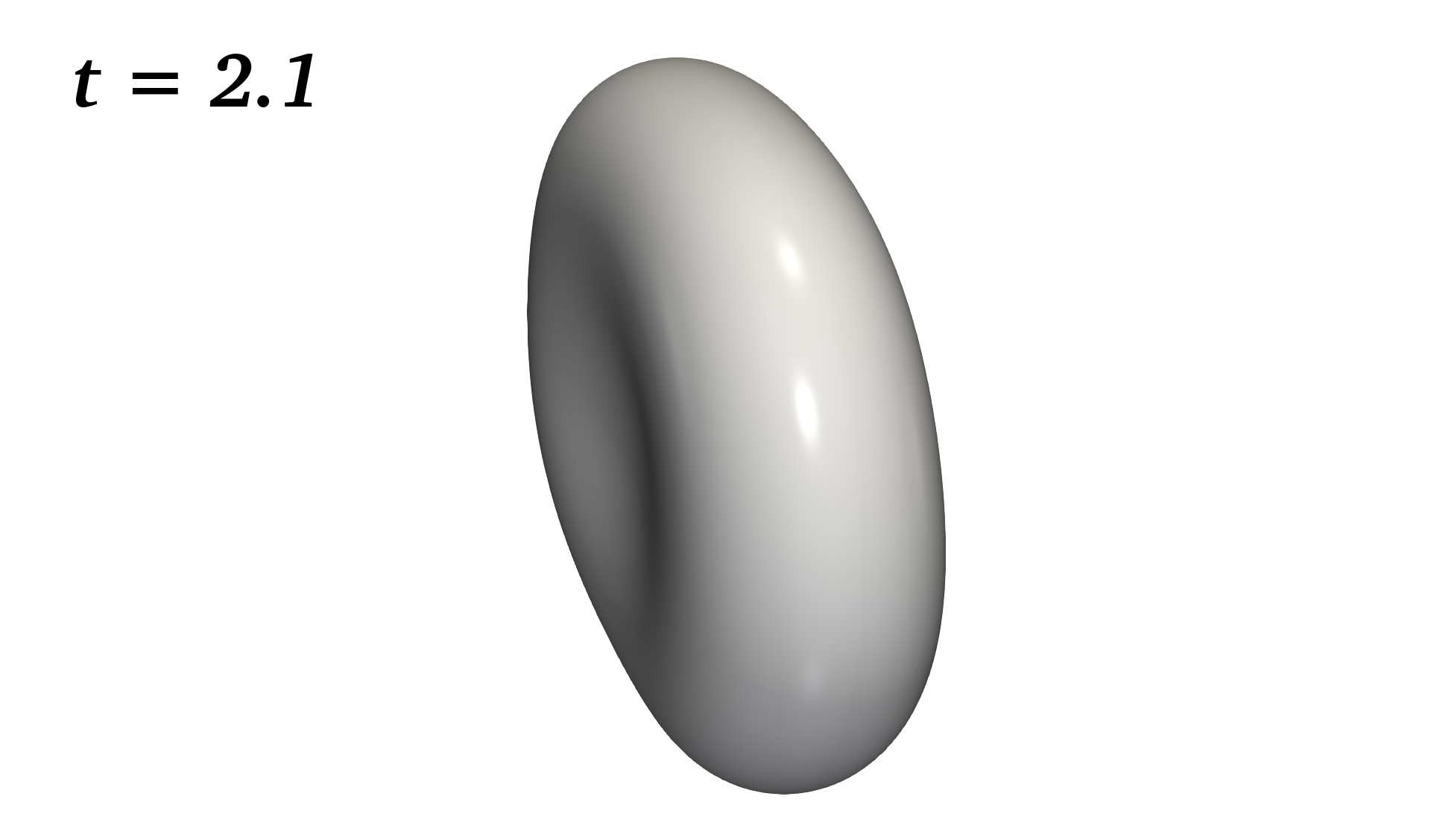}    \includegraphics[width=0.18\linewidth]{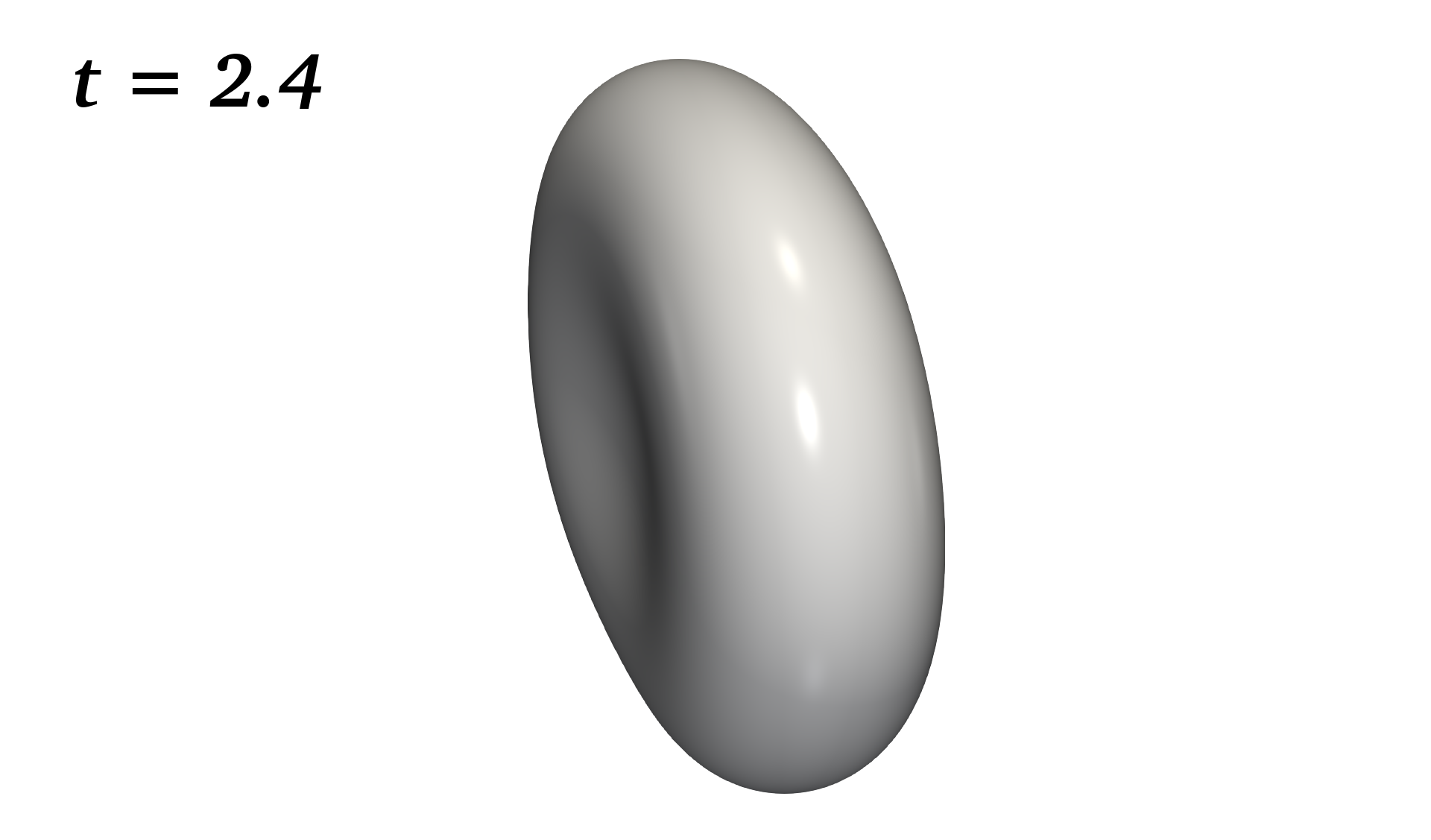}
    \includegraphics[width=0.18\linewidth]{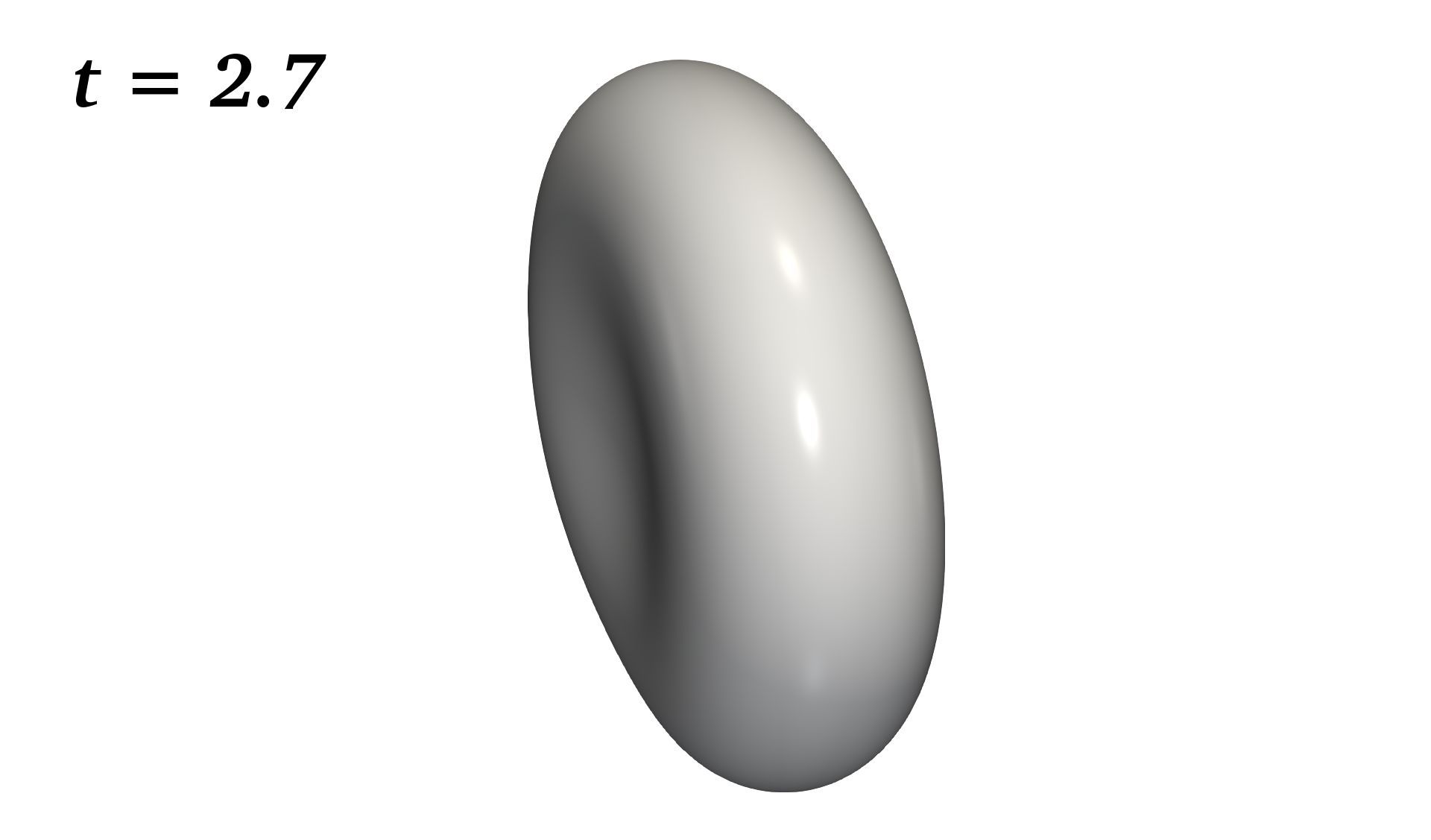}
    \includegraphics[width=0.18\linewidth]{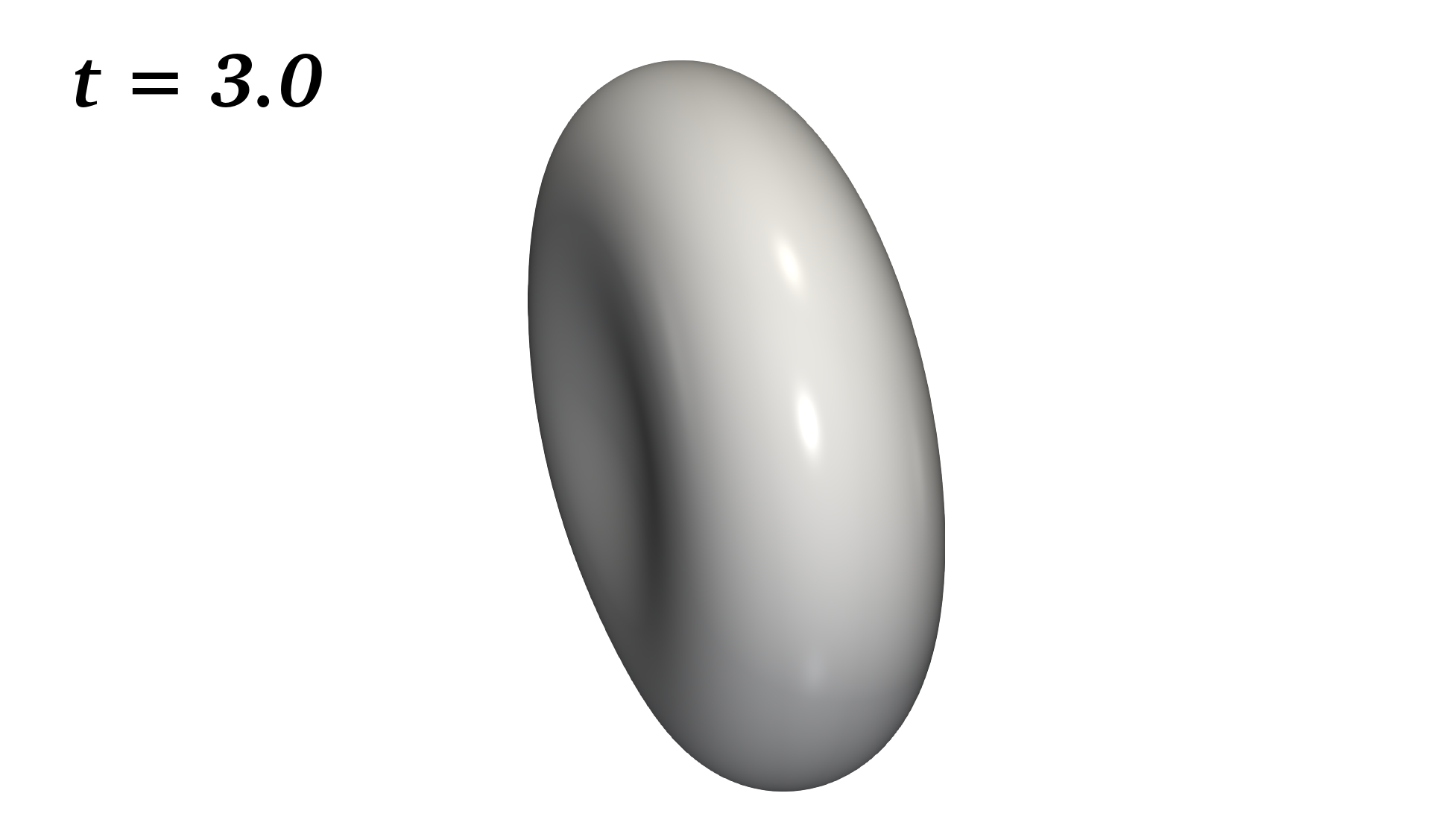}

    \includegraphics[width=0.2\linewidth]{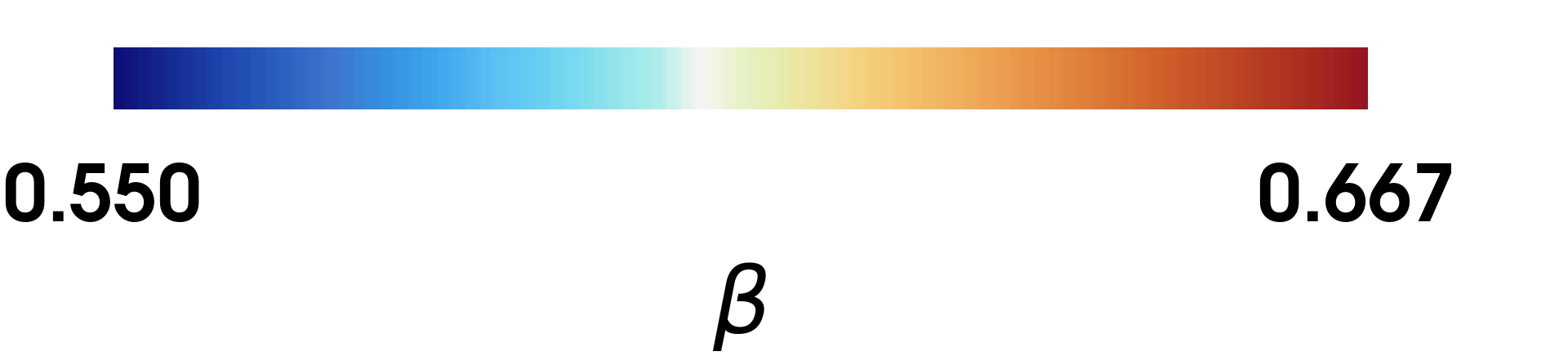}
    \caption{Comparison of evolution of the Surface Beris--Edwards--Helfrich (SBEH) model (top) and the Surface Stokes--Helfrich (SSH) model (bottom). Shown is the shape $\surf(t)$ and the liquid crystal order $\beta(t)$ (for SBEH) at selected time instances, starting from a perturbed sphere and converging to an oblate-like shape.}
    \label{fig:evolution}
\end{figure}

In order to further highlight the differences in evolution, we consider as initial shape $\surf(0)$ a sphere of radius $1$ perturbed by a spherical harmonic $\surf(0) = \{r+r_0Y^m_l(\phi,\vartheta): \phi \in [0,\pi],\vartheta\in[-\pi,\pi] \}$. We consider the case $l = 5, m = 3, r = 1, r_0 = 0.5$, which leads to a reduced volume of $V_r  = 0.738$, and again set $\beta(0) = \frac{2}{3}$. As before, we compare the dynamics of the Surface Beris--Edwards--Helfrich (SBEH) model \cref{eq:model,eq:model2} to the Surface Stokes--Helfrich (SSH) model \cref{eq:u-p,eq:variationaderivative}.
In Fig. \ref{fig:evolution} the shapes together with $\beta$ (for the Surface Beris--Edwards--Helfrich (SBEH) model) are shown at selected time instances. Both converge to an oblate-like shape and evolve through various symmetry breaking events. While qualitatively similar, differences in shape evolution are already visible, see, e.g. the shapes at $t=2.1$ and $t=2.4$. For the Surface Beris--Edwards--Helfrich (SBEH) model we also observe a decrease in $\beta$ at high curvature regions, which is in accordance with the results obtained for the stationary shapes in \cref{fig:PhaseDiagram}. Especially the reduced order along the edges of the threefold-structure at intermediate times (between $t = 0.6$ and $t = 2.4$) shows the impact on the evolution. For the Surface Beris--Edwards--Helfrich (SBEH) model it takes longer to break this symmetry and evolve towards the oblate-like shape.

\begin{figure}[tbp]
    \centering
    \input{images/Comparison}
    \caption{Comparison of various quantities, computed for the evolutions shown in \cref{fig:evolution}. (left) Energy evolutions of the Surface–Beris–Edwards–Helfrich (SBEH) model and the Surface–Stokes–Helfrich (SSH) model. (middle) Evolution of Minkowski functional $W_2(\surf)$ as a shape measure for both models. (right) Evolution of the standard deviation as a measure for the variation of $\beta$ for both models.}
    \label{fig:minkowski}
\end{figure}

In order to quantify these differences, we examine the evolution of the free energy $\potenergy$. For the Surface Stokes-Helfrich model, this again corresponds to the sum of $\energyH$, with $\kappa = \kappa(\frac{2}{3})$, $\kappa_G = \kappa_G(\frac{2}{3})$ and $\meanc_0 = 0$, and $\energyTH$ with $\beta = \frac{2}{3}$. Other quantities characterizing the shape are the Minkowski functionals $W_0(\surf) = V, W_1(\surf) = A$ and $W_2(\surf) = \int_\surf \meanc \, d \surf$ \cite{Mecke2000}. As the enclosed volume and the surface area are conserved quantities, only the curvature-weighted integral $W_2(\surf)$ is of interest. As a measure for $\beta$, we consider its standard deviation $\operatorname{std}(\beta) = \frac{1}{A}\int_\surf (\beta - \Bar{\beta})^2\dS$. All three quantities are shown in \cref{fig:minkowski}. Again the free energy $\potenergy$ is decreasing over time. While for the final state the Surface Beris-Edwards-Helfrich (SBEH) model also leads to a lower energy, during the evolution this is not always the case. Significant morphological changes occur earlier in time for the Surface Stokes-Helfrich (SSH) model and reduce the energy faster. This shift in time is also clearly shown for the shape measure $W_2(\surf)$. The differences in both curves correlate with the variations in $\beta$. The standard deviation strongly increases until the first morphological change. During this change it decreases and reaches a plateau. This plateau corresponds to the low values for $\beta$ at the edges of the threefold shape (between $t= 0.6$ and $t= 2.6$). The standard deviation is further decreased during the following morphological change of this shape towards an oblate. The final increase corresponds to the rearrangement of $\beta$ and the emerging lower values at the rim of the oblate. For the Surface Stokes-Helfrich (SSH) model the standard deviation is zero as $\beta \equiv \frac{2}{3}$. Also these dynamics for $\beta$, as well as for the shape $\surf$ provide experimentally testable results.

\section*{Discussion}

We have extended Helfrich's \cite{Helfrich_ZfNC_1973} classical curvature-elasticity free energy and various dynamic approaches based on it by explicitly treating the liquid crystal structure of the fluid lipid membrane. This replaces the Helfrich energy \cref{eq:helfrichenergy} by an elastic energy \cref{eq:energya} for a surface Q-tensor field aligned with the surface normal field. The dynamic equations are derived using the Lagrange--d'Alembert principle considering this energy and a thermotropic energy \cref{eq:energyb} and various flux potentials, such as the nematic viscous flux potential, the external friction-damping flux potential, the immobility flux potential, as well as constraints on inextensibility and enclosed volume. The derived system of equations is a special case of the Surface Beris--Edwards--Helfrich model \cite{nitschke2025beris}. However, it is significantly simpler, as instead of the full surface Q-tensor $\Qb$, only the scalar order field $\beta$ is considered. The structure of \cref{eq:model,eq:model2} is thus similar to the Surface Stokes--Helfrich (SSH) model \cref{eq:u-p,eq:variationaderivative} with parameters depending on $\beta$, additional terms containing $\nabla \beta$, and an additional evolution equation for $\beta$. Fixing $\beta$ constant leads to the Surface Stokes--Helfrich (SSH) model \cref{eq:u-p,eq:variationaderivative}. In the friction-dominated limit we obtain the $L^2$-gradient flow  \cref{eq:L2Helfrich,eq:variationaderivative}. The proposed Surface Beris--Edwards--Helfrich (SBEH) model \cref{eq:model,eq:model2} significantly differs from these special cases. It no longer treats the fluid lipid membrane as a homogeneous, isotropic thin material structure, but introduces a scale coupling between the microscopic details of lipid order and the mesoscopic length and time scales of shape evolution. This coupling is in both directions and thus provides a mechanical feedback mechanism. Locally varying liquid crystal order leads to local variations in bending rigidities and surface viscosity and thus influences the shape evolution, and the geometric properties of the shape influence the liquid crystal order.

Numerical solutions of the model confirm these multiscale couplings. Within the considered parameter space high curvature regions lead to reduced liquid crystal order. In turn, this reduced order locally reduces the bending rigidity and in principle thus allows for even higher curvature. Only the constraints on local inextensibility and conservation of enclosed volume restrict this feedback mechanism from amplifying.

As a consequence, the equilibrium shapes differ from minimizers of the Helfrich energy \cref{eq:helfrichenergy} under constraints on local inextensibility or global area and conservation of the enclosed volume. While for the classical problem the shape only depends on the reduced volume $V_r$, it now depends also on the elastic parameters. As a result of the interplay between the elastic and thermotropic energy, curvature and liquid crystal order can be balanced, leading to different shapes with varying liquid crystal order. The overall energy is reduced for these new shapes. The differences to the classical Seifert shapes (shown in red in \cref{fig:Shapes}) enhance for decreasing reduced volumes $V_r$ and increasing elastic parameter $L$. We have explored this only for the one-constant approximation $L = L_1$, $L_2 = L_3 = 0$. Going beyond this special case will introduce additional weighting between the different curvature contributions $\kappa(\beta) \meanc^2$ and $\kappa_G(\beta) \gaussc$, which can stabilize different shapes \cite{elliott2010modeling,barrett2017finite,sischka2025two,sischka2025influence}. Also the dynamics is affected by these mechanisms, essentially leading to a delay in shape evolution (shown in \cref{fig:minkowski} for $\potenergy$ and $W_2(\surf)$). While explored systematically only for the equilibrium shapes, also the considered dynamic shapes indicate a strong dependency of liquid crystal order $\beta$ on the curvature of the fluid lipid membrane $\norm{\shop}^2$. For equilibrium shapes a linear decrease, with slope depending on $L$, is extracted. All these general features, modified equilibrium shapes with varying liquid crystal order, as well as a decreasing liquid crystal order with curvature, call for experimental validation. The general trends shown in \cref{fig:Shapes} and \cref{fig:PhaseDiagram} can guide such experiments to confirm the impact of liquid crystalline order also in real systems.

The proposed model provides a multiscale coupling between the local membrane mechanics on the level of the lipids and the mesoscopic length and time scales of biological functions. Here, we considered only symmetric fluid lipid membranes and a thermotropic energy with a minimum at the fully ordered state $\beta = \frac{2}{3}$ and a metastable critical point at $\beta = 0$. However, the model allows for a general local potential $\dwpotential(\beta)$. A key step towards more quantitative results would certainly be a modification of this potential tailored to lipid bilayers, and the determination of the functional form and coefficients either experimentally or by theoretical deduction from molecular interactions. Early attempts in this direction have been considered in \cite{Seelig_QRoB_1977_Dmrtaatlm,Jaehnig_1981}. They rely on order parameters and their influence on fluid lipid membranes. Such lipid-related order parameters exist at multiple scales, see \cite{MorvanGrelardDufourc_BeBAB-B_2025_OpimFJSp} for a review. We here neglect intramolecular parameters (e.g. orientational order of specific covalent bonds or conformational changes) and only consider molecular parameters (e.g. orientation of the lipid molecules) and collective parameters (the overall contribution to membrane deformation). These order parameters are experimentally accessible and allow for tuning the thermotropic energy and validation of its impact on shape evolution for specific lipid molecules.

With further modifications of $\dwpotential(\beta)$, also phase separation becomes feasible, providing another mesoscale property critically influencing protein localization, signaling, and trafficking \cite{simons2011membrane}. Combinations with coarse-grained two-phase models \cite{julicher1996shape,jiang2000phase,kumar2001budding,lowengrub2009phase,elliott2010modeling,barrett2017finite,bachini2023derivation,sischka2025two} or their extensions to multi-phase systems can account for compositional diversity, which gives rise to lateral heterogeneity and even richer phase behavior. As already pointed out in \cite{NitschkeSischkaVoigt_JoFM_2026_Hlcmflb}, an extension to asymmetric fluid lipid membranes can be realized by additional degrees of freedom which couple the surface Q-tensor $\Qb$ and the surface normal $\normal$ to break the symmetry. Flexoelectric polarization \cite{GoversVertogen_PRA_1984,BarberoDozovPalierneDurand_PRL_1986,Alexe-Ionescu_PLA_1993} realizes such couplings and allows to express the spontaneous curvature $\meanc_0$ in terms of these coupling terms and the liquid crystal order $\beta$. More quantitative results and presumably stronger deviations from the Surface Stokes-Helfrich (SSH) model probably also require to take area changes into account, as the area per lipid also depends on the liquid crystalline order $\beta$. This requires weakening the local inextensibility constraint, e.g. by accounting for variations in thickness, as addressed in \cite{sischka2026}, and/or area growth/shrinkage, as addressed in \cite{krause2024wrinkling,PBV25}. In both cases, the variations will need to depend on $\beta$. Such terms are not considered in the Landau-de Gennes thermotropic energy $\energyTH$ in \cref{eq:energyTH}. Other possible extensions consider the fluid lipid membrane as a composite structure made of two distinct monolayers coupled by inter-monolayer friction \cite{USeifert_1993,torres2019modelling}. Combining this mechanism with explicit treatment of liquid crystal order of both leaflets provides additional mechanisms to couple local membrane mechanics with biological functions.


\section*{Appendix}

\subsection*{Notation}

As we are dealing with tensor-fields on surfaces and the evolution of these surfaces, some notation from differential geometry is required. We assume that the time-dependent moving surface $ \surf=\surf(t)\subset\R^3 $ is sufficiently smooth and parameterizable into the 3-dimensional~Euclidean space,
\eg, by a parameterization $\para(t)=\para(t,\cdot,\cdot):\R^2\supset\mathcal{U}\rightarrow\surf\vert_{t}$,
or more generally, by a collection of parameterizations providing an open cover of $\surf$.
Let be $\xb(t)\in\surf(t)$, characterized by the existence of $(y^1,y^2)\in\mathcal{U}$ such that $\xb(t) = \para(t,y^1,y^2)$ holds.
Throughout this paper, we assume that $\surf$ has no boundaries.

Tensor fields in $ \tangentR[^n] $ are considered exclusively on the surface.
Coordinate invariance, and the resulting freedom in the choice of frame, allow for an interpretation as a Cartesian frame for the tensor fields.
For instance, we could write $ \Wb = W^A\eb_A $ (Einstein summation) with $ A\in\{x,y,z\} $ for a vector, \resp\ 1-tensor, field $ \Wb\in\tangentR := \tangentR[^1] $.
A special vector field is the normal field $ \normal\in\tangentR $ perpendicular to the surface and normalized.
It spans a one-dimensional subvector field space, whose orthogonal complement consists of the tangential vector fields $ \tangentS < \tangentR $.
The associated projection is given by the surface identity tensor field $ \IdS $, \ie\ it is $\tangentS = \IdS\tangentR  $ valid.
This concept readily scales to $n$-tensor fields $ \tangentR[^n] $ and tangential $n$-tensor fields $ \tangentS[^n] < \tangentR[^n] $.
The surface identity tensor field $ \IdS\in\tangentS[^2] $ is such a tangential tensor field and could be represented  by $ \IdS = \Id - \normal\otimes\normal $, where $ \Id\in\tangentR[^2] $ is the usual Euclidean identity tensor field, \eg\ given by $ \delta^{AB}\eb_A\otimes\eb_B $ \wrt\ a Cartesian frame.
Furthermore, our framework also involves Q-tensor fields in $ \tangentQR := \{ \Qb\in\tangentR[^2] \mid \Qb = \Qb^T \text{ and } \Tr\Qb=0 \} $
and tangential Q-tensor fields in $ \tangentQS := \{ \qb\in\tangentS[^2] \mid \qb = \qb^T \text{ and } \Tr\qb=0 \}\le \tangentQR $, also known as flat-degenerated Q-tensor fields.
A complete orthogonal decomposition of $ \tangentQR $ is given in \cref{eq:QDecomposition}.

The local inner product is denoted by angle brackets:
$ \inner{}{\cdot,\cdot}: \mathcal{V}\times\mathcal{V} \rightarrow\tangentScal $,
where $ \mathcal{V} \le \tangentR[^n] $.
A specification of the frame is not needed as coordinate invariance and the properties of subvector spaces guarantee that the results remain consistent.
Note that we also use the dot notation for the inner product of vector fields,
\ie, $\Vb\cdot\Wb = \inner{}{\Vb,\Wb}$ for all $\Vb,\Wb\in\tangentR$.
All norms are defined as those induced by the inner products:
$ \normsq{}{\cdot} := \inner{}{\cdot,\cdot} $.

The covariant derivative $\nabla:\tangentS[^n]\rightarrow\tangentS[^{n+1}]$ is defined in the usual manner, \eg, by use of Christoffel symbols if a local metric is defined.
The covariant divergence $\div:\tangentS[^{n+1}]\rightarrow\tangentS[^n]$ is minus the $L^2$-adjoint of $\nabla$ and it holds $\div = -\nabla^* = \Tr\circ\nabla$.
As a consequence, the Laplace-Beltrami operator is given by $\Delta=\div\circ\nabla:\tangentScal\rightarrow\tangentScal$ on scalar fields.
We use the componentwise derivative $ \nablaC: \tangentR[^n] \rightarrow \tangentR[^n]\otimes\tangentS < \tangentR[^{n+1}] $ as the starting point for derivatives on tensor fields, since it captures all possible first-order derivative information \wrt\ the ambient Euclidean space.
In principle, it is the ordinary $ \R^3 $-derivative modulo the normal derivative, \ie\ $ \nablaC = (\IdS\nabla_{\R^3})\vert_{\surf} $, should one wish to consider extensions in the normal direction, although $ \nablaC $ is invariant \wrt\ any sufficiently smooth normal extension.
If we use an arbitrary tangential frame $ \{\partial_i\para\mid i=1,2\} $ we could define $ \nablaC\Rb := g^{ik}(\partial_k\Rb)\otimes\partial_i\para $ for all $ \Rb\in\tangentR[^n] $,
where $ g^{ik} $ yields the contravariant proxy of the metric tensor, also known as inverse metric tensor, \wrt\ the chosen frame.
Equivalently, it holds $ [\nablaC\Rb]^{A_1 \ldots A_n B} \eb_B = \nabla R^{A_1 \ldots A_n}$ for $ \Rb= R^{A_1 \ldots A_n} \eb_{A_1}\otimes\ldots\otimes \eb_{A_n}$ in a Cartesian frame.
The shape operator $ \shop\in\tangentS[^2] $, also known as the second fundamental form, follows directly from this via $ \shop = -\nablaC\normal $.
Other curvature related quantities derived from this are the mean curvature $ \meanc=\Tr\shop = \IdS\dbdot\shop $
and the Gaussian curvature $ \gaussc = \frac{1}{2}( \meanc^2 - \normsq{}{\shop} ) $, the two scalar-valued invariants of the shape operator.
On scalar fields the componentwise and covariant derivative are the same, \ie\ $ \nablaC=\nabla:\tangentScal\rightarrow\tangentS $.
This is not generally valid, not even for tangential tensor fields.
For a vector field $ \Wb=\wb+\wnor\normal\in\tangentR $, \eg\, holds
$ \nablaC\Wb = \nabla\wb - \wnor\shop + \normal\otimes\left( \nabla\wnor + \shop\wb\right) $.
As the divergence operator, we use the component-wise trace divergence $ \DivC := \Tr\circ\nablaC $, where the trace applies on the two rear column dimensions, including the derivative acting on the last index.
It should be noted that the trace divergence equals the $ \hil $-adjoint $ -\nablaC^{*} $ only for right-tangential tensor fields,
since the correct relation is $ \DivC\Rb = - ( \nablaC^{*}\Rb + \meanc\Rb\normal ) $ for all $ \Rb\in\tangentR[^n] $.
Fortunately, this is true for fluid stress fields in this paper.
However, at least for pressure gradients, we need the $ \hil $-adjoint of the trace divergence.
This gives rise to the adjoint gradient $ \GradC:=-\DivC^{*} $.
The key representations in the covariant differential calculus relevant to this paper are
\begin{align*}
    \forall\sigmab\in\tangentS[^2],\etab\in\tangentS&:
    &\DivC\left( \sigmab + \normal\otimes\etab \right)
        &=\div\sigmab - \shop\etab + \left( \div\etab + \shop\dbdot\sigmab \right)\normal\formComma\\
    \forall\wb\in\tangentS,\wnor\in\tangentScal&:
    &\DivC\left(\wb+\wnor\normal\right) &= \div\wb - \wnor\meanc\formComma\\
    \forall p\in\tangentScal&:
    &\GradC p &= \DivC(p \IdS) = \nabla p + p \meanc \normal\formPeriod
\end{align*}
The material derivative on scalar fields $\beta\in\tangentScal$ is given by
\begin{align*}
	\dot{\beta}
		&:= \ddt\beta
		  = \partial_t\beta + \inner{}{\vb-\vb_{\ofrak},\nabla\beta}
		  = \partial_t\beta + \inner{}{\Vb-\Vb_{\ofrak},\nablaC\beta}\formComma
\end{align*}
where $\Vb=\vb+\vnor\normal\in\tangentR$ is the material velocity and $\Vb_{\ofrak}=\vb_{\ofrak}+\vnor\normal\in\tangentR$ is an arbitrary observer velocity, defined as the velocity of the local moving coordinate system with respect to which $\beta$ is evaluated, see \cite{NITSCHKE2022104428,nitschke2023tensorial,bachini2023derivation}.
For instance, $\vb_{\ofrak}=\vb$ yields the Lagrangian perspective, or, $\vb_{\ofrak}=\nullb$ yields the tangential Eulerian, \resp\ transversal, perspective.

\subsection*{Model derivation}

We derive the equations of motion using the Lagrange--d'Alembert principle, following the same approach as in \cite{NitschkeSischkaVoigt_JoFM_2026_Hlcmflb}.
For this purpose we consider the potential, \resp\ free, energy
\begin{align*}
	\potenergy
		&:= \energyEL + \energyTH\formComma
\end{align*}
which describes the energetic (instantaneous) state of the system, and comprises the elastic energy in \cref{eq:energyEL} and thermotropic energy in \cref{eq:energyTH}.
In addition, we include the flux potential
\begin{align*}
	\fluxpotential
		&:= \energyNV + \energyFD + \energyIM + \energyIE + \energyVP\formComma
\end{align*}
which governs the process of energy exchange with the environment, typically through heat dissipation,
and contains the (nematic) viscous flux potential in \cref{eq:energyNV}, external friction-damping flux potential in \cref{eq:energyFD},
immobility flux potential in \cref{eq:energyIM}, Lagrange-multiplier inextensibility flux potential in \cref{eq:energyIE}, and Lagrange-multiplier volume-preserving flux potential in \cref{eq:energyVP}.
Globally on the time interval $\Tcal\subset\R$ and surface $\surf\subset\R^3$, the Lagrange--d'Alembert principle reads
\begin{align}\label{eq:LDA_global}
	\inner{\Tcal\times\surf}{\deltafrac{\Acal}{\para}, \Wb} + \inner{\Tcal\times\surf}{\deltafrac{\Acal}{\beta}, \psi}
		&= \int_{\Tcal} \inner{\surf}{\deltafrac{\fluxpotential}{\Vb}, \Wb}
						+\inner{\surf}{\deltafrac{\fluxpotential}{\dot{\beta}}, \psi}
						+\inner{\surf}{\deltafrac{\energyIE}{p}, \theta}
						+ \left(\ddfrac[\energyVP]{\lambda}\right)\phi  \dt
\end{align}
for all virtual displacements $\Wb\in\tangentR$, $\psi,\theta\in\tangentScal$, and $\phi:\Tcal\rightarrow\R$,
where the variations are to be taken \wrt\ the corresponding $L^2$-inner products, whose domains of integration are specified by the subscripts,
\ie, $ \inner{\Tcal\times\surf}{\bullet,\bullet} = \int_{\Tcal}\int_{\surf}\inner{}{\bullet,\bullet}\dS\dt $,
and $ \inner{\surf}{\bullet,\bullet} = \int_{\surf}\inner{}{\bullet,\bullet}\dS $.
The action functional and Lagrangian are given by
\begin{align*}
	\Acal
		&= \int_{\Tcal}\Lcal\dt\formComma
	&\Lcal
		&= 	\int_{\surf} \frac{\rho}{2}\normsq{}{\Vb}\dS  - \potenergy	\formPeriod
\end{align*}
State variables are the surface and the statistical order of the lipid molecules, which are aligned apolar with the normal direction $\normal$ on average.
The surface is represented by $\para$, which could be a (set of) parameterization(s).
The statistical lipid order is represented by the scalar field $\beta\in\tangentScal$, see below.
Genuine process variables are the material velocity $\Vb\in\tangentR$ and  lipid order rate $\dot{\beta}\in\tangentScal$.
The material velocity $\Vb= \vb + \vnor\normal$ has a tangential component $\vb\in\tangentS$ and a normal component $\vnor\in\tangentScal$,
where $\Vb=\partial_t\para$ holds if $\para$ parameterized not only the surface $\surf$ geometrically but also the material.
Auxiliary process variables are $p\in\tangentScal$ and $\lambda:\Tcal\rightarrow\R$, which serve as Lagrange function and Lagrange parameter to enforce local inextensibility of the surface and preservation of the enclosed volume, respectively.
By adapting the procedure used in \cite{nitschke2025beris} to the present setting, we localize \cref{eq:LDA_global} in time and space.
We neglect inertial forces, since lipid layers have a very low areal mass density and are subject to substantial viscous damping from the surrounding medium on the relevant time scales.
This yields the governing equations:
\begin{align}\label{eq:LDA_local}
	\fb_{\potenergy} + \fb_{\fluxpotential} &= \nullb \formComma
	&\omega_{\potenergy} + \omega_{\fluxpotential} &= 0 \formComma
	&c_{\IE} &= 0 \formComma
	&c_{\VP} &= 0 \formComma
\end{align}
where the generalized applied forces are weakly given as follows:
For brevity, we omit force contributions that vanish identically, as shown below.
For the fluid equation, the Lagrangian force $\fb_{\potenergy}\in\tangentR$ and flux force $\fb_{\fluxpotential}\in\tangentR$ read
\begin{align}\label{eq:fluid_forces}
	\begin{alignedat}{4}
		\fb_{\potenergy}
			&= \fbEL + \fbTH &&:\quad
		&\inner{\surf}{\fbEL,\Wb}
			&= -\inner{\surf}{\deltafrac{\energyEL}{\para},\Wb}\formComma\quad
		&\inner{\surf}{\fbTH,\Wb}
			&= -\inner{\surf}{\deltafrac{\energyTH}{\para},\Wb}\formComma\\
		\fb_{\fluxpotential}
			&= \fbNV + \fbFD + \fbIE + \fbVP &&:\quad
		&\inner{\surf}{\fbNV,\Wb}
			&= -\inner{\surf}{\deltafrac{\energyNV}{\Vb},\Wb}\formComma\quad
		&\inner{\surf}{\fbFD,\Wb}
			&= -\inner{\surf}{\deltafrac{\energyFD}{\Vb},\Wb}\formComma\\
   &&&&\inner{\surf}{\fbIE,\Wb}
			&= -\inner{\surf}{\deltafrac{\energyIE}{\Vb},\Wb}\formComma\quad
		&\inner{\surf}{\fbVP,\Wb}
			&= -\inner{\surf}{\deltafrac{\energyVP}{\Vb},\Wb}\formComma
	\end{alignedat}
\end{align}
for all $\Wb\in\tangentR$.
For the lipid order equation, the Lagrangian force $ \omega_{\potenergy} $ and flux force $ \omega_{\fluxpotential} $ read
\begin{align}\label{eq:order_forces}
\begin{alignedat}{4}
	\omega_{\potenergy}
  		&= \omegaEL + \omegaTH &&:\quad
	&\inner{\surf}{\omegaEL,\psi}
		&= -\inner{\surf}{\deltafrac{\energyEL}{\beta},\psi}\formComma\quad
	&\inner{\surf}{\omegaTH,\psi}
		&= -\inner{\surf}{\deltafrac{\energyTH}{\beta},\psi}\formComma\\
  	\omega_{\fluxpotential}
  		&= \omegaNV + \omegaIM &&:\quad
  	&\inner{\surf}{\omegaNV,\psi}
  		&= -\inner{\surf}{\deltafrac{\energyNV}{\dot{\beta}},\psi}\formComma\quad
  	&\inner{\surf}{\omegaIM,\psi}
  		&= -\inner{\surf}{\deltafrac{\energyIM}{\dot{\beta}},\psi}\formComma\quad
\end{alignedat}
\end{align}
for all $\psi\in\tangentScal$.
The inextensibility constraint force $ c_{\IE}\in\tangentScal$ and volume-preserving constraint force $ c_{\VP}:t\mapsto\R$ read
\begin{align*}
	\inner{\surf}{c_{\IE}, \theta}
		&= -\inner{\surf}{\deltafrac{\energyIE}{p}, \theta}\formComma
	&c_{\VP}
		&= -\ddfrac[\energyVP]{\lambda}\formComma
\end{align*}
for all $\theta\in\tangentScal$.
Note that the variation \wrt\ the surface is not uniquely determined unless the behavior of $\beta$ under this variation is specified,
\ie\ unless it is specified in what sense $\beta$ is independent of $\para$, see \cite{nitschke2023tangential}.
The equations derived here, however, are invariant under this choice, since any admissible choice of the corresponding independence relation cancels out in the governing equations.
For simplicity, we therefore present the applied fluid forces below under the convention that the direct variation of $\beta$ \wrt\ $\para$ vanishes,
\ie, with a slight abuse of notation, we set ``$\deltafrac{\beta}{\para}=0$''.

A reasonable orthogonal decomposition for Q-tensor fields $ \Qb\in\tangentQR $ \cite{nitschke2025beris} is provided by
\begin{align}\label{eq:QDecomposition}
	\Qb &= \qb + \etab\otimes\normal + \normal\otimes\etab + \beta\left( \normal\otimes\normal - \frac{1}{2}\IdS \right)\formComma
\end{align}
where $ \qb\in\tangentQS $ yields the flat degenerated part, $ \etab\in\tangentS $ the surface non-conforming part, and $ \beta\in\tangentScal $ the uniaxial normal part.
Since we restrict our attention to lipid molecules that are, on average, oriented normal to the surface, we set $\qb=\nullb$ and $\etab=\nullb$, and consider only uniaxial Q-tensor fields aligned with the surface normal field:
\begin{align}\label{eq:lipid_ansatz}
	\Qb &= \beta\left( \normal\otimes\normal - \frac{1}{2}\IdS \right) = S \left( \normal\otimes\normal - \frac{1}{3}\Id \right)\formComma
\end{align}
where $ S = \frac{3}{2}\beta $ is the scalar order field, \ie, $\beta$ ranges from 0 (disordered state) to $\frac{2}{3}$ (ordered state) locally.
Thus, modulo an arbitrary sufficiently smooth extension in the normal direction $\normal$, the spatial derivative of $\Qb$ is given by
\begin{align*}\label{eq:nablaCQAnsatz}
	\nablaC\Qb
	&= \left( \normal\otimes\normal - \frac{1}{2}\IdS \right) \otimes\nabla\beta
	-\frac{3}{2}\beta\left( \normal\otimes\shop + (\normal\otimes\shop)^{T_{(1\, 2)}}\right) \formComma
\end{align*}
which serves as the basis for the following elastic energy contributions:
\begin{align*}
	\frac{L_1}{2}\normsq{}{\nablaC\Qb} 
		&= \frac{3 L_1}{4}\left( \normsq{}{\nabla\beta} + 3 \beta^2 \left( \meanc^2 - 2\gaussc \right)\right)\formComma \\
	\frac{L_2}{2}\normsq{}{\DivC\Qb} 
		&= \frac{L_2}{8}\left( \normsq{}{\nabla\beta} + 9 \beta^2 \meanc^2 \right)\formComma\\
	\frac{L_3}{2} \inner{}{\nablaC\Qb, \nablaC^{T}\Qb} 
		&=\frac{L_3}{8}\left( \normsq{}{\nabla\beta} + 9 \beta^2 \left( \meanc^2 - 2\gaussc \right) \right)\formComma
\end{align*}
where we assume the lipid molecules to be achiral and therefore omit the $L_4$ term. Since the lipid molecules can exhibit only splay and saddle-splay characteristics, the degenerate splay-bend state in three-dimensional liquid crystals does not exist. Consequently, we also omit the $L_6$ term and any other cubic terms \cite{SchieleTrimper_pssb_1983,BerremanMeiboom_PRA_1984}. All these terms can be derived as a thin film limit from the classical elastic energy contributions in the three-dimensional theory \cite{nitschke2018nematic}. The elastic energy yields
\begin{align}\label{eq:energyEL}
	\energyEL
		&:= \frac{1}{2}\int_{\surf} L_1 \normsq{}{\nablaC\Qb} + L_2 \normsq{}{\DivC\Qb} + L_3 \inner{}{\nablaC\Qb, \nablaC^{T}\Qb} \dS
		 = \int_{\surf} \frac{l_1}{2} \normsq{}{\nabla\beta} + \frac{\meanelastic(\beta)}{2}\meanc^2 + \gausselastic(\beta)\gaussc \dS
\end{align}
with order distortion coefficient $l_1$, and $\beta$-dependent curvature-elastic moduli $\meanelastic(\beta)$ and $ \gausselastic(\beta)$, given by
\begin{align*}
	l_1
		&= \frac{1}{4}\left(6 L_1 + L_2 + L_3  \right) \formComma
	&\meanelastic(\beta)
		&= \frac{9}{4}\beta^2 (2 L_1 + L_2 + L_3) \formComma
	&\gausselastic(\beta)
		&= - \frac{9}{4}\beta^2 (2 L_1  + L_3)
\end{align*}
in terms of elastic parameters $L_\bullet$.
If we consider the fully ordered state ($\beta=\frac{2}{3}$), we recover the Helfrich energy \cref{eq:helfrichenergy} for symmetric surfaces,
\ie\ it is $ \energyEL\vert_{\beta=2/3} = \energyH\vert_{\meanc_0=0}$ valid. If we assume $L_2=0$, the curvature-elastic moduli are no longer mutual independent as $\kappa(\beta) = - \kappa_G(\beta)$, and we obtain
$ \energyEL\vert_{L_2=0} =  \int_{\surf} \frac{l_1}{2} \normsq{}{\nabla\beta} + \frac{\meanelastic(\beta)}{2}\left( \meanc^2 - 2\gaussc \right)\dS $,
where $ \meanc^2 - 2\gaussc= \normsq{}{\shop}$.
The commonly used one-constant approximation in liquid crystal theory yields $\energyEL\vert_{L_1=L, L_2=L_3=0} = \int_{\surf} \frac{3L}{4} \normsq{}{\nabla\beta} + \frac{9L}{4}\beta^2\left( \meanc^2 - 2\gaussc \right)\dS$.
Due to the $\beta$-dependent curvature-elastic moduli, in contrast with $\energyH$, even without topological changes, the Gaussian curvature term can no longer be neglected. The dependency on $\beta$ furthermore requires to consider the variational derivative $\frac{\delta \energyEL}{\delta \para}$ not only with respect to normal but also tangential variations. For a detailed discussion we refer to \cite{NITSCHKE2022104428,nitschke2023tensorial,nitschke2023tangential,backofen2026scalar}. Following \cite{NitschkeSischkaVoigt_JoFM_2026_Hlcmflb} we obtain
\begin{align*}
	\fbEL
		&= \GradC\left(\frac{l_1}{2} \normsq{}{\nabla\beta} + \frac{\meanelastic(\beta)}{2}\meanc^2 \right)
			-\DivC\Big( l_1\nabla\beta\otimes\nabla\beta + \meanelastic(\beta)\meanc\shop
						+\normal\otimes\Big( \nabla(\meanelastic(\beta)\meanc) + \meanc\nabla\gausselastic(\beta) - \shop\nabla\gausselastic(\beta)  \Big)\Big)\\
		&= \frac{l_1}{2} \GradC\normsq{}{\nabla\beta}
			+ \left(\frac{\meanelastic'(\beta)}{2}\meanc^2 + \gausselastic'(\beta)\gaussc \right)\nabla\beta\\
		&\quad
			- \left(  \Delta(\meanelastic(\beta)\meanc) + \frac{\meanelastic(\beta)}{2}\meanc\left( \meanc^2 - 4\gaussc \right)
					+\div\left( \meanc\nabla\gausselastic(\beta) - \shop\nabla\gausselastic(\beta) \right)\right)\normal\\
		&= -\omegaEL\nabla\beta
		   - \left( l_1\inner{}{\shop\nabla\beta,\nabla\beta} -  l_1\frac{\meanc}{2}\normsq{}{\nabla\beta}
		           +\Delta(\meanelastic(\beta)\meanc) + \frac{\meanelastic(\beta)}{2}\meanc\left( \meanc^2 - 4\gaussc \right)
		           +\div\left( \meanc\nabla\gausselastic(\beta) - \shop\nabla\gausselastic(\beta) \right)\right)\normal \formComma\\ 
	\omegaEL
		&= l_1\Delta\beta - \frac{\meanelastic'(\beta)}{2}\meanc^2 - \gausselastic'(\beta)\gaussc
\end{align*}
which contains normal and tangential components.
The Landau--de Gennes thermotropic energy yields a lipid order state potential:
\begin{align}\label{eq:energyTH}
    \energyTH
        	&= \int_{\surf} a \Tr\Qb^2 + \frac{2b}{3}\Tr\Qb^3 + c\Tr\Qb^4 \dS
           = \int_{\surf} \dwpotential(\beta) \dS\formComma
    &\text{with}\quad\dwpotential(\beta)
          &= \frac{\beta^2}{8}\left( 12a + 4b\beta + 9c\beta^2 \right)\formPeriod
\end{align}
The original purpose of this energy is to drive the nematic fields towards either the isotropic state or specific uniaxial states.
For lipid systems, however, other state potentials may be considered.
We therefore treat $\dwpotential(\beta)$ as a general local potential throughout the following.
Variation \wrt\ the surface and scalar order result in
\begin{align*}
    \fbTH
      &= \GradC\dwpotential(\beta)
       = -\omegaTH\nabla\beta + \dwpotential(\beta)\meanc\normal \formComma
    &\omegaTH
      &= -\dwpotential'(\beta)\formPeriod
\end{align*}
The nematic viscous flux potential \cite{nitschke2025beris} measures the temporal distortion of an anisotropic nematic metric $ \Id - \xi\Qb $ by its lower-convected rate $ \Dlow\left(\Id - \xi\Qb \right) $ \cite{nitschke2023tensorial}.
Ansatz \cref{eq:lipid_ansatz} reduces this to a lipid order viscous flux potential:
\begin{align}
	\energyNV
		&= \frac{\coeffIF}{4}\int_{\surf} \normsq{}{\Dlow\left(\Id - \xi\Qb \right)}\dS
     = \frac{\coeffIF}{4}\int_{\surf} \normsq{}{\Dlow\left(\left( 1 + \frac{\xi}{2}\beta \right)\IdS + (1-\xi\beta)\normal\otimes\normal \right)}\dS\notag\\
		&= \frac{\coeffIF}{4}\int_{\surf} \left( 1 + \frac{\xi}{2}\beta \right)^2\normsq{}{\IdS\nablaC\Vb + (\IdS\nablaC\Vb)^T}
		 								  +\frac{3}{2}\xi^2\dot{\beta}^2
		                                  +2\xi\left( 1 + \frac{\xi}{2}\beta \right)\dot{\beta}\DivC\Vb\dS\formComma \label{eq:energyNV}
\end{align}
where $\coeffIF\ge 0$ is the reference viscosity and $\xi\in\R$ dimensionless order-viscosity coupling parameter.
Variation \wrt\ the material velocity and order rate yield
\begin{align*}
	\fbNV
		&= \frac{\coeffIF\xi}{2} \GradC\left( \left( 1 + \frac{\xi}{2}\beta \right)\dot{\beta} \right)
			+ \coeffIF\DivC\left( \left( 1 + \frac{\xi}{2}\beta \right)^2 \left( \IdS\nablaC\Vb + (\IdS\nablaC\Vb)^T \right)\right)\\
		&= \coeffIF\div\left( \left( 1 + \frac{\xi}{2}\beta \right)^2 \left(\nabla\vb + \nabla^T\vb - 2\vnor \shop \right) \right)
		+ \frac{\coeffIF\xi}{2} \nabla \left( \left( 1 + \frac{\xi}{2}\beta \right)\dot{\beta} \right)\\
		&\quad +\left( \coeffIF \left( 1 + \frac{\xi}{2}\beta \right)^2 \inner{}{\shop, \nabla\vb + \nabla^T\vb - 2\vnor \shop}
           + \frac{\coeffIF\xi}{2}\left( 1 + \frac{\xi}{2}\beta \right)\dot{\beta}\meanc \right)\normal\formComma\\
	\omegaNV
		&= 	-\frac{3\coeffIF\xi^2}{4}\dot{\beta}
		    -\frac{\coeffIF\xi}{2}\left( 1 + \frac{\xi}{2}\beta \right)	\DivC\Vb
		 = -\frac{3\coeffIF\xi^2}{4}\dot{\beta}
		   -\frac{\coeffIF\xi}{2}\left( 1 + \frac{\xi}{2}\beta \right)\left(\div\vb - \vnor\meanc \right)\formComma
\end{align*}
where the original anisotropic nematic viscous force degenerates to an isotropic $\beta$-depending viscous force modulo pressure terms.
Other equivalent representations of this force can be found in \cite{backofen2026scalar}.
Another mechanism contributing to energy dissipation in the overall system is the motion of the surface itself, which we control through the
external friction-damping flux potential:
\begin{align}\label{eq:energyFD}
    \energyFD
      &= \int_{\surf} \frac{\gamma_{\parallel}}{2} \normsq{}{\vb} + \frac{\gamma_{\bot}}{2}\vnor^2 \dS
       = \frac{\gamma}{2} \int_{\surf} \normsq{}{\Vb} \dS\formComma
\end{align}
where, for the sake of simplicity, we set the surface-friction equal the normal-damping, \ie, $\gamma := \gamma_{\parallel} = \gamma_{\bot} $.
By variation, we obtain the fluid force
\begin{align*}
    \fbFD &= -\gamma\Vb\formComma
\end{align*}
without any additional general forces in the order-parameter equation, \ie\, $\omegaFD=0$.
In general, the associated fluid force would result in $ \fbFD = -(\gamma_{\parallel}\vb + \gamma_{\bot}\vnor\normal)$.
Analogously, but with respect to a Q-tensor rate rather than the surface rate, we consider the
Jaumann immobility flux potential:
\begin{align}\label{eq:energyIM}
    \energyIM
      &= \frac{M}{2}\int_\surf\normsq{}{\Djau\Qb}\dS
       = \frac{M}{2}\int_\surf \normsq{}{\dot{\beta}\left( \normal\otimes\normal - \frac{1}{2}\IdS \right)}\dS
       = \frac{3M}{4}\int_\surf \normsq{}{\dot{\beta}}\dS\formComma
\end{align}
where $\Djau:\tangentQR \rightarrow \tangentQR$ is the Jaumann, \resp\ corotational, rate on Q-tensor fields, see \cite{nitschke2023tensorial}.
Here, we consider only Jaumann immobility, since the lipid molecules,  in a fixed layer medium, are approximately invariant under rigid-body rotations of the surface.
For material immobility, see \cite{backofen2026scalar}.
Variation \wrt\ the order rate yields
\begin{align*}
    \omegaIM &= -\frac{3M}{2}\dot{\beta}\formPeriod
\end{align*}
No additional fluid forces are obtained from this flux potential, \ie\ $\fbIM=\nullb$.
To enforce local inextensibility of the surface, we use the inextensibility flux potential as the associated Lagrange-multiplier term:
\begin{align}\label{eq:energyIE}
    \energyIE
      &= -\int_{\surf} p \DivC\Vb \dS
\end{align}
with Lagrange parameter $p\in\tangentScal$.
Variation \wrt\ $\Vb$ and $p$, and noting that $p$ enters the model only through this term, yields
\begin{align*}
    \fbIE &= -\GradC p\formComma
    &0 &= \DivC\Vb = \div\vb - \vnor\meanc\formComma
\end{align*}
without any additional general forces in the order-parameter equation, \ie\, $\omegaIE = 0$.
To ensure conservation of the enclosed volume, we introduce the volume-preserving flux potential as the associated Lagrange-multiplier term:
\begin{align}\label{eq:energyVP}
    \energyVP
      &= \lambda\int_{\surf} \Vb \cdot \normal \dS
\end{align}
with Lagrange parameter $\lambda: t \mapsto \R$.
Variation \wrt\ $\Vb$ and $\lambda$, and noting that $\lambda$ enters the model only through this term, yields
\begin{align*}
    \fbVP &= - \lambda\normal\formComma
	&0 &=  \int_{\surf} \Vb \cdot \normal \dS\formComma
\end{align*}
without any additional general forces in the order-parameter equation, \ie\, $\omegaVP = 0$.

As a final step in this section, we consider the energy rate resulting from the equations in \cref{eq:LDA_local}.
Since, as argued above, inertial forces have been neglected, we disregard the kinetic energy rate and consider exclusively the potential/free energy rate $ \ddt\potenergy$.
Following \cite{nitschke2023tangential} and substituting \cref{eq:LDA_local}, we obtain
\begin{align}\label{eq:energyrate_I}
	\ddt\potenergy
		&= \inner{\surf}{\frac{\partial\potenergy}{\partial\para},\Vb} + \inner{\surf}{\frac{\partial\potenergy}{\partial\beta},\dot{\beta}}
		 = \inner{\surf}{\deltafrac{\potenergy}{\para}, \Vb} + \inner{\surf}{\deltafrac{\potenergy}{\beta}, \dot{\beta}}
		 = -\int_{\surf} \inner{}{\fb_{\potenergy}, \Vb} + \omega_{\potenergy}\dot{\beta}\dS
		 = \int_{\surf} \inner{}{\fb_{\fluxpotential}, \Vb} + \omega_{\fluxpotential}\dot{\beta}\dS\formComma
\end{align}
where we have adopted the same independence assumption between $\beta$ and $\para$ as above,
under which the partial and total variations \wrt\ $\para$ coincide.
We emphasize, however, that this assumption is without loss of generality.
The identity \cref{eq:energyrate_I} is obtained for any admissible choice of the independence relation between $\beta$ and $\para$.
See \cite{nitschke2025beris}, where a similar result is established for the Surface Beris-Edwards-Helfrich model without imposing a specific independence assumption.
Taking into account the composition of the flux forces in \cref{eq:fluid_forces,eq:order_forces} and substituting the inextensible and volume-preserving constraints,
which are also part of \cref{eq:LDA_local}, we finally obtain
\begin{align}\label{eq:energyrate_final}
	\ddt\potenergy
		&= \int_{\surf} \inner{}{\fbNV + \fbFD + \fbIE + \fbVP, \Vb} + \left( \omegaNV + \omegaIM \right)\dot{\beta}\dS
		 = -2\left( \energyNV + \energyFD + \energyIM \right)
		 \le 0 \formPeriod
\end{align}

\subsection*{Model}
Including all previously derived generalized applied forces, the governing equations in \cref{eq:LDA_local} lead to \cref{eq:model,eq:model2}. Substituting all relevant contributions into the energy rate \cref{eq:energyrate_final}, we obtain
\begin{align*}
	\ddt\potenergy
		&= - \int_{\surf} \gamma \normsq{}{\Vb}
						+ \frac{3}{2}\left( M + \frac{\coeffIF\xi^2}{2} \right)\normsq{}{\dot{\beta}}
						+ \frac{\coeffIF}{2}\left( 1 + \frac{\xi}{2}\beta \right)^2\normsq{}{\IdS\nablaC\Vb + (\IdS\nablaC\Vb)^T}
			\dS
		\le 0 \formComma
\end{align*}
\ie\, the model is guaranteed to be dissipative.

\subsection*{Numerics}

\cref{eq:model,eq:model2} are discretized using an Arbitrary-Lagrangian-Eulerian (ALE) Surface Finite Element Method (SFEM) combining the approaches considered in \cite{bachini2023derivation,sischka2025two, NitschkeSischkaVoigt_JoFM_2026_Hlcmflb}. For completeness, we summarize the approach here. Note, that for simplicity, we restrict ourselves to the one-constant approximation $L_1 = L, L_2 = L_3 = 0$.

\subsubsection*{Spatial discretization of an evolving surface}
The discrete structure mirrors the setup described in the previous sections. We start with a fixed reference grid $\mathcal{M}_\meshparam$ with sphere topology, on which we define a discrete parametrization $\para_h(t) \in \DiscSpaceVector{k}(\mathcal{M}_\meshparam)$. Here $\DiscSpaceVector{k}(\mathcal{N}_h)$ is the vectorial $k$th order Lagrange Finite Element space on the triangulation $\mathcal{T}_{\mathcal{N}_\meshparam}$ of the grid $\mathcal{N}_h$, that is $\DiscSpace{k}(\mathcal{N}_h) :=\{ \psi \in C^0(\mathcal{N}_h) \,\vert \, \psi\vert_\Cell \in \mathcal{P}_{k}(\Cell) \,\forall \Cell \in \mathcal{T}_{\mathcal{N}_h}\}$ and $\DiscSpaceVector{k} := \DiscSpace{k}^3$. The discrete, now evolving in time, surface $\surf_\meshparam(t)$ is given by the image of $\para_\meshparam(t)$, inheriting the grid topology from $\mathcal{M}_\meshparam$, i.e. its triangulation is given by $\mathcal{T}_{\surf_\meshparam} = \{\para_\meshparam(T) \vert T \in \mathcal{T}_{\mathcal{M}_\meshparam}\}$. Note, that the elements of this triangulation are $k$th order curved triangles. All problems we solve for a given time $t$ are solved on $\surf_\meshparam$, using Finite Element spaces $\DiscSpace{k}(\surf_\meshparam)$ and $\DiscSpaceVector{k}(\surf_\meshparam)$.
To solve the Stokes part of the system, we use the Taylor-Hood element, such that $\Vb_h \in \DiscSpaceVector{k}(\surf_\meshparam)$ and $p_h \in \DiscSpace{k-1}(\surf_\meshparam)$, and similarly $\beta_h \in \DiscSpace{k}(\surf_\meshparam)$. Note, that the former mandates $k \geq 2$. Throughout this paper, we set $k=3$.
\subsubsection*{Curvature treatment}
In principle, one can compute approximations of the curvature quantities $\shop$, $\meanc$, and $\gaussc$ directly from the curved grid $\surf_\meshparam$. Due to the important role of $\meanc$ in the problem \cref{eq:model}, we use the well known `BGN' method \cite{barrett2008parametric} to provide an semi-implicit representation of $\meanc$. To this end, we extend the system \cref{eq:model} by the contracted Gauss-Weingarten equation
\begin{align*}
    \DivC(\nablaC \para) = \meanc \normal,
\end{align*}
where the main idea is to replace $\para$ with the new surface in the time discretization, see below. Note that in the continuous setting, this leaves the tangential part of $\partial_t \para$ free, but in the time discrete setting introduces a beneficial tangential part. The system is then closed by requiring the change in the parametrization $\partial_t \para$ to equal the material velocity $\Vb$ on normal components, that is $\partial_t \para \cdot \normal = \Vb \cdot \normal$,
which formalizes the ALE approach. The additional variables are taken in isoparametric Finite Element spaces, that is $\meanc_h \in \DiscSpace{k}(\surf_\meshparam)$ and the surface update (see below) $\paramUpdate_\meshparam \in \DiscSpaceVector{k}(\surf_\meshparam)$.

\subsubsection*{Time discretization}
We use a semi-implicit approach in time, since a fully implicit scheme would require to solve on an unknown surface. To avoid this nonlinearity, we consider a given surface $\surf_{\meshparam}(t)$ and consider all differential operators, Finite Element spaces, inner products, and geometric quantities \wrt\ $\surf_{\meshparam}(t)$. Moreover, we decouple the evolution equation for $\beta$ from the system. Treating all linear terms implicitly, that is we solve for the pullback of those quantities from the unknown next surface onto $\surf_\meshparam(t)$, and using a simple semi-implicit treatment for the  terms nonlinear in $\beta$, we obtain the following discrete scheme: At time $t_{n-1}$, given a surface $\surf_\meshparam(t_{n-1})$ described by $\para^{n-1}$, a timestep $\tau$, and $(\Vb_\meshparam^{n-1}, \beta_h^{n-1}, \paramUpdate^{n-1})\in[\DiscSpaceVector{3}\times \DiscSpace{3} \times \DiscSpaceVector{3}](\surf_\meshparam^{n-1})$, first find $\beta_h^n \in \DiscSpace{k}(\surf_\meshparam^{n-1})$ such that
\begin{subequations}
\begin{align}
    \frac{3}{2}\left(M + \frac{\mu \xi^2}{2} \right) \InnerApprox{\beta_h^{n} + \nabla \beta_h^{n} \cdot (\Vb - \frac{1}{\tau}\paramUpdate}{\psi_h} + \frac{3}{2}L \InnerApprox{\nabla \beta_h^n}{\nabla \psi_h} &\nonumber\\ - \frac{9}{2}L\InnerApprox{\beta_h^n \norm{\shop}}{\psi_h} + \frac{3}{2}\InnerApprox{(2a + (b + 3c \beta_h^{n-1})\beta^{n-1}_h)\beta_h^n}{\psi_h} &= \InnerApprox{\beta_h^{n-1}}{\psi_h}
\end{align}
for all $\psi_h \in \DiscSpace{k}(\surf_h(t_{n-1})$, then find $(\Vb_\meshparam^{n},p_\meshparam^{n}, \meanc_h^{n},\paramUpdate^{n}, \lambda)\in[\DiscSpaceVector{3}\times \DiscSpace{2}\times \DiscSpace{3} \times \DiscSpaceVector{3}\times \R]$, such that
\begin{align}
      \coeffIF \InnerApprox{
      \tsigmab(\Vb_\meshparam^{n})}{\nablaC\Wb_\meshparam} + \gamma \InnerApprox{\Vb_\meshparam^n}{\Wb_\meshparam}\nonumber
    + \InnerApprox{p_h^n}{\DivC\Wb_\meshparam} &\\
    + \lambda \InnerApprox{\normal_h^{n-1}}{\Wb_\meshparam}
    - \frac{9}{2}L\InnerApprox{\left(\beta^n_h\right)^2\left(\meanc^n_h \shop  + \normal \otimes \nabla \meanc^n_h \right)}{\nablaC \Wb_h}&= \frac{3}{2}L\InnerApprox{\nabla \beta^n_h \otimes \nabla \beta^n_h + 2 \beta^n_h \normal \otimes \shop \nabla \beta^n_h}{\nablaC \Wb_h}
          \\
      \InnerApprox{\Vb_\meshparam^{n} \cdot \normal_h^{n-1}}{\mu} &= 0
      \\
      \InnerApprox{\DivC\Vb_\meshparam^{n}}{q_\meshparam} &= 0
            \\
      \InnerApprox{\paramUpdate^n_h \cdot \normal^{n-1}_h}{\mathcal{G}_\meshparam} - \tau\InnerApprox{\Vb_\meshparam^n\cdot \normal^{n-1}_h}{\mathcal{G}_\meshparam} &= 0
    \\
      \label{eq::discStructuralEquation}
     \InnerApprox{\meanc_h^n \normal_h^{n-1}}{\paramUpdateTest_h}
      + \InnerApprox{\nablaC \paramUpdate^n}{\nablaC \paramUpdateTest_h}
      &=
    - \InnerApprox{\nablaC \para^{n-1}}{\nablaC \paramUpdateTest_h}
    \end{align}
    for all $(\Wb_\meshparam, q_\meshparam, \mathcal{G}_\meshparam, \paramUpdateTest_h, \mu )\in[\DiscSpaceVector{3}\times \DiscSpace{2}\times \DiscSpace{3}\times\DiscSpaceVector{3}\times \R](\surf_h^{n-1})$.
    The new parametrization is then given by
\begin{equation}
    \para_\meshparam^n=\para_\meshparam^{n-1} + \paramUpdate_\meshparam^n,
\end{equation}
\end{subequations}
which has been exploited in \eqref{eq::discStructuralEquation} using the `BGN` method \cite{barrett2008parametric}. Above, $\InnerApprox{\cdot}{\cdot}$ denotes the $\mathrm{L}^2$ inner product on the discrete surface $\surf_h$.

\section*{Author Contributions}

A.V. designed the research. I.N. derived the model. M.P. developed and implemented the algorithms and carried out all simulations and analyzed the data. I.N. M.P. and A.V. wrote the article.

\section*{Acknowledgments}

This work was supported by the German Research Foundation (DFG) through the research unit FOR3013, ``Vector- and Tensor-Valued Surface PDEs,'' within project TP01, ``Numerical methods for surface fluids'' (project number VO 899/29-2). We acknowledge computing resources provided by JSC within project MORPH and by ZIH within project WIR.

\printbibliography

@Article{NitschkeSischkaVoigt_JoFM_2026_Hlcmflb,
  author    = {Nitschke, Ingo and Sischka, Jan Magnus and Voigt, Axel},
  title     = {Hydrodynamic liquid crystal models for lipid bilayers},
  journal   = {Journal of Fluid Mechanics},
  year      = {2026},
  volume    = {1034},
  pages     = {A5},
  month     = Apr,
  issn      = {1469-7645},
  doi       = {10.1017/jfm.2026.11474},
  publisher = {Cambridge University Press (CUP)},
}

@article{torres2019modelling,
  title={Modelling fluid deformable surfaces with an emphasis on biological interfaces},
  author={Torres-S{\'a}nchez, Alejandro and Mill{\'a}n, Daniel and Arroyo, Marino},
  journal={Journal of Fluid Mechanics},
  volume={872},
  pages={218--271},
  year={2019},
  publisher={Cambridge University Press}
}

@article{arroyo2009relaxation,
  title={Relaxation dynamics of fluid membranes},
  author={Arroyo, Marino and DeSimone, Antonio},
  journal={Physical Review E},
  volume={79},
  number={3},
  pages={031915},
  year={2009},
  publisher={APS}
}

@article{reuther2020numerical,
  title={A numerical approach for fluid deformable surfaces},
  author={Reuther, Sebastian and Nitschke, Ingo and Voigt, Axel},
  journal={Journal of Fluid Mechanics},
  volume={900},
  pages={R8},
  year={2020},
  publisher={Cambridge University Press}
}

@article{krause2023numerical,
  title={A numerical approach for fluid deformable surfaces with conserved enclosed volume},
  author={Krause, Veit and Voigt, Axel},
  journal={Journal of Computational Physics},
  volume={486},
  pages={112097},
  year={2023},
  publisher={Elsevier}
}

@article{olshanskii2023equilibrium,
  title={On equilibrium states of fluid membranes},
  author={Olshanskii, Maxim A},
  journal={Physics of Fluids},
  volume={35},
  number={6},
  year={2023},
  pages={062111},
  publisher={AIP Publishing}
}

@article{nitschke2025beris,
  title={Beris-Edwards models on evolving surfaces: A Lagrange-D'Alembert approach},
  author={Nitschke, Ingo and Voigt, Axel},
  journal={Advances in Differential Equations},
  volume={30},
  number={5/6},
  pages={335--420},
  year={2025},
  publisher={Khayyam Publishing}
}

@article{PBV25,
  title = {Self-Avoiding Fluid Deformable Surfaces},
  author = {Porrmann, Maik and Bartels, S{\"o}ren and Voigt, Axel},
  year = {2026},
  month = {oct},
  journal = {Interfaces and Free Boundaries, Mathematical Analysis, Computation and Applications},
  issn = {1463-9963, 1463-9971},
  doi = {10.4171/ifb/583},
  urldate = {2026-10-29},
  langid = {english}
}

@article{garcke2025parametric,
  author = {Garcke, Harald and Nürnberg, Robert},
  title = {A Parametric Finite Element Method for the Incompressible Navier–Stokes Equations on an Evolving Surface},
  journal = {Communications in Computational Physics},
  year = {2026},
  volume = {40},
  number = {4},
  pages = {992-1018},
  doi = {10.4208/cicp.OA-2025-0221}
}

@article{sauer2025curvilinear,
  title={A curvilinear surface ALE formulation for self-evolving {N}avier--{S}tokes manifolds--general theory and analytical solutions},
  author={Sauer, Roger A},
  journal={Journal of Fluid Mechanics},
  volume={1016},
  pages={A34},
  year={2025},
  publisher={Cambridge University Press}
}

@article{sahu2025arbitrary,
  title={Arbitrary {L}agrangian--{E}ulerian finite element method for lipid membranes},
  author={Sahu, Amaresh},
  journal={Journal of Fluid Mechanics},
  volume={1020},
  pages={A46},
  year={2025},
  publisher={Cambridge University Press}
}

@article{NV24,
    author = {Nitschke, Ingo and Voigt, Axel},
    doi = {10.1098/rspa.2024.0380},
    journal = {Proc. Roy. Soc. A},
    pages = {20240380},
    title = {Active nematodynamics on deformable surfaces},
    volume = {481},
    year = {2025}
}

@article{igel2026streamfunctionformulationsurfacestokeshelfrich,
      title={A stream-function formulation for the Surface Stokes-Helfrich model - Surface Finite Element discretization and validation},
      author={Enno Igel and Maik Porrmann and Axel Voigt},
      year={2026},
      journal = {arXiv preprint arXiv:2909.02347}
      }

@Article{Helfrich_ZfNC_1973,
  author    = {Helfrich, W},
  title     = {Elastic properties of lipid bilayers: Theory and possible experiments},
  journal   = {Zeitschrift für Naturforschung C},
  year      = {1973},
  volume    = {28},
  number    = {11–12},
  pages     = {693--703},
  month     = dec,
  issn      = {0939-5075},
  doi       = {10.1515/znc-1973-11-1209},
  publisher = {Walter de Gruyter GmbH},
}

@Article{Alexe-Ionescu_PLA_1993,
  author    = {Alexe-Ionescu, A.L.},
  title     = {Flexoelectric polarization and second order elasticity for nematic liquid crystals},
  journal   = {Physics Letters A},
  year      = {1993},
  volume    = {180},
  number    = {6},
  pages     = {456--460},
  month     = sep,
  issn      = {0375-9601},
  doi       = {10.1016/0375-9601(93)90299-f},
  publisher = {Elsevier BV},
}

@Article{BarberoDozovPalierneDurand_PRL_1986,
  author    = {Barbero, G. and Dozov, I. and Palierne, J. F. and Durand, G.},
  title     = {Order electricity and surface orientation in nematic liquid crystals},
  journal   = {Physical Review Letters},
  year      = {1986},
  volume    = {56},
  number    = {19},
  pages     = {2056--2059},
  month     = may,
  issn      = {0031-9007},
  doi       = {10.1103/physrevlett.56.2056},
  publisher = {American Physical Society (APS)},
}

@Article{GoversVertogen_PRA_1984,
  author    = {Govers, E. and Vertogen, G.},
  title     = {Elastic continuum theory of biaxial nematics},
  journal   = {Physical Review A},
  year      = {1984},
  volume    = {30},
  number    = {4},
  pages     = {1998--2000},
  month     = oct,
  issn      = {0556-2791},
  doi       = {10.1103/physreva.30.1998},
  publisher = {American Physical Society (APS)},
}

@article{deserno2015fluid,
  title={Fluid lipid membranes: From differential geometry to curvature stresses},
  author={Deserno, Markus},
  journal={Chemistry and Physics of Lipids},
  volume={185},
  pages={11--45},
  year={2015},
  publisher={Elsevier}
}

@article{chakraborty2020cholesterol,
  title={How cholesterol stiffens unsaturated lipid membranes},
  author={Chakraborty, Saptarshi and Doktorova, Milka and Molugu, Trivikram R and Heberle, Frederick A and Scott, Haden L and Dzikovski, Boris and Nagao, Michihiro and Stingaciu, Laura-Roxana and Standaert, Robert F and Barrera, Francisco N and others},
  journal={Proceedings of the National Academy of Sciences},
  volume={117},
  number={36},
  pages={21896--21905},
  year={2020},
  publisher={National Academy of Sciences}
}

@article{BASSEREAU201447,
title = {Bending lipid membranes: Experiments after W. Helfrich's model},
journal = {Advances in Colloid and Interface Science},
volume = {208},
pages = {47-57},
year = {2014},
author = {Patricia Bassereau and Benoit Sorre and Aurore Lèvy},
}

@article{PhysRevE.80.021931,
  title = {Effect of cholesterol on structural and mechanical properties of membranes depends on lipid chain saturation},
  author = {Pan, Jianjun and Tristram-Nagle, Stephanie and Nagle, John F.},
  journal = {Phys. Rev. E},
  volume = {80},
  issue = {2},
  pages = {021931},
  numpages = {12},
  year = {2009},
  month = {Aug},
}

@article{dimova2014recent,
  title={Recent developments in the field of bending rigidity measurements on membranes},
  author={Dimova, Rumiana},
  journal={Advances in Colloid and Interface Science},
  volume={208},
  pages={225--234},
  year={2014},
  publisher={Elsevier}
}

@article{gracia2010effect,
  title={Effect of cholesterol on the rigidity of saturated and unsaturated membranes: Fluctuation and electrodeformation analysis of giant vesicles},
  author={Gracia, Ruben Serral and Bezlyepkina, Natalya and Knorr, Roland L and Lipowsky, Reinhard and Dimova, Rumiana},
  journal={Soft Matter},
  volume={6},
  number={7},
  pages={1472--1482},
  year={2010},
  publisher={The Royal Society of Chemistry}
}

@article{doktorova2017determination,
  title={Determination of bending rigidity and tilt modulus of lipid membranes from real-space fluctuation analysis of molecular dynamics simulations},
  author={Doktorova, M and Harries, D},
  journal={Physical Chemistry Chemical Physics},
  volume={19},
  number={25},
  pages={16806--16818},
  year={2017},
  publisher={The Royal Society of Chemistry}
}

@article{rupp2026gradient,
  title={Gradient flow dynamics for cell membranes in the Canham--Helfrich model},
  author={Rupp, Fabian and Scharrer, Christian and Schlierf, Manuel},
  journal={Archive for Rational Mechanics and Analysis},
  volume={250},
  number={5},
  pages={81},
  year={2026},
  publisher={Springer}
}

@article{kusner2023canham,
  title={On the Canham problem: Bending energy minimizers for any genus and isoperimetric ratio},
  author={Kusner, Robert and McGrath, Peter},
  journal={Archive for Rational Mechanics and Analysis},
  volume={247},
  number={1},
  pages={10},
  year={2023},
  publisher={Springer}
}

@article{mondino2020existence,
  title={Existence and regularity of spheres minimising the Canham--Helfrich energy},
  author={Mondino, Andrea and Scharrer, Christian},
  journal={Archive for Rational Mechanics and Analysis},
  volume={236},
  number={3},
  pages={1455--1485},
  year={2020},
  publisher={Springer}
}

@article{du2004phase,
  title={A phase field approach in the numerical study of the elastic bending energy for vesicle membranes},
  author={Du, Qiang and Liu, Chun and Wang, Xiaoqiang},
  journal={Journal of Computational Physics},
  volume={198},
  number={2},
  pages={450--468},
  year={2004},
  publisher={Elsevier}
}

@article{barrett2008parametric,
  title={Parametric approximation of Willmore flow and related geometric evolution equations},
  author={Barrett, John W and Garcke, Harald and N{\"u}rnberg, Robert},
  journal={SIAM Journal on Scientific Computing},
  volume={31},
  number={1},
  pages={225--253},
  year={2008},
  publisher={SIAM}
}

@article{elliott2010modeling,
  title={Modeling and computation of two phase geometric biomembranes using surface finite elements},
  author={Elliott, Charles M and Stinner, Bj{\"o}rn},
  journal={Journal of Computational Physics},
  volume={229},
  number={18},
  pages={6585--6612},
  year={2010},
  publisher={Elsevier}
}

@article{aland2014diffuse,
  title={Diffuse interface models of locally inextensible vesicles in a viscous fluid},
  author={Aland, Sebastian and Egerer, Sabine and Lowengrub, John and Voigt, Axel},
  journal={Journal of Computational Physics},
  volume={277},
  pages={32--47},
  year={2014},
  publisher={Elsevier}
}

@article{PhysRevE.47.461,
  title = {Shape equations of the axisymmetric vesicles},
  author = {Jian-Guo, Hu and Zhong-Can, Ou-Yang},
  journal = {Phys. Rev. E},
  volume = {47},
  issue = {1},
  pages = {461--467},
  numpages = {0},
  year = {1993},
  month = {Jan},
  publisher = {American Physical Society},
}

@article{PhysRevE.49.4728,
  title = {Shape equations for axisymmetric vesicles: A clarification},
  author = {J\"ulicher, Frank and Seifert, Udo},
  journal = {Phys. Rev. E},
  volume = {49},
  issue = {5},
  pages = {4728--4731},
  numpages = {0},
  year = {1994},
  month = {May},
  publisher = {American Physical Society},
}

@article{seifertShapeTransformationsVesicles1991,
  title = {Shape transformations of vesicles: {{Phase}} diagram for spontaneous-curvature and bilayer-coupling models},
  author = {Seifert, U. and Berndl, K. and Lipowsky, R.},
  year = {1991},
  journal = {Physical Review A},
  volume = {44},
  pages = {1182--1202},
}

@article{faizi2022vesicle,
  title={A vesicle microrheometer for high-throughput viscosity measurements of lipid and polymer membranes},
  author={Faizi, Hammad A and Dimova, Rumiana and Vlahovska, Petia M},
  journal={Biophysical Journal},
  volume={121},
  number={6},
  pages={910--918},
  year={2022},
  publisher={Elsevier}
}

@article{honerkamp2013membrane,
  title={Membrane viscosity determined from shear-driven flow in giant vesicles},
  author={Honerkamp-Smith, Aurelia R and Woodhouse, Francis G and Kantsler, Vasily and Goldstein, Raymond E},
  journal={Physical Review Letters},
  volume={111},
  number={3},
  pages={038103},
  year={2013},
  publisher={APS}
}

@article{bachini2023derivation,
  title={Derivation and simulation of a two-phase fluid deformable surface model},
  author={Bachini, Elena and Krause, Veit and Nitschke, Ingo and Voigt, Axel},
  journal={Journal of Fluid Mechanics},
  volume={977},
  pages={A41},
  year={2023},
  publisher={Cambridge University Press}
}

@article{sischka2025two,
  title={Two-phase fluid deformable surfaces with constant enclosed volume and phase-dependent bending and Gaussian rigidity},
  author={Sischka, Jan Magnus and Voigt, Axel},
  journal={Computer Methods in Applied Mechanics and Engineering},
  volume={445},
  pages={118166},
  year={2025},
  publisher={Elsevier}
}

@article{vey2007amdis,
  title={{AMDiS}: adaptive multidimensional simulations},
  author={Vey, S. and Voigt, A.},
  journal={Comput. Vis. Sci.},
  volume={10},
  pages={57--67},
  year={2007},
}

@article{witkowski2015software,
  title={Software concepts and numerical algorithms for a scalable adaptive parallel finite element method},
  author={Witkowski, T. and Ling, S. and Praetorius, S. and Voigt, A.},
  journal={Adv. Comput. Math.},
  volume={41},
  pages={1145--1177},
  year={2015},
}

@misc{AMDiS:2.10,
  key    = {AMDiS},
  title  = {{AMDiS}: {A}daptive {M}ulti-{D}imensional {S}imulations},
  note   = {v2.10},
  howpublished = {\url{https://gitlab.com/amdis/amdis}},
}

@article{Dune2.10,
  title = {The {{Distributed}} and {{Unified Numerics Environment}} ({{DUNE}}), {{Version}} 2.10},
  author = {Blatt, M. and Burbulla, S. and Burchardt, A. and Dedner, A. and Engwer, C. and Gr{\"a}ser, C. and Gr{\"u}ninger, C. and Kl{\"o}fkorn, R. and Koch, T. and Ospina De Los R{\'i}os, S. and Praetorius, S. and Sander, O.},
  year = {2025},
  journal={arXiv preprint arXiv:2506.23558},
  doi = {10.48550/arXiv.2506.23558},
}

@Article{SchieleTrimper_pssb_1983,
  author    = {Schiele, K. and Trimper, S.},
  title     = {On the elastic constants of a nematic liquid crystal},
  journal   = {physica status solidi (b)},
  year      = {1983},
  volume    = {118},
  number    = {1},
  pages     = {267--274},
  month     = Jul,
  issn      = {1521-3951},
  doi       = {10.1002/pssb.2221180132},
  publisher = {Wiley},
}

@Article{BerremanMeiboom_PRA_1984,
  author    = {Berreman, Dwight W. and Meiboom, Saul},
  title     = {Tensor representation of Oseen-Frank strain energy in uniaxial cholesterics},
  journal   = {Physical Review A},
  year      = {1984},
  volume    = {30},
  number    = {4},
  pages     = {1955--1959},
  month     = Oct,
  issn      = {0556-2791},
  doi       = {10.1103/physreva.30.1955},
  publisher = {American Physical Society (APS)},
}

@article{nitschke2018nematic,
  title={Nematic liquid crystals on curved surfaces: A thin film limit},
  author={Nitschke, Ingo and Nestler, Michael and Praetorius, Simon and L{\"o}wen, Hartmut and Voigt, Axel},
  journal={Proc. Roy. Soc. A},
  volume={474},
  number={2214},
  pages={20170686},
  year={2018}
}

@article{NITSCHKE2022104428,
title = {Observer-invariant time derivatives on moving surfaces},
journal = {Journal of Geometry and Physics},
volume = {173},
pages = {104428},
year = {2022},
author = {Ingo Nitschke and Axel Voigt},
}

@article{nitschke2023tangential,
  title={Tangential tensor fields on deformable surfaces—how to derive consistent $L^2$-gradient flows},
  author={Nitschke, Ingo and Sadik, Souhayl and Voigt, Axel},
  journal={IMA Journal of Applied Mathematics},
  volume={88},
  number={6},
  pages={917--958},
  year={2023},
  publisher={Oxford University Press}
}

@article{nitschke2023tensorial,
    author = {Nitschke, Ingo and Voigt, Axel},
    doi = {10.1016/j.geomphys.2023.105002},
    journal = {J. Geometry Physics},
    pages = {105002},
    title = {Tensorial time derivatives on moving surfaces: General concepts and a specific application for surface Landau-de Gennes models},
    volume = {194},
    year = {2023}
}

@article{backofen2026scalar,
  author  = {Backofen, Rainer and Nitschke, Ingo and Voigt, Axel},
  title   = {Scalar Truesdell time derivative and $(L^{2},H^{-1})$--surface gradient flows},
  journal = {arXiv preprint arXiv:2604.08186},
  year    = {2026},
}

@article{sischka2026,
      title={Fluid deformable surfaces with variable thickness - a Surface Shallow-Water-Helfrich model},
      author={Jan Magnus Sischka and Axel Voigt},
      year={2026},
      journal={arXiv preprint arXiv:2610.11903},
}

@article{dune-curvedGrid,
  title = {Dune-{{CurvedGrid}} - {{A Dune}} module for surface parametrization},
  author = {Praetorius, S. and Stenger, F.},
  year = {2022},
  journal = {Archive of Numerical Software},
  volume = {6},
  doi = {10.11588/ans.2022.1.75917},
}

@book{duneBook,
  title = {{{DUNE}} --- {{The Distributed}} and {{Unified Numerics Environment}}},
  author = {Sander, O.},
  year = {2020},
  series = {Lecture {{Notes}} in {{Computational Science}} and {{Engineering}}},
  volume = {140},
  publisher = {Springer International Publishing},
  address = {Cham},
  doi = {10.1007/978-3-030-59702-3},
}

@article{dune-alugrid,
  title = {The {{DUNE-ALUGrid module}}.},
  author = {Alk{\"a}mper, M. and Dedner, A. and Kl{\"o}fkorn, R. and Nolte, M.},
  year = {2016},
  journal = {Archive of Numerical Software},
  volume = {4},
  pages = {1--28},
  doi = {10.11588/ans.2016.1.23252},
}

@article{dune-functions,
  title = {Concepts for {{Composing Finite Element Function Space Bases}}},
  author = {Engwer, C. and Gr{\"a}ser, C. and M{\"u}thing, S. and Praetorius, S. and Sander, O.},
  year = {2025},
  journal={arXiv preprint arXiv:2508.10125},
  doi = {10.48550/arXiv.2508.10125},
}

@MISC{eigenweb,
  author = {G. Guennebaud and B. Jacob and others},
  title = {Eigen v3},
  howpublished = {http://eigen.tuxfamily.org},
  year = {2010}
 }

@manual{intel2025oneapi,
  title        = {Intel oneAPI},
  author       = {{Intel Corporation}},
  year         = {2025},
  note         = {\url{https://www.intel.com/content/www/us/en/developer/tools/oneapi/overview.html}}
}

@book{Paraview,
  author = {Ayachit, Utkarsh},
  title = {The ParaView Guide: A parallel visualization application},
  year = {2015},
  isbn = {1930934300},
  publisher = {Kitware, Inc.},
  address = {Clifton Park, NY, USA},
}

@article{simons2011membrane,
  title={Membrane organization and lipid rafts},
  author={Simons, Kai and Sampaio, Julio L},
  journal={Cold Spring Harbor perspectives in biology},
  volume={3},
  number={10},
  pages={a004697},
  year={2011},
  publisher={Cold Spring Harbor Laboratory Press}
}

@article{USeifert_1993,
year = {1993},
month = {jul},
publisher = {},
volume = {23},
number = {1},
pages = {71},
author = {U. Seifert and S. A. Langer},
title = {Viscous modes of fluid bilayer membranes},
journal = {Europhysics Letters},
}

@article{jiang2000phase,
  title={Phase separation and shape deformation of two-phase membranes},
  author={Jiang, Y. and Lookman, T. and Saxena, A.},
  journal={Physical Review E},
  volume={61},
  pages={R57},
  year={2000},
  publisher={APS}
}

@article{julicher1996shape,
  title={Shape transformations of vesicles with intramembrane domains},
  author={J{\"u}licher, F. and Lipowsky, R.},
  journal={Physical Review E},
  volume={53},
  pages={2670-2693},
  year={1996},
  publisher={APS}
}

@article{kumar2001budding,
  title={Budding dynamics of multicomponent membranes},
  author={Kumar, P. B. S. and Gompper, G. and Lipowsky, R.},
  journal={Phys. Rev. Lett.},
  volume={86},
  pages={3911},
  year={2001},
  publisher={APS}
}

@article{lowengrub2009phase,
  title={Phase-field modeling of the dynamics of multicomponent vesicles: Spinodal decomposition, coarsening, budding, and fission},
  author={Lowengrub, John S and R{\"a}tz, Andreas and Voigt, Axel},
  journal={Physical Review E},
  volume={79},
  number={3},
  pages={031926},
  year={2009},
  publisher={APS}
}

@article{barrett2017finite,
  title={Finite element approximation for the dynamics of fluidic two-phase biomembranes},
  author={Barrett, John W and Garcke, Harald and N{\"u}rnberg, Robert},
  journal={ESAIM: Mathematical Modelling and Numerical Analysis},
  volume={51},
  number={6},
  pages={2319--2366},
  year={2017}
}

@InProceedings{Mecke2000,
author="Mecke, Klaus R.",
editor="Mecke, Klaus R.
and Stoyan, Dietrich",
title="Additivity, Convexity, and Beyond: Applications of Minkowski Functionals in Statistical Physics",
booktitle="Statistical Physics and Spatial Statistics",
year="2000",
publisher="Springer Berlin Heidelberg",
address="Berlin, Heidelberg",
pages="111--184",
isbn="978-3-540-45043-6"
}

@article{sischka2025influence,
  title={The influence of higher order geometric terms on the asymmetry and dynamics of membranes},
  author={Sischka, Jan Magnus and Nitschke, Ingo and Voigt, Axel},
  journal={Faraday Discussions},
  volume={259},
  pages={454--474},
  year={2025},
  publisher={The Royal Society of Chemistry}
}

@article{Jaehnig_1981,
  title={Critical effects from lipid-protein interaction in membranes. I. Theoretical description},
  author={F. Jähnig},
  journal={Biophysical Journal},
  volume={36},
  pages={329--345},
  year={1981},
}

@article{krause2024wrinkling,
  title={Wrinkling of fluid deformable surfaces},
  author={Krause, Veit and Voigt, Axel},
  journal={Journal of the Royal Society Interface},
  volume={21},
  number={216},
  pages={20240056},
  year={2024}
}

@Article{Seelig_QRoB_1977_Dmrtaatlm,
  author    = {Seelig, Joachim},
  title     = {Deuterium magnetic resonance: theory and application to lipid membranes},
  journal   = {Quarterly Reviews of Biophysics},
  year      = {1977},
  volume    = {10},
  number    = {3},
  pages     = {353--418},
  month     = Aug,
  issn      = {1469-8994},
  doi       = {10.1017/s0033583500002948},
  publisher = {Cambridge University Press (CUP)},
}

@Article{MorvanGrelardDufourc_BeBAB-B_2025_OpimFJSp,
  author    = {Morvan, Estelle and Grélard, Axelle and Dufourc, Erick J.},
  title     = {Order parameters in membranes: Following Joachim Seelig’s path},
  journal   = {Biochimica et Biophysica Acta (BBA) - Biomembranes},
  year      = {2025},
  volume    = {1867},
  number    = {8},
  pages     = {184449},
  month     = Dec,
  issn      = {0005-2736},
  doi       = {10.1016/j.bbamem.2025.184449},
  publisher = {Elsevier BV},
}



\end{document}